\documentclass{elsart}

\usepackage{epsfig}

\usepackage{graphicx}
\usepackage{amssymb}
\usepackage[figuresright]{rotating}

\begin{document}


\begin{frontmatter}

\title{Electromagnetic Probes at RHIC-II}

\author{G.~David$^{a}$,
R.~Rapp$^{b}$
and Z.~Xu$^{a}$}
\address[a]{Physics Department, Brookhaven National Laboratory, Upton, NY
11793-5000, USA}
\address[b]{Cyclotron Institute and Physics Department, 
Texas A\&M University, College Station, TX 77843-3366, USA}


\vspace{1in}

\begin{abstract}
We summarize how future measurements of
electromagnetic (EM) probes at the Relativistic Heavy Ion Collider
(RHIC), in connection with theoretical analysis, can advance our
understanding of strongly interacting matter at high energy densities
and temperatures.
After a brief survey of the important role that EM probes data have
played at the Super Proton Synchrotron (SPS, CERN) and RHIC to date,
we identify key physics objectives and observables that remain to be
addressed to characterize the (strongly interacting) Quark-Gluon Plasma
(sQGP) and associated transition properties at RHIC. These include
medium modifications of vector mesons via low-mass dileptons, a
temperature measurement of the hot phases via continuum radiation,
as well as $\gamma$-$\gamma$ correlations to characterize early source
sizes.
We outline strategies to establish microscopic matter and
transition properties such as the number of degrees of freedom in
the sQGP, the origin of hadron masses and manifestations of
chiral symmetry restoration, which will require accompanying but
rather well-defined advances in theory.
Increased experimental precision, an order of magnitude higher statistics
than currently achievable, as well as a detailed scan of colliding
species and energies are mandatory to
discriminate between theoretical interpretations.
This increased precision can be achieved through hardware upgrades to the 
large RHIC detectors (PHENIX and STAR) along with at least a factor of ten 
increase in luminosity over the next few years, as envisioned for RHIC-II.
\end{abstract}

\begin{keyword}
direct photon \sep isolated photon \sep quark-gluon plasma \sep
relativistic heavy ion collision

\PACS 25.75.-q \sep 25.75.Nq \sep 12.38.Mh \sep 13.85.Qk
\end{keyword}

\end{frontmatter}

\newpage

\tableofcontents

\newpage

\section{INTRODUCTION}
\subsection{Toward Discovery and Characterization of Hot and Dense QCD Matter}
One of the key goals of ultrarelativistic heavy-ion collisions
is the creation of hot and dense strongly interacting matter that:
(i) resembles the conditions in the early universe;
(ii) can be related to the phase diagram of the underlying
theory (Quantum Chromodynamics (QCD)); and
(iii) enables the discovery of new phases.  Recent surveys 
of results from the first three years of data taking by the
four experiments at the Relativistic Heavy Ion Collider
(RHIC)~\cite{Arsene:2004fa,Back:2004je,Adams:2005dq,Adcox:2004mh}
are unanimous in their conclusion that a new form of
matter has been created. This matter is very dense, opaque and
exhibits a high level of collectivity which has largely been attributed
to the expansion of a partonic phase. 
It is inconsistent with naive expectations based on
a weakly-interacting (gas-like)
Quark-Gluon Plasma (wQGP), while it is best described in terms
of a so-called strongly interacting QGP (sQGP) constituting an almost
perfect fluid. Thus, a discovery has been made in a {\it qualitative}
sense, but the properties of this new state of matter remain
under intense debate.

A closer look indeed reveals that we are still quite far from a
{\it coherent} and {\it quantitative} description of the sQGP at RHIC.
The wealth and precision of new data from Run-4 and Run-5~\cite{QM05} is a
first step in this direction.
Measurements of previously inaccessible signals, such as semileptonic
electron-decay spectra, $J/\psi$ production and three-hadron correlations,
and improvements on the statistical and systematic
errors, as well as the range of the existing data, have been achieved. 
In addition, new and ongoing analyses of SPS data, most notably from
NA60~\cite{Damjanovic:2005ni,Arnaldi:2006jq}
(including low- and intermediate-mass dileptons and $J/\psi$ production),
have reached unprecedented levels of precision that now can distinguish
between model predictions which were consistent with earlier data sets.
Furthermore, issues have been raised again, that
seemed to have essentially been settled a few years ago, 
{\em e.g.} the energy-loss mechanism of jet quenching.
It therefore appears fair to say that whereas the {\it existence} of
the new form of matter has been established, we neither understand
its {\it microscopic properties} nor deduced from it convincing signals of
the {\it phase transition} itself. While theory has also made substantial
progress in the last few years by moving from more signal-specific
explanations to a coherent description by connecting different
phenomena and improving theoretical tools, a widely accepted
``grand scheme" encompassing both bulk and microscopic components
has not yet been realized.
As ideas get refined (largely steered by
data), different theoretical predictions often approach each other,
thus increasing the demand for higher quality measurements to
differentiate between them.

 \begin{table}[h]
  \begin{tabular}{|l|cccc|cc|c|} \hline
   Upgrades & High $T$ & QCD...& QGP &  & Spin  & &  Low-$x$ \\ \hline
            &  $e+e-$ & heavy &   jet &  quarkonia &   W &  $\Delta G/G$  & \\
            &   &     flavor &  tomog.  &             &      &         & \\ \hline
   PHENIX   &   &            &          &             &      &         & \\ \hline
   Hadron blind detector
            &   X &         &          &               &     &         & \\
   Vertex Tracker
            &   X   &   X   &     O   &    O  &       &  X   &      O     \\
   Muon Trigger
            &       &       &         &    O  &  X    &      &           \\
   Forward Cal (NCC)
            &       &       &     O   &    O  &  O    &      &      X     \\ \hline
   STAR   &   &            &          &             &      &         & \\ \hline   Time of Flight (ToF)
            &   X    &   O   &     X   &    X  &       &      &           \\
   Heavy Flavor Tracker
            &   X    &   X   &         &    X  &       &      &           \\
   Forward Tracker
            &       &   O   &         &       &  X    &  O   &           \\
   Forward Cal (FMS)
            &       &       &         &       &       &  O   &    X       \\
   DAQ 1000
            &   O    &   O   &     X   &    X  &  O    &  O   &    O      \\ \hline
   RHIC Luminosity   &  O  &  O  &   X   &   X   & O  &  O  & O  \\ \hline
  \end{tabular}
  \vspace{0.3cm}
  \caption{ {\label{tab:axel_table}} Matrix of detector or accelerator
    upgrades {\em vs} physics measurements.  X = upgrade critical for
    measurement.  O = upgrade important for measurement.}
 \end{table}

After very successful operation of RHIC\footnote{By 2004 (Run-4)
the accelerator exceeded design luminosities by a factor of 2.5
both in Au+Au and $pp$ collisions.  RHIC collided 4 different systems
including asymmetric d+Au interactions, collisions 
at 6 different center-of-mass
energies, among others at 19 GeV (Au+Au) and 22 GeV (Cu+Cu), 
establishing the first overlap with the CERN SPS.  In 2007 (Run-7) it
delivered 3260 $\mu$b$^{-1}$ integrated Au+Au luminosity at 200 GeV,
in 2008 (Run-5) the first tests with 9\,GeV and 5 GeV per nucleon
(below injection energy) are planned with Au+Au.}
and its detectors over the first five years of data taking,
we are approaching a point where further
progress requires improved experimental capabilities.
Table~\ref{tab:axel_table} lists important physics topics
that are either beyond our current reach or would be significantly
enhanced with the indicated detector and luminosity upgrades.
These should be put into context with overarching questions in the
investigation of the sQGP and chiral/deconfinement transitions,
emphasizing those for which
electromagnetic probes are particularly relevant, including:
\begin{itemize}
\item{What are the temperatures and corresponding system sizes of
      thermalized matter at its early stages?}
\item{How do hadron properties change in hot and dense matter and how
      are hadron masses generated? How do the medium modifications
      depend on temperature and net/total baryon density?}
\item{Can we deduce signatures of Chiral Symmetry Restoration ($\chi$SR)
      and, if so, how is it realized?}
\item{What are the relevant degrees of freedom in the sQGP?
      Does it harbor bound states and/or resonances?}
\item{How do the medium properties change if the net baryon density is
      increased? Can a QCD critical point be found?}
\item{How does the system reach (local) equilibrium on the apparently short
      time scales required by hydrodynamics?}
\item{What exactly causes jet quenching?}
\end{itemize}
In the remainder of this introductory section, we will first elaborate
in somewhat more detail on the features and strongholds of EM probes
in heavy-ion collisions (Sec.~\ref{sec_unique}).  In particular, we
will discuss the achievements and shortcomings
of the EM probes program at the SPS (Sec.~\ref{sec_sps}) which
will help us to sharpen the case for future RHIC measurements.
In Sec.~\ref{sec_theory}
we lay out the theoretical framework for describing and interpreting
EM probes.  We will also provide predictions for observables and
formulate strategies for deducing information on key properties
of the medium that may not follow from an immediate interpretation
of experimental data.
In Sec.~\ref{sec_obs} we give a brief overview on the current status
of EM observables at RHIC. In Sec.~\ref{sec_future} we then focus
on how future detector and accelerator upgrades can be used and geared toward
answering the above questions.
In Sec.~\ref{sec_concl} we reiterate the main points of this document.

\subsection{Where EM Probes Are Unique or Very Important}
\label{sec_unique}



\begin{figure}
\centering{
\includegraphics[width=0.45\linewidth]{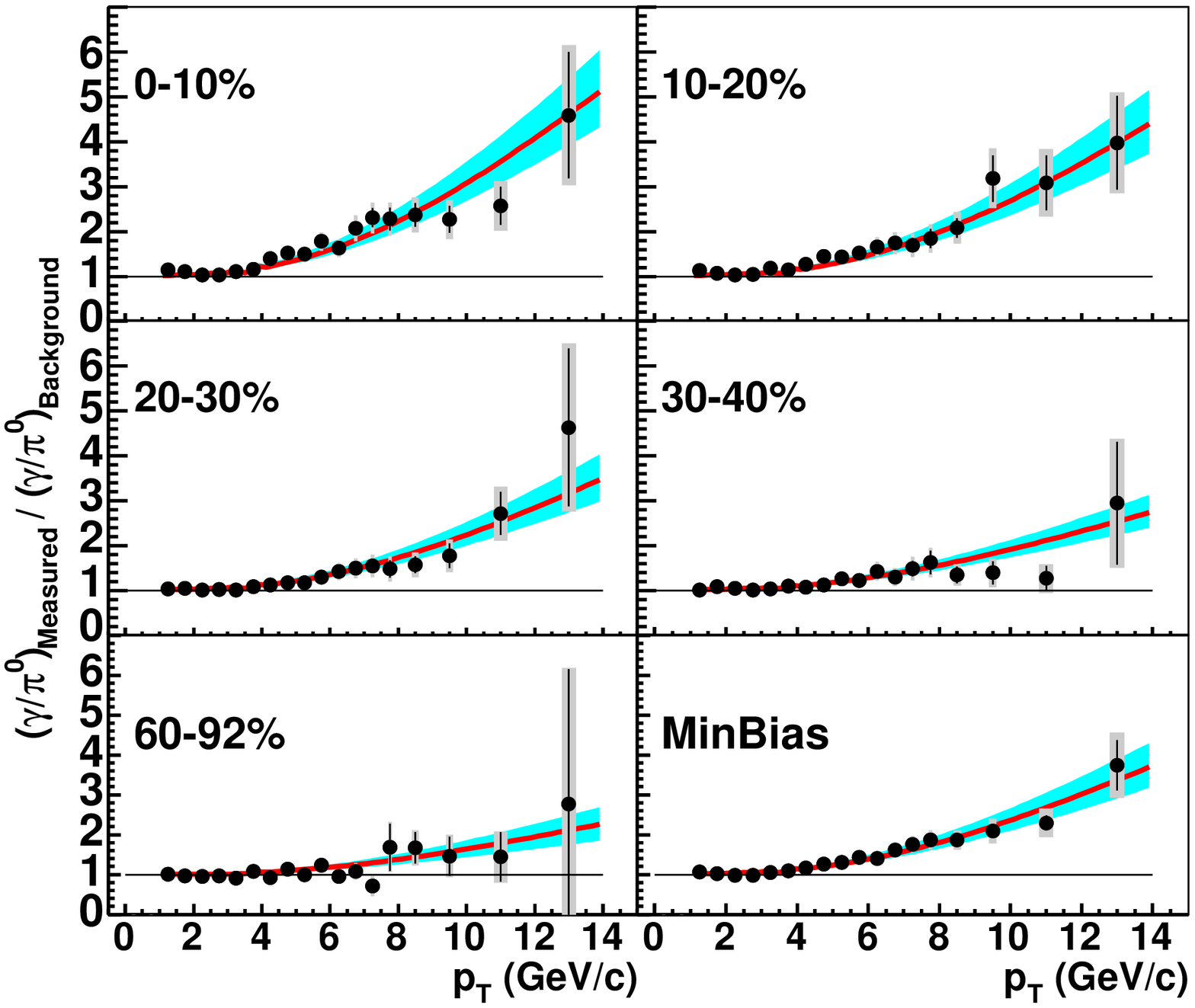}
\includegraphics[width=0.45\linewidth]{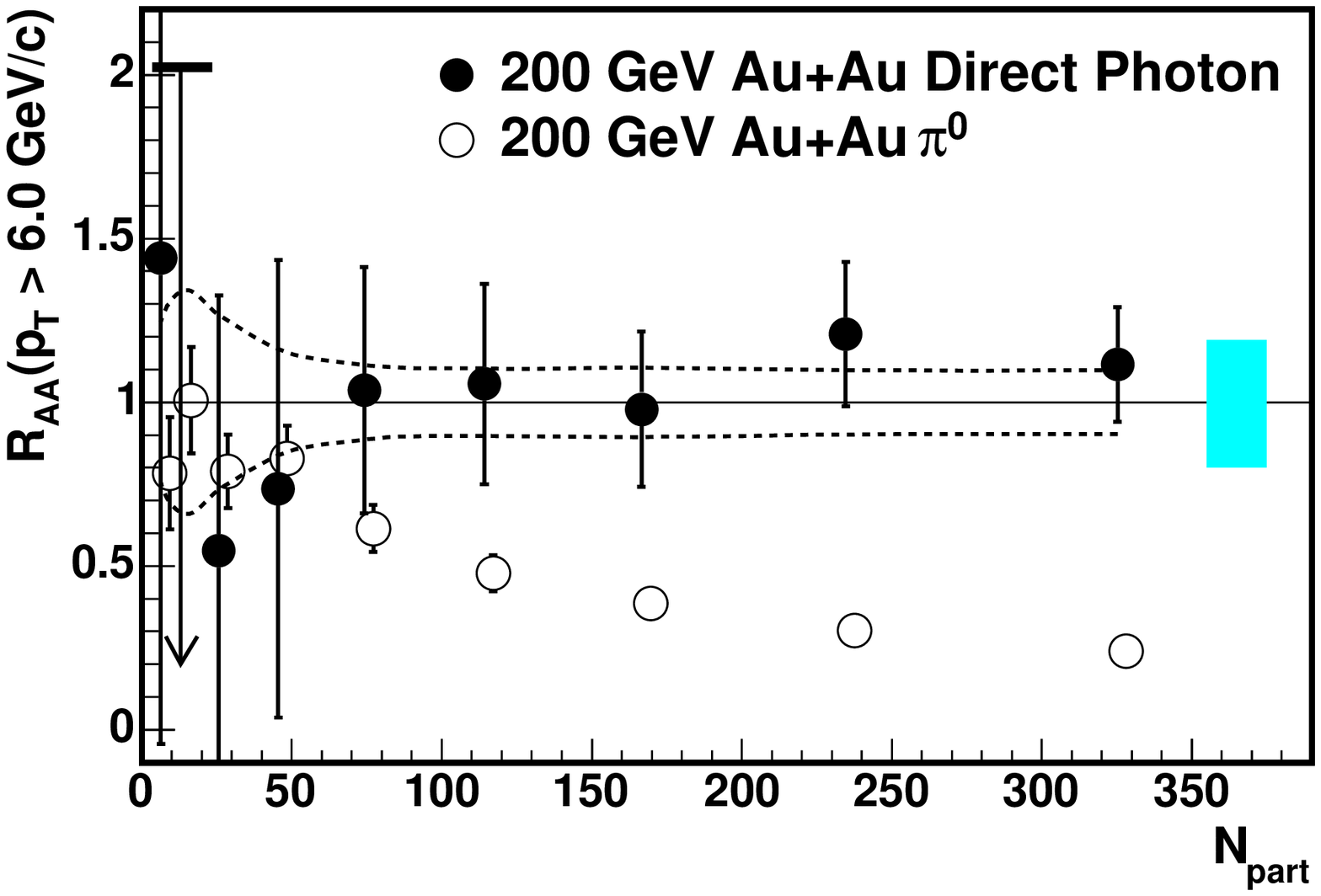}}
\caption[]{Left panel: direct photon excess ratios measured by PHENIX in
$\sqrt{s_{_{NN}}} =200$ GeV Au+Au collisions~\cite{ppg042}.
The curves are the ratios predicted by NLO pQCD~\cite{pQCDgamma}.
Right panel: integrated $R_{AA}$ for photons and $\pi^0$s
as a function of the number of
participants, $N_{\rm part}$, measured by PHENIX in 
$\sqrt{s_{_{NN}}} =200$ GeV Au+Au collisions~\cite{ppg042}.}
\label{fig:ppg042}
\end{figure}

Electromagnetic probes are real, $\gamma$, and virtual photons
(dileptons), $\gamma^* \rightarrow l^+l^-$ where $l=e$ and $\mu$.
They are emitted from the entire reaction volume throughout
the evolution of a heavy-ion collision, from
first impact in primordial (hard) $NN$ collisions until
decays of long-lived hadrons long after strong interactions
have ceased.  
Once created, most of them leave the interaction volume
unchanged due to their negligible final-state interaction with the 
strongly-interacting medium.  Following the usual terminology we
define {\em direct photons} as those that are not decay products of
final state hadrons.  Accordingly, direct photon spectra are defined 
as the spectra remaining after 
subtraction/elimination\footnote{This can be done in a number of ways
  like statistical subtraction, ``tagging'', {\em etc.}}
of photons coming from final-state 
decays (``background'' or ``decay photons''), whereas in
measured dilepton spectra pairs from final state decays
are often included and their estimated contribution shown
separately ("hadronic decay cocktail").  The broad category of
``direct'' photons is then often subdivided according to their
source such as ``prompt'' photons (from hard scattering) ``thermal''
photons, ``jet conversion'' photons, {\em etc.}

Unfortunately, the same property that allows photons to escape,
$\alpha << \alpha_s$, also leads to major experimental
challenges: low rates and large backgrounds from the above-mentioned
late hadron decays ({\em e.g.} Dalitz-decays for dileptons and 
$\pi^0\rightarrow\gamma\gamma$, $\eta\rightarrow\gamma\gamma$, 
{\em etc}. for real photons).

The fact that direct photons from initial hard scatterings escape
the system~\cite{ppg042} is demonstrated in Fig.~\ref{fig:ppg042} where
the measured direct photon excess ratio\footnote{The ratio of
inclusive photons over hadron decay photons. For more 
details on the direct photon excess ratios and $R_{AA}$ see 
Sec~\ref{sec:photonraa}.}
at sufficiently high
transverse momentum, $p_T$, is shown to be consistent with
next-to-leading order (NLO) perturbative QCD (pQCD) calculations,
even in the most central Au+Au collisions.

The interesting physics we are primarily concerned with here is at 
significantly lower $p_T$ and dilepton
invariant masses, $M$ than inferred from
Fig.~\ref{fig:ppg042}.  Thus, in this review, we stress the
importance of precision data at masses and momenta below $\sim 3-4$~GeV.

Electromagnetic probes are {\bf unique}.
\begin{itemize}
\item
They give direct access to the in-medium modifications of hadronic states
(the $\rho$(770), $\omega$(782), and $\phi$(1020) vector mesons) via dilepton
invariant-mass spectra, which can illuminate the nature of hadron mass
generation and thus the origin of $\sim 98$\% of the visible mass in the
universe as well as related changes in the structure of the QCD vacuum,
including chiral symmetry restoration.
Detecting novel nonperturbative (resonant) correlations in the sQGP
at masses above $\sim 1$ GeV is also possible.
\item
They can be used to infer the temperature of the system during its hottest 
phases via direct thermal photon and dilepton radiation.  In addition,
HBT interferometry of thermal photons offers the cleanest
measurement of early system sizes.
\end{itemize}
Indirect consequences of the above studies include
insights into the mechanism of rapid thermalization, {\em e.g.} via resonance
formation in the sQGP.
The effective degrees of freedom can be determined 
if a temperature measurement is complemented with independent
information on the energy or entropy density.
Although not unique, EM~probes are valuable for
(i) disentangling the energy loss mechanism of jets (jet quenching)
by inferring photon radiation off energetic quarks and
establishing the jet energy scale in $\gamma$-hadron jets and
(ii) obtaining complementary information on (early) matter flow by
investigating 
photon elliptic flow, $v_2$,
which, in turn, discriminates between thermal photons and
those radiated off jets. 

In the following sections, we will elaborate on these statements
together with the requirements to measure associated observables
with sufficient accuracy.

\subsection{Theory {\em vs} Experiment at the CERN-SPS}
\label{sec_sps}
\begin{figure}[!t]
\vspace{-0cm}
\begin{minipage}{0.48\linewidth}
\vspace{0.2cm}
\begin{center}
\includegraphics[width=0.93\linewidth]{dlsQH69PbyT250-956.eps}
\end{center}
\end{minipage}
\begin{minipage}{0.48\linewidth}
\vspace{-1.7cm}
\hspace{-0.4cm}
\includegraphics[width=0.88\linewidth]{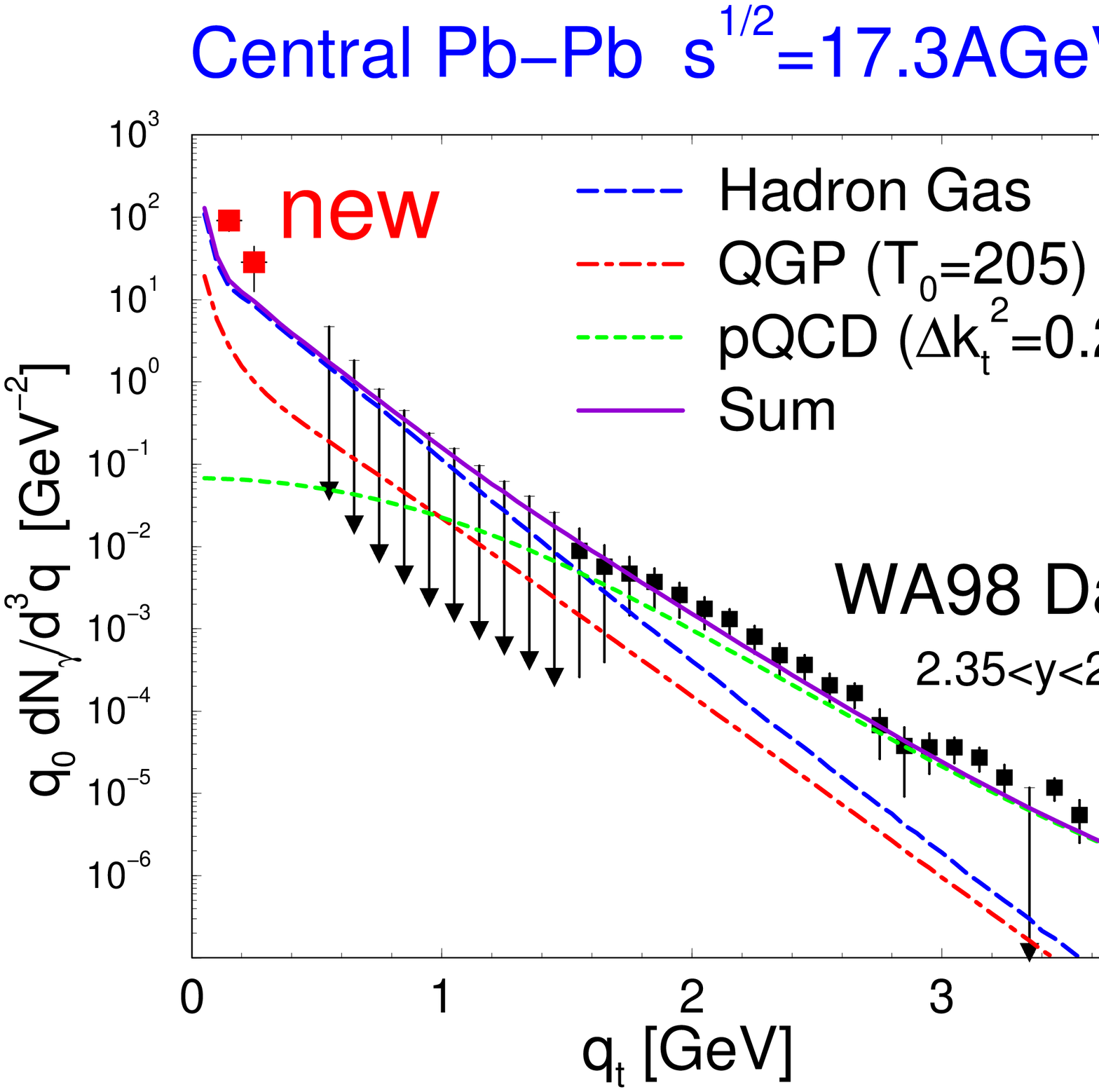}
\end{minipage}
\begin{minipage}{0.48\linewidth}
\begin{center}
\vspace{0.2cm}
\includegraphics[width=0.92\linewidth]{dlPb38031na50.eps}
\end{center}
\vspace{-0.2cm}
\caption{Dilepton spectra at the SPS compared to
calculations~\cite{RW99,RS00} within an expanding thermal
fireball model.
Upper panel: CERES/NA45 low-mass dielectrons~\cite{Agakichiev:2005ai};
lower panel: NA50 intermediate-mass dimuons~\cite{na50-00}. }
\label{fig_spsdilep}
\end{minipage}
\hspace{0.2cm}
\begin{minipage}{0.48\linewidth}
\vspace{-0.1cm}
\hspace{-0cm}
\includegraphics[width=0.98\linewidth,angle=0]{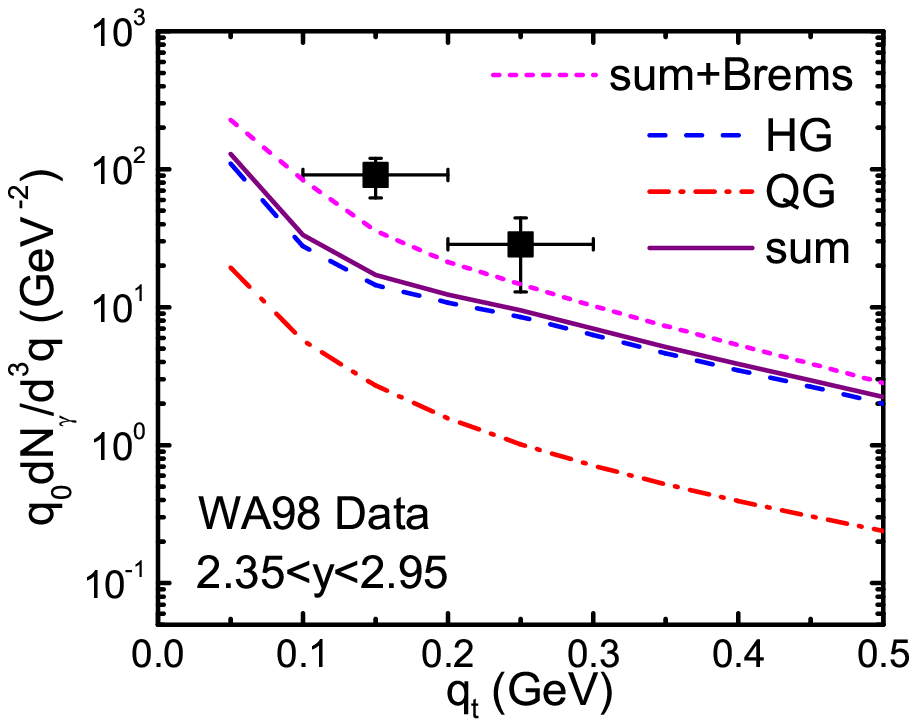}
\vspace{-0.8cm}
\caption{WA98 direct photon spectra~\cite{wa98-00,wa98-lowpt}
compared to thermal emission within the same fireball
model as in Fig.~\ref{fig_spsdilep}
with the same EM~correlator used for dileptons extrapolated 
to the photon point~\cite{Turbide:2003si,Liu:2006zy}.  The lower panel
expands the low $q_t$ region.}
\label{fig_spsphot}
\end{minipage}
\end{figure}

Before beginning a more detailed discussion of the theoretical
aspects underlying a comprehensive electromagnetic probes program
for future RHIC runs, we briefly review the main achievements
at the SPS prior to the new NA60
data~\cite{Damjanovic:2005ni,Arnaldi:2006jq}, discussed
later.  An example of one theoretical approach to electromagnetic emission
simultaneously applied to semi-/central Pb+Au and Pb+Pb data is compiled 
in Figs.~\ref{fig_spsdilep}
and \ref{fig_spsphot}.
The data suggest a common
thermal source with an initial temperature of
$T_0\simeq 210\pm 30$~MeV and a lifetime of about $12\pm 3$~fm/$c$ with a 
thermal freezeout temperature of around $100-120$~MeV.
Importantly, the low-mass dilepton enhancement in the CERES $e^+e^-$
data~\cite{Agakichiev:2005ai} (upper panel of Fig.~\ref{fig_spsdilep}) 
requires substantial medium effects over a sufficiently long
lifetime, on the $\rho$ spectral function.  However, decisive 
discrimination between a dropping-mass scenario~\cite{Brown:1995qt} and a 
substantially broadened
spectral function~\cite{RW99} was not possible since the 
QGP contribution is small, around $10-15$\%. The sensitivity to QGP
radiation increases in the intermediate-mass dimuon spectra of
NA50~\cite{na50-00} (lower panel) where the observed 
factor of two excess over the baseline
charm and Drell-Yan sources can be reasonably accommodated with
thermal radiation containing a $30-50$\% \cite{RS00,Kampf00,KGS02} 
QGP component, the main
evidence for the $T_0$  quoted above. A very similar
decomposition is found in the WA98 direct photon spectra~\cite{wa98-00}
at $q_t\simeq 2$~GeV (upper panel of Fig.~\ref{fig_spsphot}) where $q$ is
the real or virtual photon momentum. 
The recently published low-momentum data~\cite{wa98-lowpt}
(lower panel), extracted via photon HBT methods,
are not easily reconciled with 
theory~\cite{Turbide:2003si,Srivastava:2004xp}.
The inclusion of soft Bremsstrahlung off $\pi\pi$ and $\pi K$
scattering appears to improve the situation~\cite{Liu:2006zy}.
The low-momentum yield is essentially emanating from the later 
stages of the fireball evolution and thus it is proportional to
the total fireball lifetime.

The SPS electromagnetic probe
program was incomplete for a number of reasons ranging from insufficient
statistics to draw conclusions to the lack of experimental cross checks.  A
partial list of open questions is given below.
\begin{itemize}
\item[(a)] There was no decisive discrimination of in-medium $\rho$
modifications (although the recent NA60 
data~\cite{Damjanovic:2005ni,Arnaldi:2006jq} has improved the situation, 
albeit in a smaller system, see below).  In particular, no
systematic excitation function was obtained through an energy
scan.  The only CERES/NA45 low-energy run at $E_{\rm lab}=40$~$A$GeV
\cite{Adamova:2003xx} indicated an increase over the
enhancement at 160~$A$GeV but with large errors.
\item[(b)]
The Cronin enhancement in the
primordial pQCD photons, essential for an accurate
assessment of the thermal photon yield in the $q_t\simeq 2$\,GeV
region, was not determined.
\item[(c)]
Charm dileptons were not explicitly identified, hampering the
temperature extraction in the intermediate-mass dimuon spectra.  However,
NA60 made substantial progress in an intermediate-size
system~\cite{Shahoyan:2007zz}.
\item[(d)]
There was no experimental redundancy to cross-check measurements and create
competition.
\item[(e)]
The QGP contribution to EM probes at the
SPS is perhaps too modest {\em in principle} to reveal itself in a significant
way.
\item[(f)]
The level of early thermalization at the SPS may be insufficient to 
justify thermal approaches in the early phases, as may be indicated
by the lack of elliptic flow in hadron data relative to hydrodynamic
models above $p_T\sim1$\,GeV.
\end{itemize}
We will argue below that {\em all} of these issues can be (are) overcome
at RHIC-II.

\section{THEORETICAL CONSIDERATIONS AND PREDICTIONS}
\label{sec_theory}
\subsection{Objectives and Framework}
\label{sec_frame}
Ample evidence for (early) thermalization of the matter formed
in (semi-/central) $\sqrt{s_{_{NN}}}=200$\,GeV Au+Au collisions
from current RHIC data provides the necessary prerequisite to
validate the study of the QCD phase diagram. So far,
most of the deduced features pertain to bulk matter properties,
including large energy densities, $\epsilon$, well above the
critical density extracted from lattice QCD~\cite{Karsch:2003jg}.

\begin{figure}[!tb]
\vspace{-1cm}
\begin{minipage}[t]{0.300\linewidth}
\hspace{-0.3cm}
\includegraphics[width=0.86\linewidth]{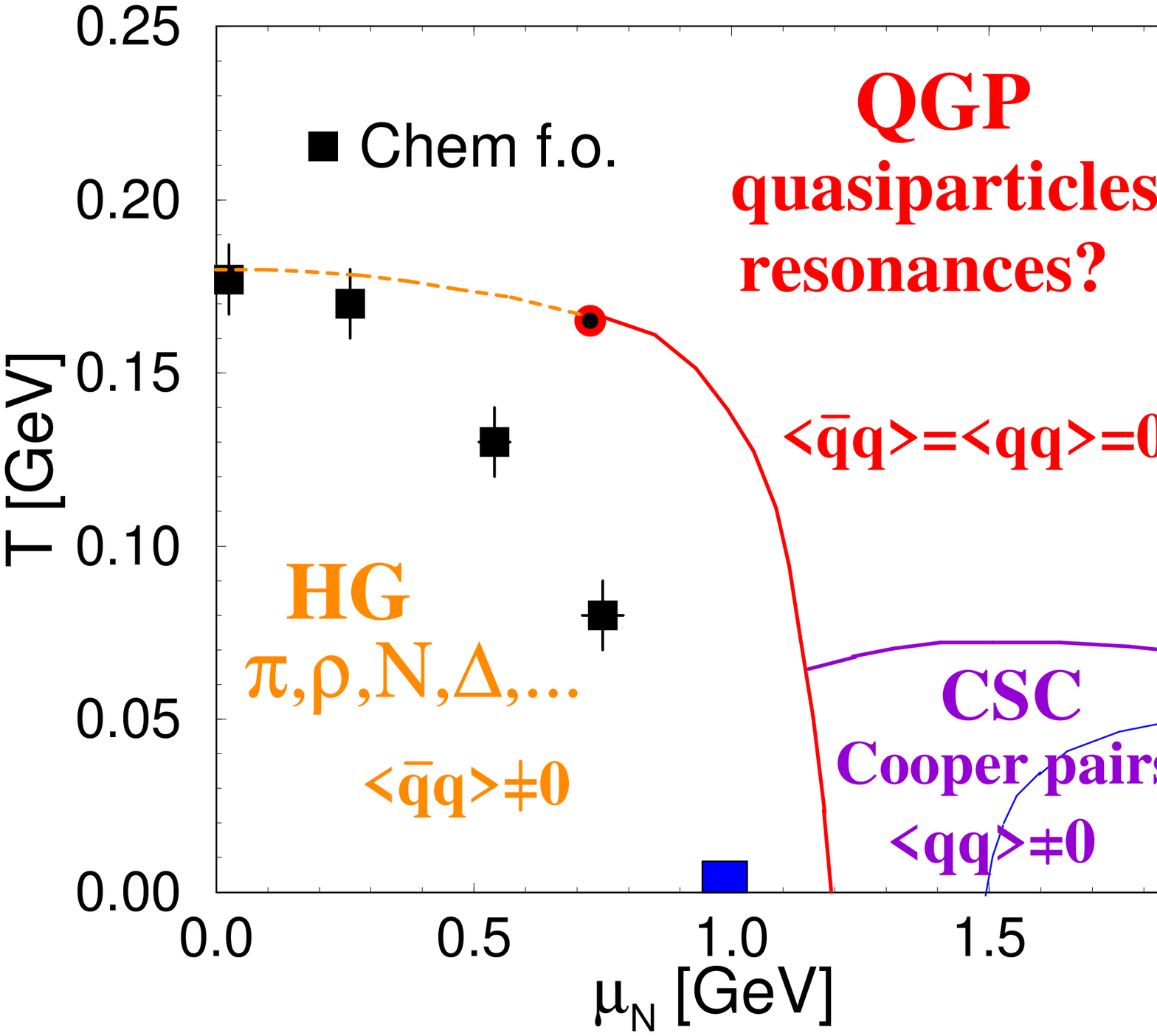}
\vspace*{-05mm}
\caption{Schematic phase diagram of QCD including temperature
``measurements" from chemical analysis of produced
hadrons~\cite{Braun-Munzinger:2003zd}.}
\label{fig_phasedia}
\end{minipage}
\hspace{0.5cm}
\begin{minipage}[t]{0.300\linewidth}
\hspace{-0.4cm}
\includegraphics[width=\linewidth,height=0.89\linewidth]{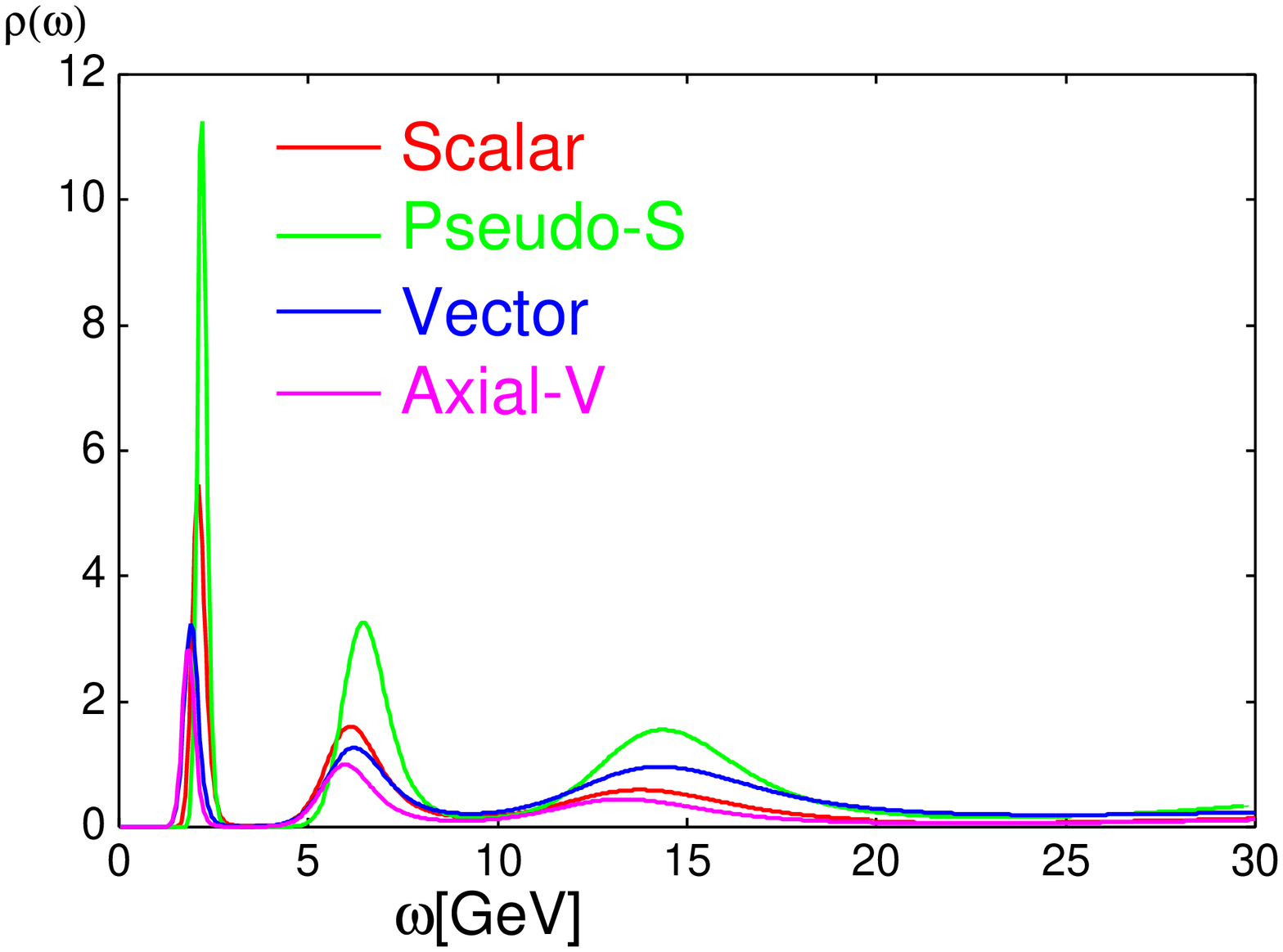}
\vspace*{-5mm}
\caption{``Light" hadronic spectral functions in the QGP
($T$=1.4$T_c$) from lattice QCD~\cite{Asakawa:2002xj}, exhibiting
resonances with approximate chiral symmetry.}
\label{fig_spectral}
\end{minipage}
\hspace{0.5cm}
\begin{minipage}[t]{0.3\linewidth}
\vspace{-2cm}
\includegraphics[width=0.86\linewidth,height=0.43\linewidth]{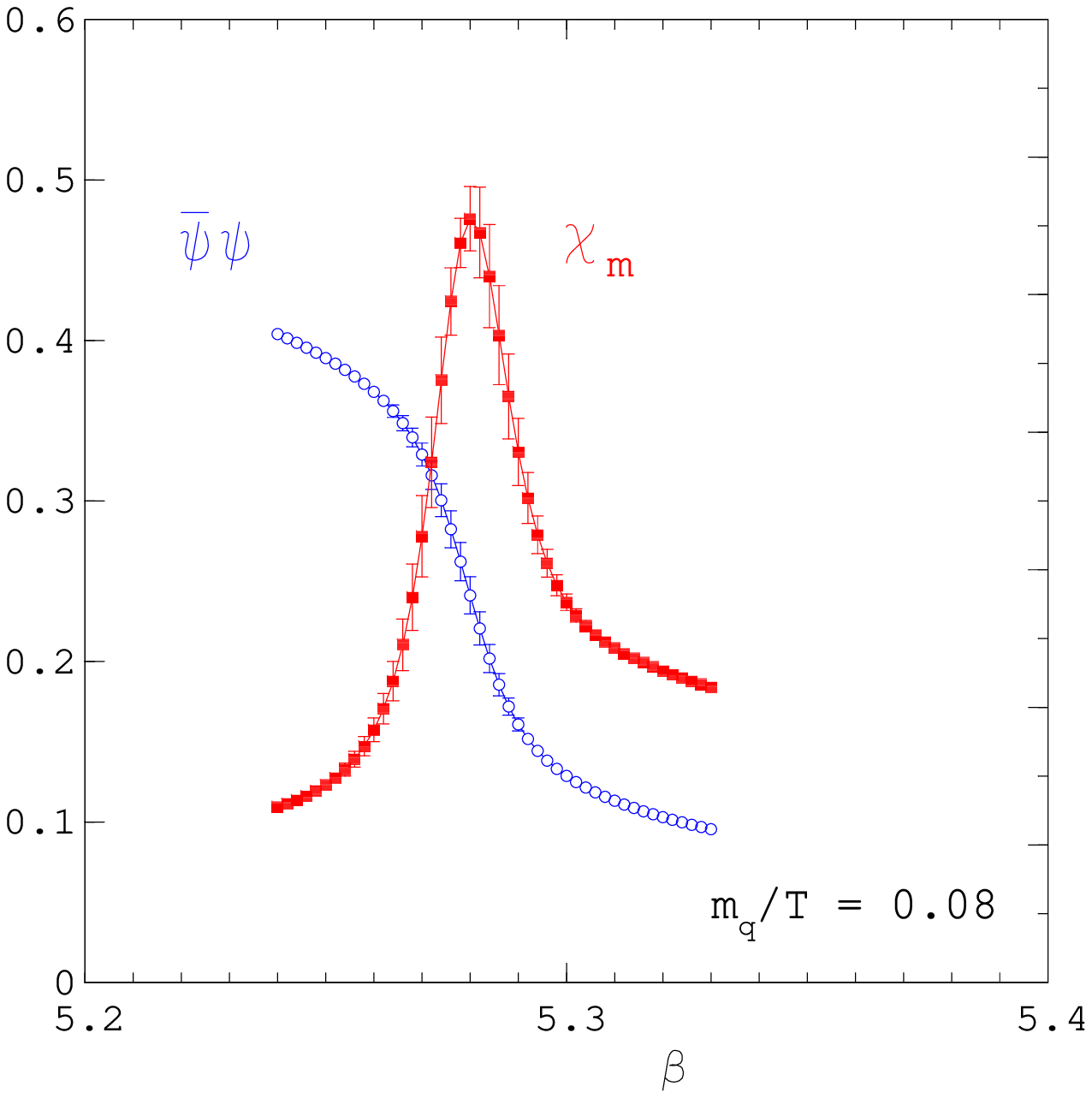}
\vspace*{-6mm}
\caption{Temperature dependence of the quark chiral condensate,
$\langle\bar qq\rangle(T)$, computed in unquenched lattice
QCD~\cite{Karsch:2001cy}.}
\label{fig_chiral}
\end{minipage}
\end{figure}
In the following subsections, we will sketch a theoretical
framework that provides the basis for gaining decisive new insights
in three areas:  temperature measurements; vector meson spectral functions; 
and the origin of mass and chiral symmetry restoration.

Explicit temperature extraction has so far been restricted to hadro-chemical 
and -thermal freezeout, see~Fig.~\ref{fig_phasedia}. 
The goal is to establish early temperatures well
above $T_c$, feasible with photons and dileptons at a typical
energy/momen\-tum/invariant-mass scale of $1<M<3$~GeV.
Together with information on the energy and/or entropy density, the 
effective number of degrees of
freedom, $d_{\rm of}$, may be accessed since $\epsilon=\frac{\pi^2}{30}
d_{\rm of}T^4$ and $s=\frac{2\pi^2}{45}d_{\rm of}T^3$
\cite{Gelis:2004ep,Muller:2005en,d'Enterria:2005vz}.

Dilepton invariant-mass spectra are invaluable and
unique means of extracting microscopic information on the constituents of the
medium. Modifications of low-mass vector mesons ($V=\rho$, $\omega$,
and $\phi$) in hot and/or dense hadronic matter have been extensively
studied theoretically (see
Refs.~\cite{Rapp:1999ej,Alam:1999sc,GH03,Harada:2003jx,Brown:2003ee}
for reviews), largely triggered by the intriguing excess radiation
observed at the CERN-SPS~\cite{Arnaldi:2006jq,Agakichiev:2005ai}. At
RHIC, for the first time, these measurements will be performed in an
environment that is close to {\em net} baryon-free, which will
provide important tests of the relevant mechanisms underlying the
predicted medium effects. In addition, the higher initial
temperatures achieved at RHIC will more directly access
radiation from the (s)QGP and thus be more sensitive to suggested
resonance/bound-state formation above $T_c$ in the vector
channel
\cite{Asakawa:2002xj,Shuryak:2003ty,Mannarelli:2005pz,Casalderrey-Solana:2004dc}
(see Fig.~\ref{fig_spectral}), relative to
more indirect probes of resonances, {\em e.g.} quantitative analyses of
energy loss~\cite{Shuryak:2004ax} and hadron elliptic flow, especially in the 
charm sector~\cite{vanHees:2005wb}), or
systematics of charmonium regeneration~\cite{Grandchamp:2003uw}.

A more ambitious goal is to infer signatures of the phase transition,
requiring the study of order parameters. For
the chiral transition, these are {\em e.g.} quark condensates, see
Fig.~\ref{fig_chiral}, the
constituent quark mass and the pion decay constant.  While none of 
these are (viable) observables, the condensate structure of the
in-medium ground state is encoded in its hadronic
excitations. It is thus necessary to establish connections
between the in-medium vector correlator measured in dilepton
spectra and order parameters.

All of the above are inevitable consequences of QGP formation,
albeit mostly nonperturbative in nature. While data interpretation
will require the application of phenomenological approaches, well-defined links
to finite-temperature
lattice QCD computations and symmetry
constraints are essential for the deduction of meaningful results.

We emphasize that
thermal production rates for photon and dilepton spectra
can be cast into a uniform theoretical framework according to
\begin{eqnarray}
q_0  \frac{dN_\gamma}{d^4xd^3q} = -\frac{\alpha}{\pi^2} \
       f^B(q_0;T) \  {\rm Im}\Pi_{\rm em}^T(q_0=q;\mu_B,T) \ ,
\label{Rphot}
\\
 \frac{dN_{e^+e^-}}{d^4xd^4q} = -\frac{\alpha^2}{M^2\pi^3} \
       f^B(q_0;T) \  {\rm Im}\Pi_{\rm em}(M,q;\mu_B,T) \ ,
\label{Rdilep}
\end{eqnarray}
where the key quantity is the retarded
electromagnetic correlation function,
$\Pi_{\rm em}$~\cite{Feinberg:1976ua,McLerran:1984ay}.
In the vacuum,
this function can be measured in $e^+e^-$ annihilation and decomposes
into two regimes: at masses above $M=\sqrt{q^2}\simeq1.5$\,GeV, the
strength of the EM~spectral function, Im\,$\Pi_{\rm em}^{\rm vac}$,
is rather accurately determined by perturbation theory, {\em i.e.}
annihilation into $q\bar q$ pairs with little impact from subsequent
hadronization.  At low mass the cross section is saturated
by the light vector mesons $\rho$, $\omega$ and $\phi$ (vector dominance), 
{\em i.e.} nonperturbative resonance formation:
\begin{equation}
{\rm Im} \Pi_{\rm em}^{\rm vac}(M) = \left\{
\begin{array}{ll}
 \sum\limits_{V=\rho,\omega,\phi} \left(\frac{m_V^2}{g_V}\right)^2 \
{\rm Im} D_V(M) &  \ M < 1.5~{\rm GeV} \, ,
\vspace{0.3cm}
\\
-\frac{M^2}{12\pi} \ (1+\frac{\alpha_s(M)}{\pi} +\dots)  \ N_c
\sum\limits_{q=u,d,s} e_q^2  &  \ M \ge 1.5~{\rm GeV} \, .
\end{array}  \right.
\label{Piem}
\end{equation}
For thermal dilepton emission this implies that, on one hand, the
low-mass region carries information on
dynamical medium effects with a relative strength of the vector mesons
approximately given by 11:1:2 for $\rho:\omega:\phi$, reflecting
the vector-dominance couplings, $m_V^4/g_V^2$, or, equivalently,
the dilepton decay widths.
On the other hand, at intermediate mass a reasonably controlled
emission strength provides the basis for probing the temperature.
The temperature and volume dependence in space-time integrated
dilepton spectra combine such that the prevalent low-mass contribution
originates from temperatures around and below $T_c$ \cite{Rapp:2004zh}.
At higher masses and energies, the exponential sensitivity of 
the Bose factor strongly biases contributions toward high temperatures.
Similar considerations apply to thermal photon transverse-momentum spectra,
corroborating the feasibility of realizing the three basic objectives listed
above.  Note that the leading-order contribution to the dilepton rate
at sufficiently large invariant mass, $M>1.5$~GeV, is 
${\mathcal{O}}(\alpha_s^0)$ while a nonzero photon rate
requires processes of at least ${\mathcal{O}}(\alpha_s)$.

In rough accord with our physics objectives, we adopt the
following classification of regimes in dilepton invariant mass, $M_{ll}$,
or photon transverse momentum, $q_t$:
\begin{itemize}
\item the low-mass region (LMR),
$M_{ll}<1.1~$GeV (vector meson decays);
\item the intermediate-mass region (IMR), $1.1 < M_{ll} < 3$ GeV 
(continuum radiation, QGP emission, resonances?);
\item the high-mass region (HMR), $M_{ll} > 3$ GeV 
(primordial emission and heavy quarkonia).
\end{itemize}

\subsection{QCD Lattice Results}
\label{sec_lat}
\begin{figure}[!t]
\begin{center}
\includegraphics[width=0.5\linewidth]{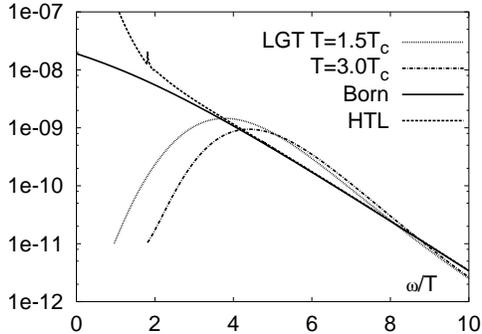}
\end{center}
\vspace*{-5mm}
\caption{Thermal dilepton production rates in the QGP as evaluated
in quenched lattice gauge theory at zero three-momentum (dotted and
dash-dotted line)~\cite{Karsch:2002wv}. The solid and dashed lines are
pQCD calculations of
the rate in the Born approximation, $q\bar q$ annihilation to leading order,
${\mathcal{O}}(\alpha_s^0)$, and with hard-thermal-loop
corrections~\cite{Braaten:1990wp} respectively.}
\label{fig:rate-lat}
\end{figure}

Thermal dilepton rates have been studied in lattice QCD within the
quenched approximation~\cite{Karsch:2002wv}. The computation of the
finite-temperature Euclidean correlators in the vector channel is
supplemented by a transformation into the time-like regime using the
maximum entropy method, after which the dilepton rate follows from
Eq.~(\ref{Rdilep}). The results for zero three-momentum at two different
temperatures above $T_c$ are compared to perturbative 
calculations~\cite{Braaten:1990wp} in Fig.~\ref{fig:rate-lat}.
The lattice rates are quite comparable to the perturbative ones at
high energy, while the enhancement at $\omega\simeq 5T$
could be related to resonance formation in the QGP. Toward small
energies, the lQCD rates drop substantially and deviate markedly
from the perturbative calculations, which, in the HTL approximation,
even diverge. Since the HTL calculation is of nontrivial 
order in $\alpha_s$,
the HTL divergence is closely related to a non-vanishing thermal
photon rate. Recent lattice computations of the vector correlator at
finite 3-momentum~\cite{Petreczky:2005,Arleo:2003gn} confirm that
the rates in Fig.~\ref{fig:rate-lat} lead to vanishing photon
emission.
If this seemingly unrealistic feature is induced by lattice
artifacts, the decrease of the dilepton rate toward small $\omega$
may not hold, leading to better agreement with HTL
extrapolations, consequently affecting the interpretation of the
``resonance" structure. The ultra-soft limit of the EM
correlator can be further studied by its relation to the electric
conductivity, $\sigma_{\rm em}$, via the Kubo formula,
\begin{equation}
\sigma_{\rm em}(T) = \frac{e^2}{3}\frac{\partial}{\partial q_0}
{\rm Im} \Pi_{\rm em}^T(q_0,q=0;\mu_B,T)  \ .
\end{equation}
The correlators found in Ref.~\cite{Karsch:2002wv} correspond to
a vanishing conductivity.  A different method employed in
Ref.~\cite{Gupta:2003zh} leads to a finite, large value
of $\sigma_{\rm em}$, $\sigma_{\rm em}(T)\simeq 7\sum e_q^2 T$, see 
Ref.~\cite{Arleo:2003gn}. It would be very interesting to
compare this result to the conductivity underlying the soft photon
emission rates used in the description of the low-momentum
WA98 data~\cite{wa98-lowpt}, see the lower panel of
Fig.~\ref{fig_spsphot}. Along these lines,
$\sigma_{\rm em}$ has been evaluated in a low-temperature pion gas
within chiral perturbation theory \cite{Fernandez-Fraile:2005ka}.

Another quantity of interest is the susceptibility
associated with conserved quantum numbers, {\em i.e.}, derivatives of the
thermodynamic free energy with respect to a chemical potential,
\begin{equation}
\chi_X = - \frac{\partial^2 \Omega}{\partial \mu_X^2} \ .
\end{equation}
These susceptibilities have been evaluated in lattice QCD for quark and isospin 
chemical potentials~\cite{Allton:2005gk}. When extrapolated
into the finite-$\mu_q$ plane, the isoscalar quark susceptibility 
develops a maximum while the isopin susceptibility remains monotonic. Since the
susceptibilities can be related to the space-like static limit of
the corresponding correlation function in the $\omega$ and $\rho$
channels, valuable information on the soft part of
the spectral functions may be inferred~\cite{Prakash:2001xm} or at
least tested for a given model. Whether the lattice results 
for the isoscalar susceptibility are related to an in-medium reduced 
$\omega$ mass {\em e.g.} via a $\sigma - \omega$ mixing mechanism close
to the QCD critical endpoint, remains to be seen.
Interestingly, recent experiments on nuclear $\omega$ photoproduction
have provided evidence of $\omega$ mass reduction
in nuclear matter~\cite{Trnka:2005ey}. A similar line of reasoning
for the $\rho$ suggests that its mass is not much affected
approaching the critical point~\cite{Rapp:2007bm}.  

\subsection{Sum Rules}
Besides direct lQCD calculations of dilepton and photon rates,
unanticipated in the immediate future,  model-independent
information and effective-model constraints are encoded in
(energy-weighted) sum rules relating (integrated moments of)
spectral functions to vacuum expectation values of composite
quark and gluon operators, ``condensates", or (partial) conservation
laws of (axial-) vector currents. These sum rules are thus
prime examples of connecting hadronic excitations to the underlying
ground-state structure (symmetry-breaking pattern), including order
parameters.  We will briefly discuss two classes of sum rules.

\subsubsection{QCD Sum Rules}
\begin{figure}[!t]
\begin{minipage}{0.5\linewidth}
\includegraphics[width=0.93\linewidth]{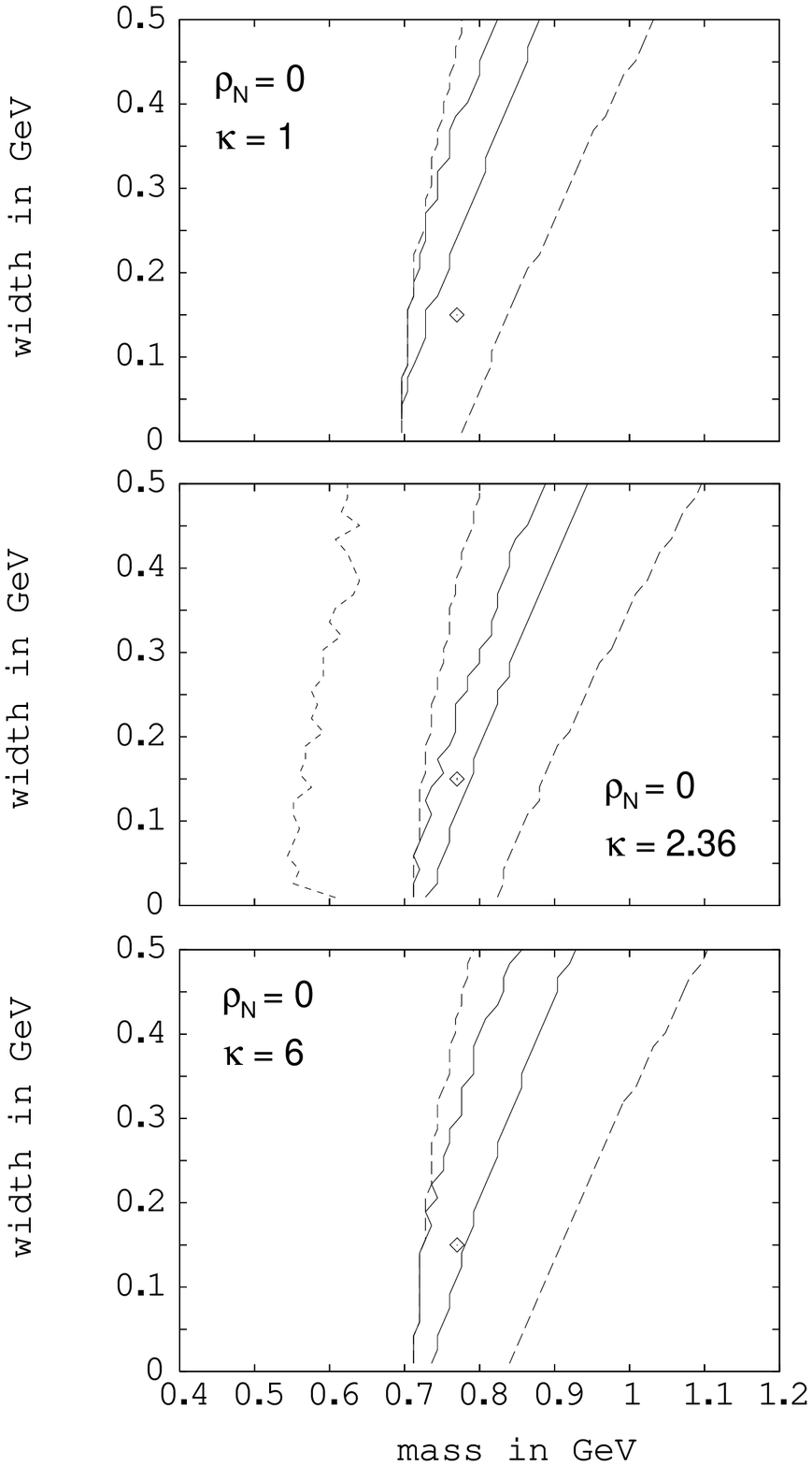}
\end{minipage}
\hspace{0.2cm}
\begin{minipage}{0.5\linewidth}
\vspace{0cm}
\includegraphics[width=0.9\linewidth]{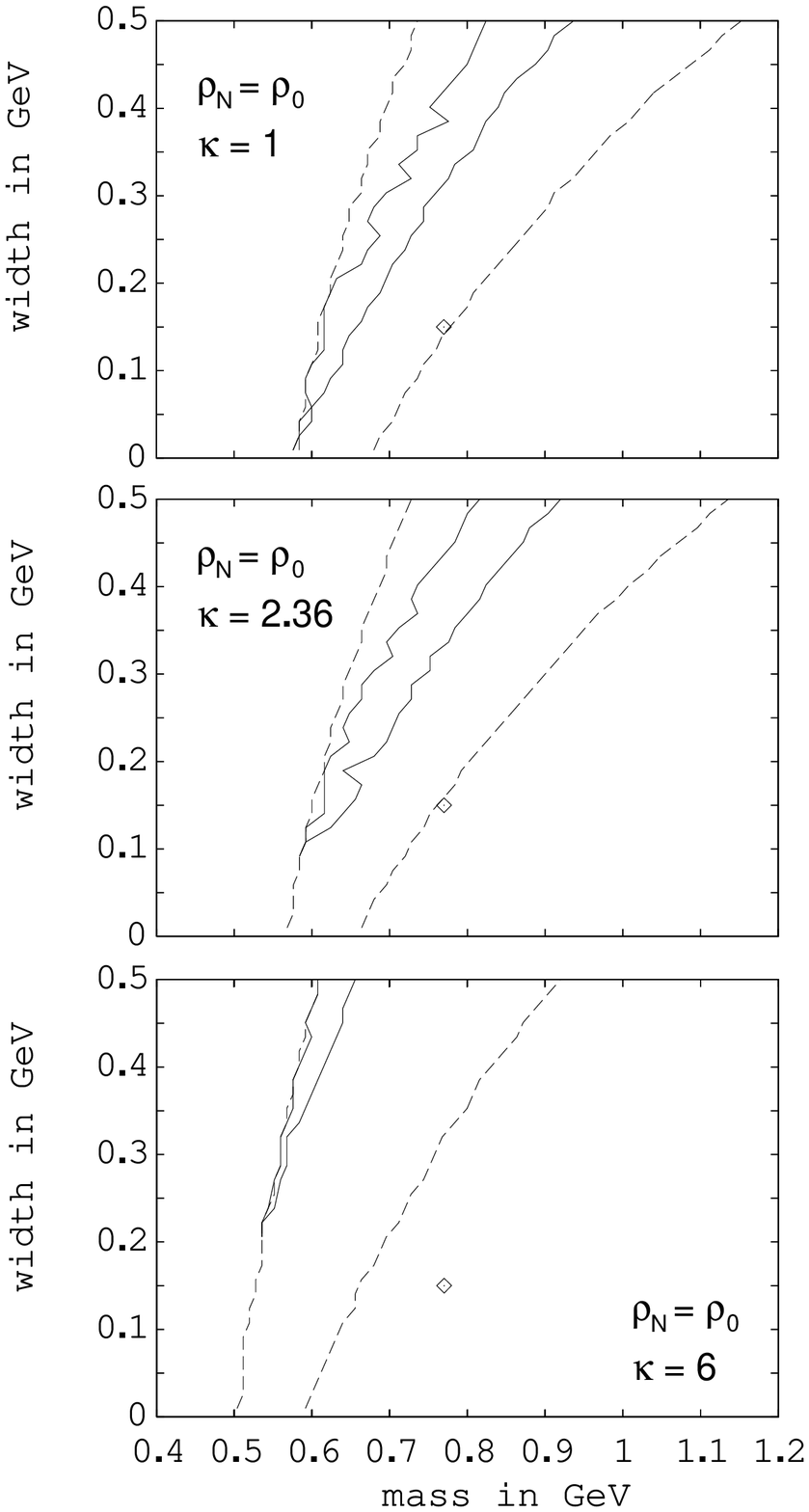}
\end{minipage}
\caption{QCD sum rule constraints on Breit-Wigner $\rho$
spectral functions in the mass-width plane~\cite{Leupold:1997dg}.
The bands indicate regions of mass and width values for
which the deviation between the left-hand and right-hand sides of
Eq.~(\ref{qcdsr}) is below 0.2\% (within the solid lines)
or below 1\% (within the dashed lines). The left column is for the
vacuum case, which is used to determine the ``correlation parameter",
$\kappa$ in the four-quark condensate while the right column is at 
nuclear-matter density (diamond: mass and width of
the free $\rho$ meson).}
\label{fig_qcdsr}
\end{figure}

QCD sum rules (QCDSRs)~\cite{Shifman:1978bx} are based on the
analyticity of correlation functions, resulting in the dispersion relation
\begin{equation}
\int d\omega\frac{{\rm Im} \Pi_{\rm em}(\omega)}{\omega-q_0}
= \sum\limits_n  \frac{C_n}{Q^{2n}}  \ .
\label{qcdsr}
\end{equation}
The left-hand-side involves an integral over a (hadronic)
spectral function in the time-like regime plus possible subtractions
not indicated here.  The right-hand-side is an expansion in
space-like momenta,
1/$Q^2$ ($Q^2\equiv -q^2 > 0$), with leading perturbative terms
and nonperturbative effects encoded in Wilson coefficients, $C_n$, via
quark and gluon condensates of dimension increasing with $n$.
When applied to the light vector mesons, $\rho$ and $\omega$, at finite
temperature and/or density~\cite{Hatsuda:1991ez}, 
it turns out that the largest sensitivity
resides in the medium dependence of the four-quark condensates,
$\langle (\bar qq)^2\rangle$. Unfortunately, rather little is so far
known about their temperature dependence from lattice QCD.
The usual assumption is to factorize the four-quark condensates into
a product of two-quark condensates corresponding to the assumption
of ``ground-state dominance" with an extra parameter, $\kappa$,
usually fixed in the vacuum, representing correlation effects.
An application of QCDSRs to the $\rho$ meson in cold nuclear
matter is shown in Fig.~\ref{fig_qcdsr}~\cite{Leupold:1997dg},
indicating that the finite-density decrease of the condensates
mandates a ``softening" of the $\rho$ spectral function. The required
low-mass enhancement can be satisfied by an increasing width, a
decreasing mass, or a suitable combination thereof as indicated by
the ``allowed regions" enclosed by the bands in the right panel of
Fig.~\ref{fig_qcdsr}, see Refs.~\cite{Asakawa:1993pq,Klingl:1997kf,Ruppert:2005id}.

\subsubsection{Chiral Sum Rules}
Chiral sum rules (CSRs)~\cite{Wei67,Das:1967ek} have been derived
prior to QCD from current algebra and chiral Ward identities.
The (partial) conservation of the (axial-) vector-isovector currents
leads to relations between the pion decay constant, $f_\pi=92$~MeV (an
order parameter of chiral symmetry breaking), to moments of differences
between pertinent spectral functions. In vacuum they
are\footnote{The form of the sum rules
in Eqs.~(\ref{xcsr0})-(\ref{xcsr3}) applies to the
chiral limit of vanishing current light-quark masses and pion mass.
Corrections to the second Weinberg Sum Rule (WSR), Eq.~(\ref{xcsr2}),
may not be small \cite{Floratos:1978jb,Narison:1981ra,Dmitrasinovic:1998uu}. 
The first WSR, Eq.~(\ref{xcsr1}), does not seem to
be affected by finite quark mass corrections.}
\begin{eqnarray}
\int\limits_0^\infty \frac {ds}{s^2}
 \left[\rho_V^{\rm vac}(s) - \rho_A^{\rm vac}(s) \right]  &=&
 f_\pi^2 \ \frac{\langle r_\pi^2 \rangle}{3} - F_A \, \, ,
\label{xcsr0}
\\
\int\limits_0^\infty \frac {ds}{s}
 \left[\rho_V^{\rm vac}(s) - \rho_A^{\rm vac}(s) \right]  &=& f_\pi^2 \, \, ,
\label{xcsr1}
\\
 \int\limits_0^\infty ds
 \left[\rho_V^{\rm vac}(s) - \rho_A^{\rm vac}(s) \right] &=&0 \,, \, ,
\label{xcsr2}
\\
\int\limits_0^\infty s ds
 \left[\rho_V^{\rm vac}(s) - \rho_A^{\rm vac}(s) \right] &=&
-2\pi \alpha_s \langle {\mathcal{O}} \rangle  \, \, ,
\label{xcsr3}
\end{eqnarray}
where $\langle r_\pi^2\rangle$ is the pion charge radius squared, $F_A$
the axial vector form factor in the radiative pion decay, $\pi\to l
\bar\nu_l\gamma$. The vacuum vector and axial vector spectral functions
are
\begin{equation}
\rho_{V,A}^{L,T} = - \frac{1}{\pi} {\rm Im} \Pi_{V,A}^{L,T} \ .
\end{equation}
In Eq.~(\ref{xcsr3}), obtained in
Ref.~\cite{Kapusta:1993hq}, $\langle {\mathcal{O}}\rangle$ denotes a four-quark
condensate.  In the factorization approximation, $\langle {\mathcal{O}}\rangle
= (16/9) \langle\bar qq \rangle^2$.
The direct connection of the CSRs to the vector correlator
renders them particularly valuable in the
context of dilepton production. The assessment of in-medium effects
requires their extension to finite temperature, 
elaborated in Ref.~\cite{Kapusta:1993hq}. Due to loss of
Lorentz invariance when specifying the thermodynamic rest frame,
the original vacuum results become energy
sum rules at fixed three-momentum and split into longitudinal ($L$) and
transverse ($T$) components of the correlators and quark condensate:
\begin{eqnarray}
 \int\limits_0^\infty \frac {dq_0^2}{(q_0^2-q^2)}
 \, \Delta \rho^L(q_0,q)   &=& 0 \,\, ,
\label{wsr1med}
\\
  \int\limits_0^\infty dq_0^2
 \, \Delta\rho^{L,T}(q_0,q) &=& 0 \,\, ,
\label{wsr2med}
\\
 \int\limits_0^\infty q_0^2 \, dq_0^2
 \left[\Delta\rho^L(q_0,q) + 2\,\Delta\rho^T(q_0,q) \right] &=&
-4\pi \alpha_s \left[
\langle\langle {\mathcal{O}_\mu^\mu} \rangle\rangle
+ 2\,\langle\langle {\mathcal{O}}^{00} \rangle\rangle \right]  \, \, ,
\label{wsr3med}
\end{eqnarray}
where $\Delta\rho\equiv\rho_V-\rho_A$ and
$\langle\langle \, \cdot \, \rangle\rangle$ denotes an in-medium
expectation value. The transverse and longitudinal components of the
spectral functions are given in terms of standard projection operators,
\begin{equation}
\rho_{V,A}^{\mu\nu} =  \rho_{V,A}^{T} P_T^{\mu\nu} +
\rho_{V,A}^{L} P_L^{\mu\nu} \ ,
\end{equation}
where the pionic piece,
$\rho_\pi^{\mu\nu} = f_\pi^2 q^2 \delta(q^2) P_L^{\mu\nu}$ in vacuum, has been
included in the longitudinal axial vector channel.  In medium, the
pion spectral function is subject to medium modifications as well.
The in-medium Weinberg-type sum-rules
(\ref{wsr1med})-(\ref{wsr3med}) impose stringent constraints on both
temperature and energy-momentum dependencies through the moments
of chiral hadronic models for vector and axial vector spectral
functions. The model-independent connection to lattice QCD can be
implemented by employing the pertinent temperature dependencies of the pion
decay constant and four-quark condensates which, in principle, are
easier to compute than full spectral functions.

We now turn to a more concrete discussion with examples of how to realize
the three main points outlined at the beginning of this Section.

\subsection{Temperature and Degrees of Freedom}
\subsubsection{Suitable Kinematic Regimes}
\label{sec_kin}
Thermal emission rates for electromagnetic radiation,
Eqs.~(\ref{Rphot}) and (\ref{Rdilep}), can in principle be used to
``infer" the temperature of thermalized matter in heavy-ion collisions
if: (i) the emission strength represented by the EM correlator is
reasonably well determined so that the $T$ dependence essentially
resides in the Bose factor and (ii) a kinematic window can be identified
where radiation from a reasonably well-defined temperature regime prevails.
Ideally, these conditions are met at the highest masses and energies where
the correlators can be reliably evaluated in pQCD and thermal emission
from the earliest phases dominates. In practice, however,
the high-mass/energy region is dominated by dileptons and photons
from primordial hard $NN$ collisions.  At lower masses, the lower temperature
contributions increase substantially.
A more comprehensive discussion of the various sources will be given
below.

Fig.~\ref{fig:qgprad} shows an example of predictions for
space-time integrated photon and dilepton spectra at intermediate masses
and transverse momenta in central $\sqrt{s_{_{NN}}}=200$~GeV Au+Au
collisions~\cite{Turbide:2003si,Rapp:2000pe}.
\begin{figure}[t]
\begin{minipage}{0.300\linewidth}
\includegraphics[width=0.965\linewidth,angle=-90]{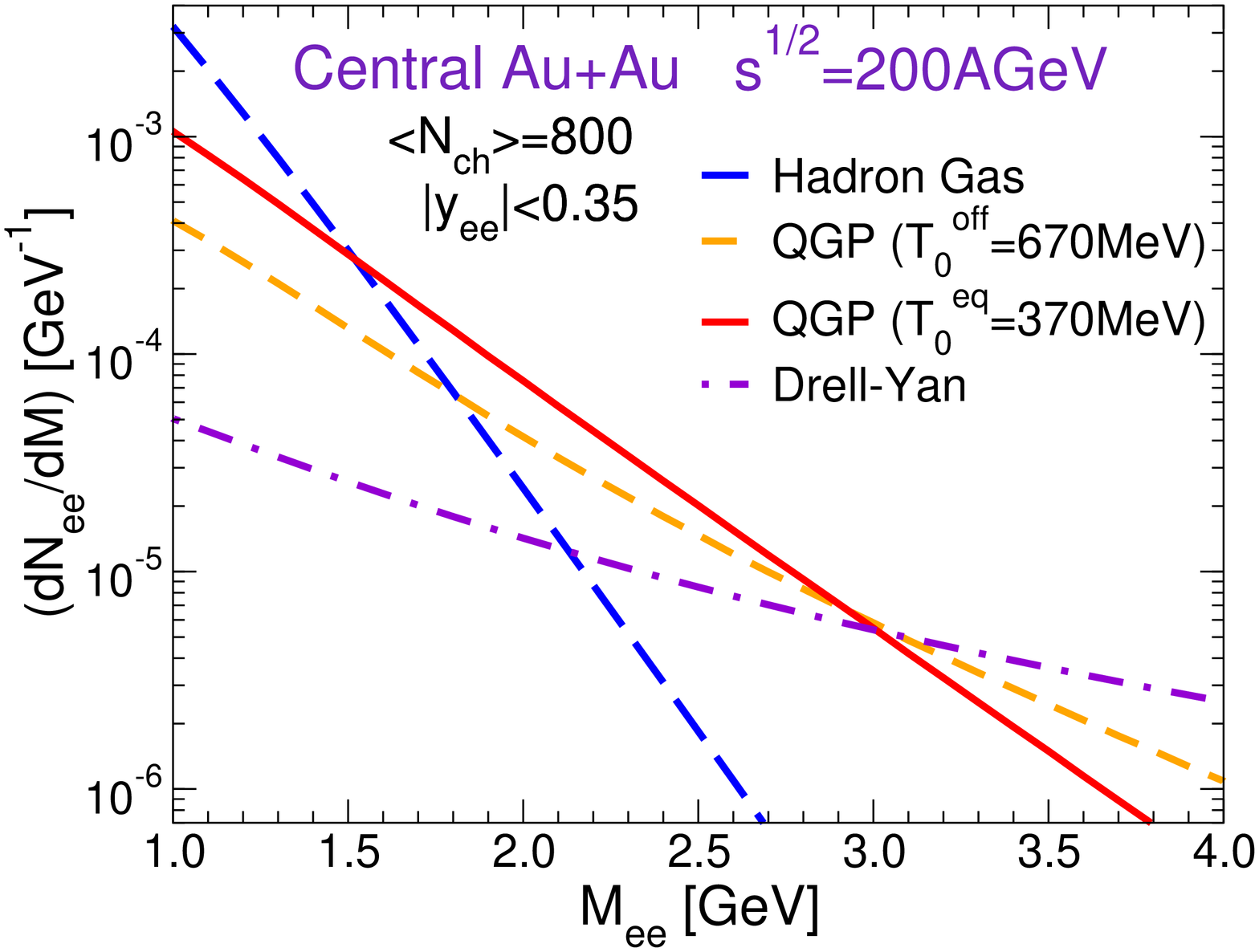}
\end{minipage}
\hspace{0.3cm}
\begin{minipage}{0.300\linewidth}
\vspace{0.2cm}
\includegraphics[width=1.06\linewidth]{phenix-all2.eps}
\end{minipage}
\hspace{0.1cm}
\begin{minipage}{0.300\linewidth}
\includegraphics[width=0.965\linewidth,angle=-90]{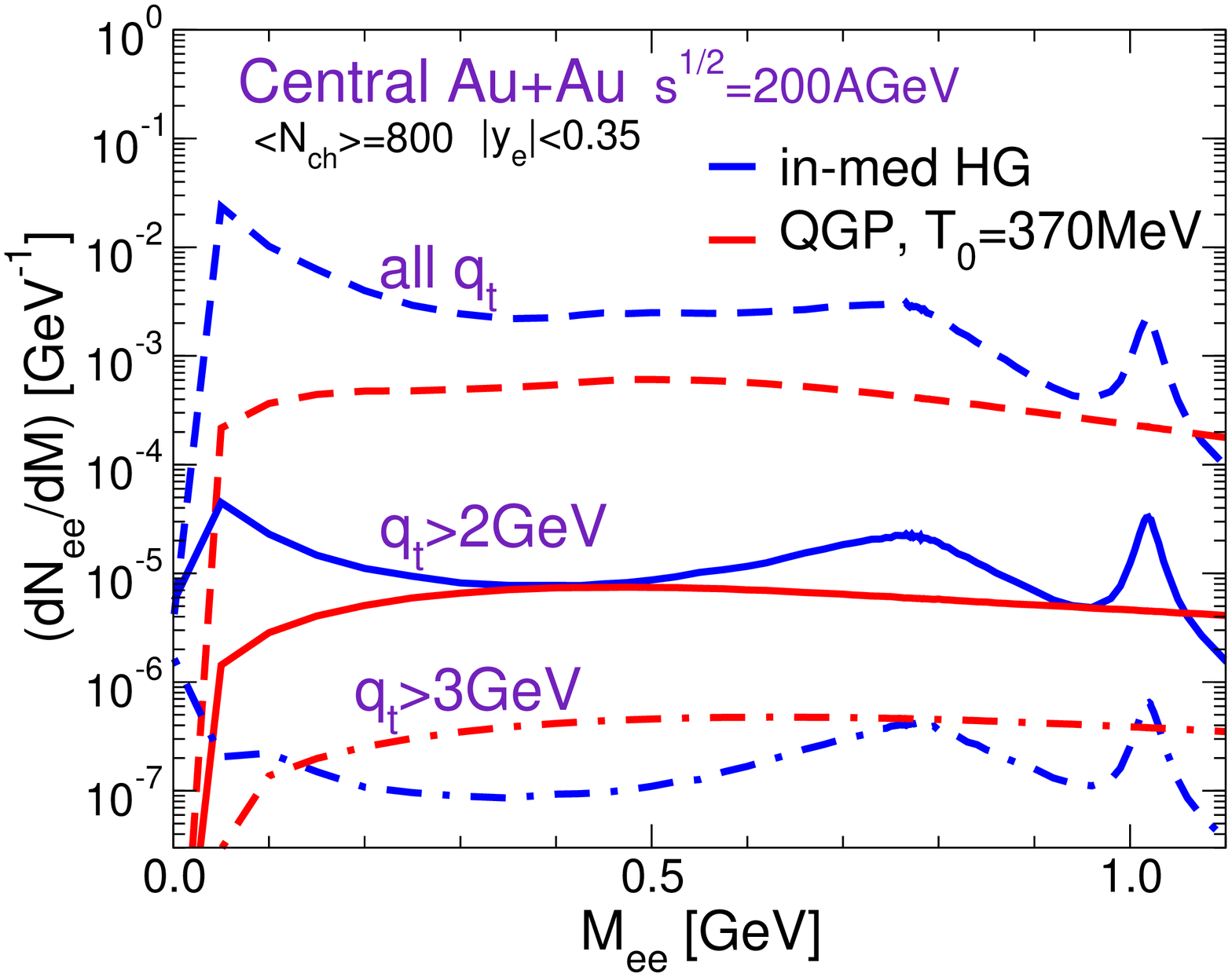}
\end{minipage}
\vspace*{-5mm}
\caption{Comparison of electromagnetic emission spectra from QGP
and HG at intermediate $M_{ll}$ and $q_t$ in central
Au+Au collisions at $\sqrt{s_{_{NN}}} = 200$ GeV. Left panel: IMR
dileptons~\cite{Rapp:2000pe} including primordial Drell-Yan
annihilation.  Middle panel: direct photons~\cite{Turbide:2003si}
including primordial pQCD contributions.  Right panel: low-mass
dileptons with pair $q_t$ cuts. Note that the
IMR dilepton spectrum (left) neither includes contributions
from the $\phi$ nor a cut on the {\em single}-electron rapidity.  The
single-electron rapidity cut has been applied to the low mass region, causing
a significant reduction in the yield.
On the other hand, the low mass spectra (right) does not include dilepton 
emission from four-pion and higher states that dominate hadronic emission
beyond the $\phi$ mass.
}
\label{fig:qgprad}
\end{figure}
The thermal spectra are decomposed into QGP and hadronic (HG) emission
from an isentropically expanding thermal fireball assuming a
critical temperature of $T_c = 180$~MeV~\cite{Rapp:2002mm} where the total
entropy is fixed to reproduce the observed hadron multiplicities at
chemical freeze-out, $\mu_N^{\rm ch}$,$T_{\rm ch}) =(25,180)$~MeV. The pQCD
photon rates, to leading order in $\alpha_s$~\cite{Arnold:2001ms},
and HTL-resummed dilepton rates~\cite{Braaten:1990wp} are convoluted
over a chemically-equilibrated QGP  assuming a formation time of
$\tau_0 =1/3$~fm/$c$, translating into $\bar T_0 =370$~MeV (670~MeV if
the initial parton densities are assumed to be undersaturated).
Uncertainties in the longitudinal expansion, affecting the QGP
lifetime, can induce changes of the QGP spectra by up to 30\%.
The sensitivity to $\tau_0$ is larger, especially at masses above
2~GeV~\cite{Rapp:2000pe}. The hadronic emission spectra include
in-medium modifications of the EM~correlator, see
Sec.~\ref{sec_spec-fct} for details, as well as chemical
off-equilibrium in the hadronic evolution until thermal freeze-out,
see Sec.~\ref{sec_lmdilep}. Three regimes emerge where QGP radiation 
outshines both HG and primordial emission, see~Fig.~\ref{fig:qgprad}):
\begin{itemize}
\item[(i)] three-momentum integrated dilepton spectra at intermediate mass,
    $1.5\leq M \leq 3$~GeV (left panel);
\item[(ii)] direct photon spectra at intermediate transverse momentum,
    $1.5 \leq q_t \leq 3$~GeV (middle panel);
\item[(iii)] low-mass dilepton spectra at
           $q_t \geq 2$~GeV (``low-virtuality" photons; right panel).
\end{itemize}
In practice a careful assessment of additional sources,
including ``pre-equilibrium" contributions and jet-plasma
interactions such as Bremsstrahlung off quark jets or Compton
scattering of gluon jets \cite{Turbide:2005fk,Turbide:2006mc}, is
mandatory before firm conclusions about the thermal component can be reached.
We note that the predictions for the thermal
photon spectra, together with jet-plasma interactions and primordial
photons extracted from $pp$ collisions, compare favorably with
preliminary RHIC direct photon data~\cite{Bathe:2005nz}, as discussed later. 
Recent calculations of the jet-plasma component suggest
that it exceeds thermal emission for real and
virtual photon transverse momenta 
$q_t\ge3-4$~GeV~\cite{Turbide:2005fk,Turbide:2006mc}.  
See Section~\ref{sec:photon_elliptic} for a possible
strategy for disentangling the different components. 

\begin{figure}[!t]
\centering{
\includegraphics[width=0.45\linewidth,
height=0.4\linewidth]{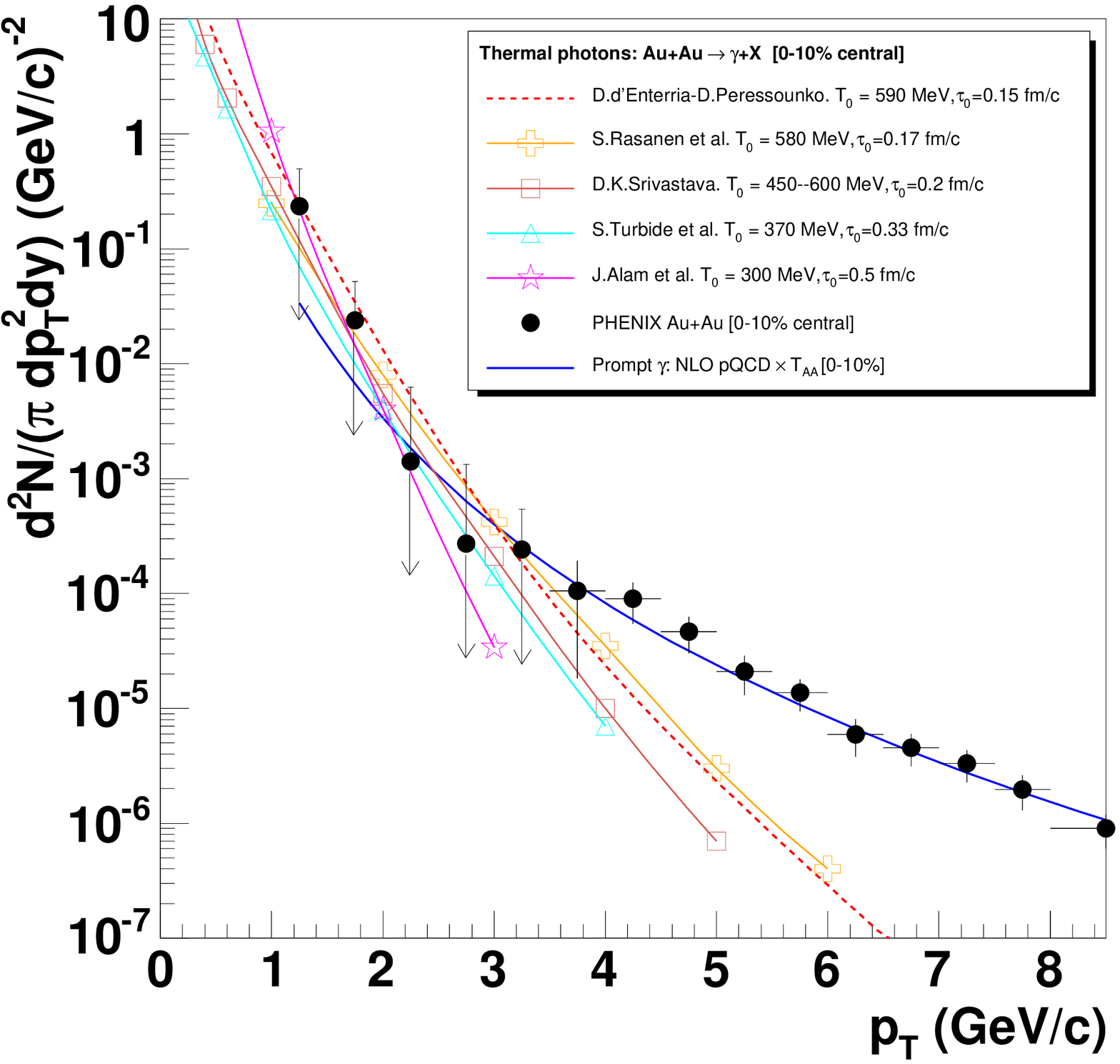}
\hspace{0.2cm}
\includegraphics[width=0.45\linewidth,height=0.4\linewidth]{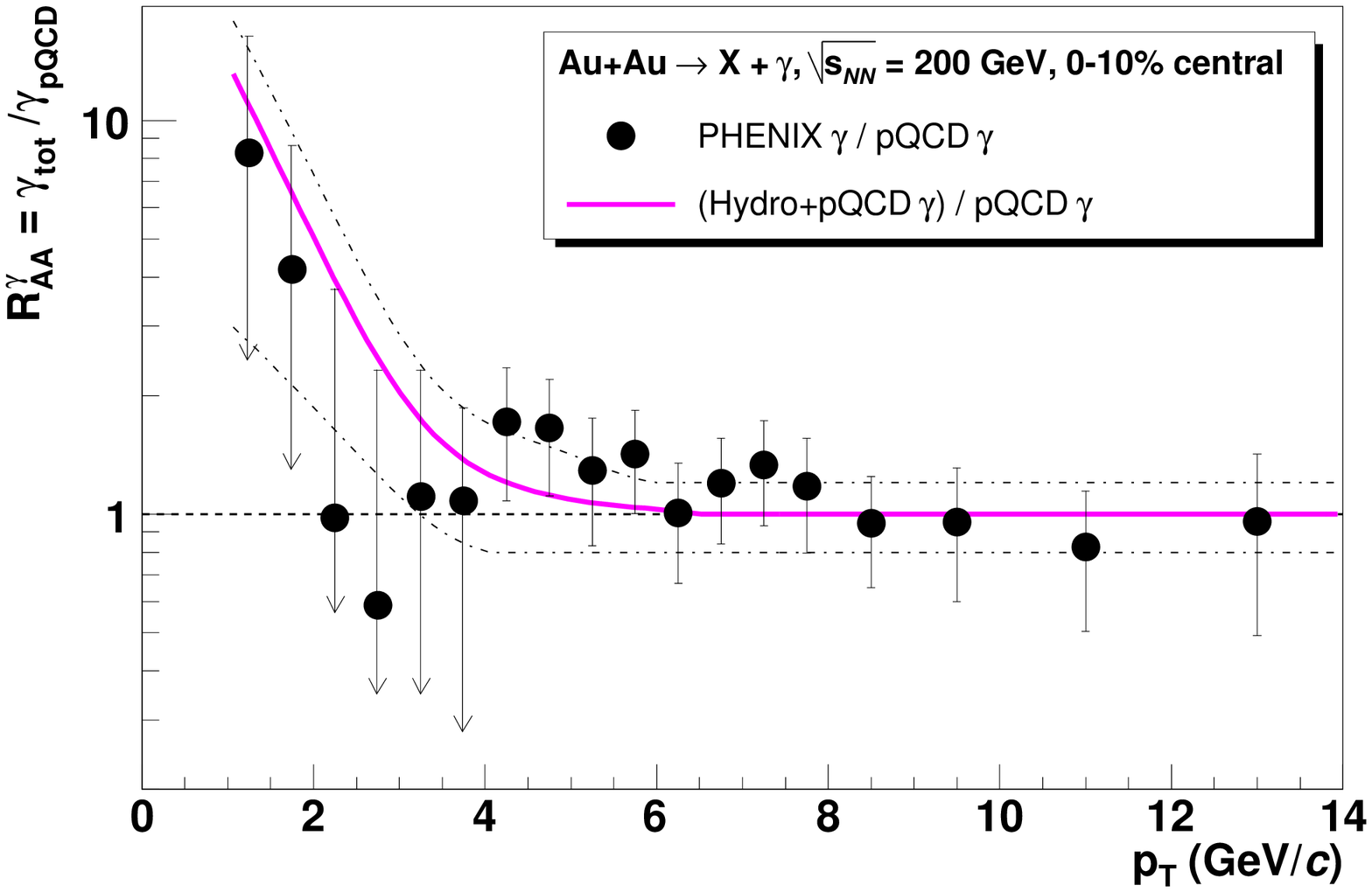}}
\caption{Left panel: Thermal photon spectra for central Au+Au reactions at
$\sqrt{s_{_{NN}}}=200$~GeV, computed within different models (see text),
compared to
the expected pQCD prompt $\gamma$ yields ($T_{AA}$-scaled NLO $pp$
calculations~\cite{pQCDgamma}: solid line, no symbols)
and to the experimental total direct photon spectrum measured by
PHENIX~\protect\cite{ppg042}.
Right panel: Direct photon ``nuclear modification factor'',
Eq.~(\ref{eq:RAA}), for the 0-10\% most central Au+Au reactions at
$\sqrt{s_{_{NN}}} =200$~GeV. The solid line is the ratio resulting from
a hydro+pQCD model~\protect\cite{d'Enterria:2005vz}. The
points show the PHENIX data~\protect\cite{ppg042} over the same NLO
yields while the dashed-dotted curves indicate the theoretical uncertainty
on the NLO calculations.}
\label{fig:thermal_photon_comparison}
\end{figure}
\subsubsection{Direct Photons and Current RHIC Data}
\label{sec_denterria}
To illustrate the uncertainties in and the required precision of a
``temperature measurement", we describe an analysis of
recent direct photon spectra in central 200~GeV Au+Au collisions at RHIC.
The left panel of Fig.~\ref{fig:thermal_photon_comparison} compiles
several model calculations of thermal photon production mostly based
on hydrodynamic evolution.  The maximum initial
temperature, $T_0$, is quoted:
Srivastava {\it et al.}~\cite{srivastava_sps_rhic},
$\tau_0 \approx 0.2$ fm/$c$ and
$450<T_0<660$~MeV; Alam {\it et al.}~\cite{alam_sps_rhic},
$\tau_0=0.5$~fm/$c$, $T_0=300$~MeV\footnote{Alam {\it et al.} have
recently~\cite{alam05} recomputed their hydrodynamic yields using
higher initial temperatures, $T_0=400$ MeV at $\tau_0 = 0.2$ fm/$c$, to
improve the agreement with the data.},
R\"as\"anen {\it et al.}~\cite{finnish_hydro}, $\tau_0=0.17$~fm/$c$ and
$T_0=580$~MeV; Turbide {\it et al.}~\cite{Turbide:2003si},
expanding fireball with $\tau_0=0.33$~fm/$c$ and $\langle T_0 \rangle=370$~MeV;
d'Enterria and Peressounko~\cite{d'Enterria:2005vz}, $\tau_0=0.15$~fm/$c$ and 
$T_0=590$~MeV; see also Steffen and Thoma~\cite{steffen_sps_rhic_lhc}, 
$\tau_0 =0.5$~fm/$c$ $\langle T_0 \rangle =300$ MeV.
For similar initial conditions, the
total thermal yields in these calculations are compatible both with the data
and each other within a
factor of $\sim 2$. While this confirms the dominant role of
thermal radiation in the window $p_T \simeq 1.5-3$~GeV, it shows that 
more quantitative, in-depth comparisons are required 
to disentangle the underlying assumptions on evolution model
(boost-invariant hydrodynamics with or without transverse expansion,
thermal fireballs, {\em etc}.) from production rates to narrow down the
viable $T_0$ range.
The excess over that expected from primordial $NN$ collisions, the 
``prompt" contribution, is
better illustrated by the nuclear modification
factor, $R_{AA}^\gamma$, the ratio of the direct
photon spectra in $AA$ collisions relative to either 
the pQCD prediction for $pp$ collisions,
scaled by the nuclear overlap interval, $T_{AA}$,
or, preferably, to $pp$ spectra measured in the same experiment, 
\begin{equation}
R_{AA}^{\gamma}(p_T)\;=\;\frac{dN_{AuAu}^{{\rm tot}\;\gamma}/dp_{T}}{T_{AA}\cdot
d\sigma_{pp}^{{\rm pQCD}\; \gamma}/dp_{T}} \,\, ,
\label{eq:RAA}
\end{equation}
shown on the right side of Fig.~\ref{fig:thermal_photon_comparison}.
The data are consistent with a significant excess over
the next-to-leading order (NLO) pQCD predictions.
However, we emphasize that below
$p_T \approx 4$~GeV it is not yet clear to what extent the NLO
predictions in the denominator of Eq.~(\ref{eq:RAA}) are
applicable\footnote{The denominator
may be modified by the ``isospin effect'' or other factors, 
see Sec.~\ref{sec:photonraa}.}. In this regime, the calculated prompt 
yields are dominated by jet Bremsstrahlung,
determined from the parametrized parton-to-photon GRV form
factor~\cite{grv_photons}, rather poorly known at the
pertinent $p_T$. The standard scale uncertainties in the NLO pQCD
calculations are $\pm 20$\% for $p_T \geq 4$~GeV but could
become as large as $^{+50}_{-200}$\% for $1 \leq p_T \leq 4$~GeV, as
indicated by the dash-dotted lines on the right side of
Fig.~\ref{fig:thermal_photon_comparison}. Obviously, precise
measurements of the direct photon baseline spectra in $pp$ and
d+Au collisions at $\sqrt{s} =200$~GeV above $p_T =1$~GeV are
essential for quantifying a thermal signal in Au+Au
collisions.

The measured slope of the resulting
thermal photon spectrum, $T_{\rm eff}$, does not directly
reflect the temperature of the hot matter since
photons are emitted throughout the space-time volume
of the evolving matter, implying varying temperatures as well as
blue shifts due to collective expansion. Nevertheless, a correlation
between the apparent photon slope and the maximum temperature
attained in the system persists, as in the recent hydrodynamical study within
\cite{d'Enterria:2005vz}.
The measured $T_{\rm eff}$ provides an
empirical link to the effective number of degrees of freedom of
the system via $d_{\rm of} = (30/\pi^2)(\epsilon/T_{\rm eff}^4)$ or
$d_{\rm of} = (45/2\pi^2)(s/T_{\rm eff}^3)$~\cite{d'Enterria:2005vz}.
The initial maximum energy and entropy densities are difficult to
access experimentally. Indeed, all observables related to
the initial energy and entropy densities such as the total transverse energy,
the total particle multiplicity, and the colored-particle density
encountered by quenched jets on their path through the medium are
related to space-time averaged quantities. Information on the
temperature dependence of $d_{\rm of}$ can, however, be obtained via centrality
and $\sqrt{s_{_{NN}}}$ dependencies. It is
possible to discriminate a QGP-like equation of state with a fixed
$d_{\rm of}$ above $T_c$ from a hadronic resonance gas
with a rapidly rising number of degrees of freedom, by establishing
the dependence of $T_{\rm eff}$ on the charged particle pseudorapidity 
density \cite{d'Enterria:2005vz}. As a further consistency check, one can
relate $d_{\rm of}$ to suitable powers of the energy and entropy
densities, $d_{\rm of} \propto s^4/\epsilon^3$~\cite{Muller:2005en}.


\subsubsection{Chemical Off-Equilibrium}
While rapid thermalization of the matter at full RHIC energy is
fairly well established, its composition in terms of quark and gluon
degrees of freedom (chemical equilibration) is much less clear, as
are the equilibration mechanisms themselves\footnote{The expansion
in hydrodynamic simulations is mostly driven by the ratio of
pressure to energy density, $P/\epsilon$, where $d_{\rm of}$ 
essentially drops out.}. From the prevalence of
gluons in the relevant $x$-range of the incoming nuclei at
midrapidity, one might expect the early matter to be a gluon plasma
(GP), as is routinely assumed {\em e.g.} in
jet-quenching by radiative energy
loss~\cite{Gyulassy:2003mc}. However, recent calculations of $q\bar
q$ pair production within the classical fields generated by the
incoming Au nuclei indicate a rather fast approach to chemical
equilibrium~\cite{Gelis:2005pb}. This rapid equilibriation could have important
consequences for disentangling the relevant fast parton energy loss
mechanism (with a significantly reduced radiative loss for quarks due to
the smaller color charge). EM probes are an obvious means of testing 
chemical equilibration since gluons carry no electric charge 
\cite{Rapp:2000pe,Shuryak:1992bt,Strickland:1994rf,Kampfer:1995mg,Srivastava:1996qd}.
LO pQCD processes ($q+g\to \gamma + X$, $q\bar q\to l^+l^-$) in
a thermally equilibrated but chemically off-equilibrium QGP suggest that
the photon and dilepton production rates scale with
$\lambda_g~\lambda_{q,\bar q}$ and $\lambda_{q}~\lambda_{\bar q}$,
respectively, where $\lambda_i$ are fugacities characterizing the
deviation of the parton densities from chemical equilibrium 
($\lambda_{q}=\lambda_{\bar q}=\lambda_g=1$).  At RHIC, a typical GP 
initial state with subsequent evolution
using inelastic pQCD reaction rates starts from
$\lambda_g\simeq1/3$, $\lambda_{q,\bar q}<0.1$ evolving to values
of $\lambda \simeq 0.5$ or larger
\cite{Srivastava:1996qd,Eskola:1995bp,Elliott:1999uz}. However,
in an isentropic expansion with fixed initial entropy,
undersaturated matter implies significantly higher initial
temperatures at otherwise identical conditions; {\em e.g.} for central
Au+Au collisions at RHIC, with $\tau_0=1/3$~fm/$c$, 
$T_0 \simeq 370$~MeV in a chemically-equilibrated system
relative to $T_0 \simeq 670$~MeV in a system off chemical equilibrium. 
In the photon spectra, reduced fugacities in the emission 
rate are largely compensated by the
higher temperatures in the QGP evolution with only a slight
hardening of the slope parameter~\cite{Gelis:2004ep}.  This 
effect appears to be more pronounced for thermal
dilepton spectra, see the left-hand side of Fig.~\ref{fig:qgprad}. 
Proving that a harder slope is evidence for a GP at RHIC is 
further complicated by the
fact that the thermal yields are still fairly sensitive to the
assumed thermalization time, $\tau_0$. A decrease in
$\tau_0$ with chemical equilibration not only decreases the slope of
the thermal spectrum but also increases the
yield. Identifying a GP or more generally, the number of degrees of
freedom above $T_c$ with thermal dileptons will thus necessarily 
involve a quantitative assessment of {\em both} slope and absolute
magnitude of the thermal spectrum, after ``removal" of non-thermal
sources including Drell-Yan dileptons, correlated
open-charm decays, as well as pre-equilibrium and jet-plasma
interactions. In the following section we briefly summarize recent
progress on the last two sources.

\subsubsection{Pre-Equilibrium and Jet-Plasma Emission}
\begin{figure}[!t]
\begin{minipage}{0.47\linewidth}
\begin{center}
\hspace{-0.2cm}
\includegraphics[width=0.98\linewidth,height=0.74\linewidth]{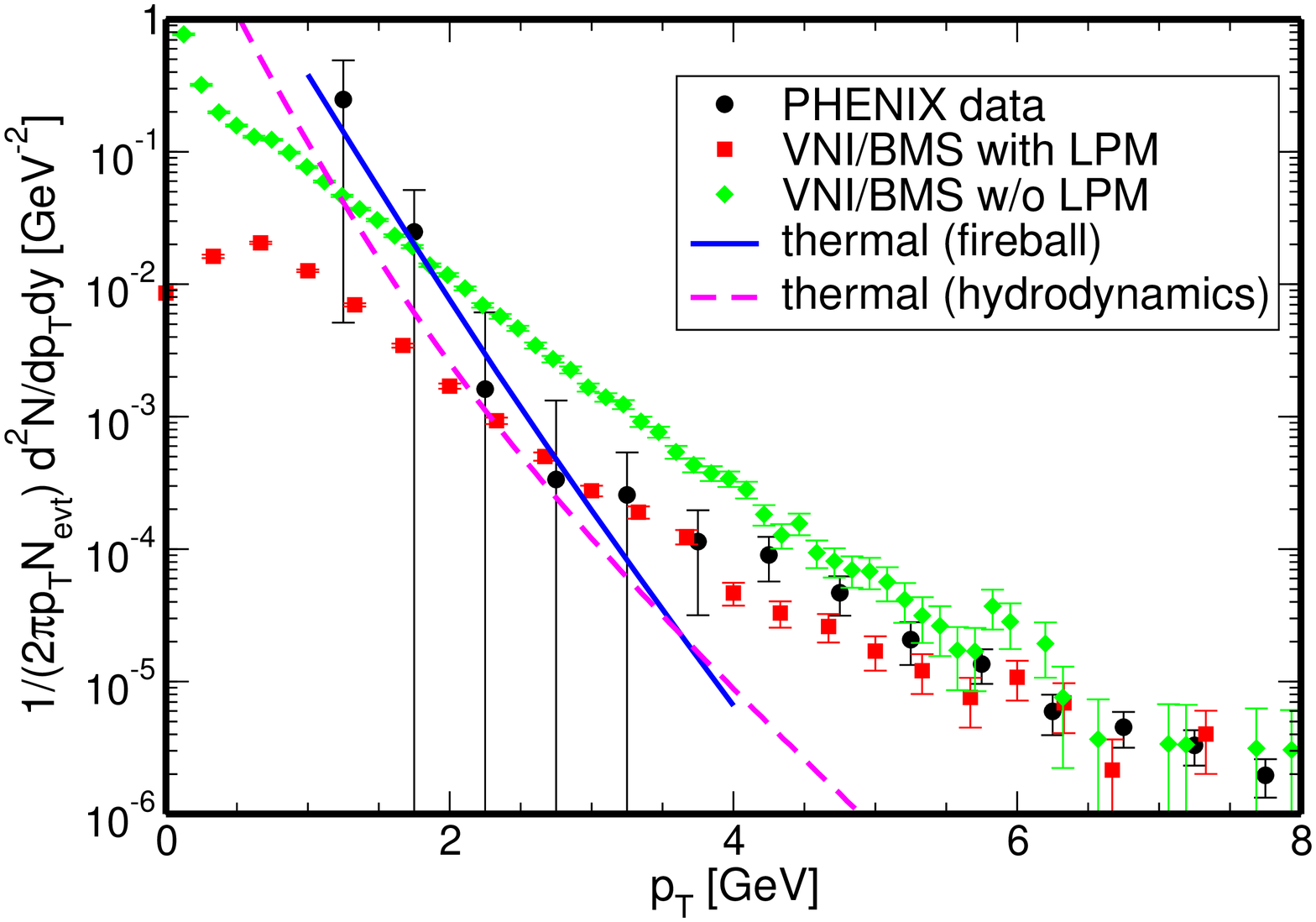}
\end{center}
\caption{Comparison of (early) photon emission from a parton
cascade for central 200 GeV Au+Au collisions \cite{Renk:2005yg}
with (squares) and without (diamond) LPM effects.
Also shown are expanding fireball~\cite{Renk:2004gy} and
hydrodynamic~\cite{Srivastava:2004xp} calculations of
thermal photons, as well as preliminary data from
PHENIX~\cite{Sakaguchi:2005hj}.}
\label{fig_pcm}
\end{minipage}
\hspace{0.8cm}
\begin{minipage}{0.47\linewidth}
\begin{center}
\includegraphics[width=0.99\linewidth]{phot-phenix-theo.eps}
\end{center}
\caption{Predictions for direct photon spectra combining
pQCD prompt photons, jet-plasma interactions~\cite{Turbide:2005fk}
and thermal radiation~\cite{Turbide:2003si} compared
to preliminary PHENIX data~\cite{Bathe:2005nz}. (Here ``Th-Th" and ``HG" are
thermal radiation from the QGP and hadron gas respectively.)}
\label{fig_jet-qgp}
\end{minipage}
\end{figure}

Dilepton emission subsequent to the initial hard $NN$ collisions
but before the assumed $\tau_0$, the so-called
pre-equilibrium contribution, can be addressed in a parton cascade
approach. A corresponding calculation~\cite{Bass:2002pm} predicted large 
emission rates, which, in fact, overestimate preliminary
PHENIX data~\cite{Sakaguchi:2005hj} (Fig.~\ref{fig_pcm}).
However, if Landau-Pomeranchuk-Migdal (LPM) interference
effects are included, the yield is appreciably suppressed~\cite{Renk:2005yg},
indicating that the combined initial pQCD plus pre-equilibrium yield is
not very different from the pQCD contribution alone (Fig.~\ref{fig_pcm}),
with the thermal yield dominating at $p_T\leq 2.5$~GeV,
similar to the center of Fig.~\ref{fig:qgprad}. Also
note that the calculated thermal spectra in Fig.~\ref{fig_pcm} 
agree reasonably well with those in the center panel of 
Fig.~\ref{fig:qgprad}. However, the
present model dependencies will have to be further reduced to
achieve enough sensitivity to discern the composition of the early
matter and realize the desired temperature measurement.

Fig.~\ref{fig_jet-qgp} shows the combined prediction for direct
photons from prompt (pQCD) production, jet-plasma 
interactions~\cite{Turbide:2005fk}, and thermal QGP and HG 
radiation~\cite{Turbide:2003si}
(see center panel of Fig.~\ref{fig:qgprad}) evaluated
within the same expanding fireball. The comparison to preliminary
low-$p_T$ PHENIX data~\cite{Bathe:2005nz} is quite encouraging.
In this calculation, jet-plasma contributions exceed the thermal
yield at $p_T\simeq2$~GeV, implying a rather narrow
``QGP window". On the other hand, at $p_T\simeq5$~GeV, where
jet-plasma radiation \cite{Turbide:2005fk} 
is still significant, the experimental direct-photon 
$R_{AA}$ in the right panel of 
Fig.~\ref{fig:thermal_photon_comparison} does not indicate much 
room for enhancement over prompt pQCD photons. 
Note, however, that thermal and jet-induced
radiation are not independent contributions since both are affected
by the lifetime and temperature
or number and energy density characterizing the QGP, imposing an additional
consistency requirements. Furthermore, as emphasized in 
Refs.~\cite{Turbide:2005fk,Turbide:2006mc}, the intensity of
electromagnetic radiation from jet-plasma interactions is intimately
related to radiative energy loss, {\it i.e.}, jet quenching via gluon
emission. In this
context, the estimates for the jet-plasma photon and dilepton
contribution in Refs.~\cite{Turbide:2005fk,Turbide:2006mc} constitute
an upper limit, since high-$p_T$ pion suppression has been entirely 
attributed to radiative energy loss whereas recent calculations find 
a significant role played by elastic energy loss~\cite{Wicks:2005gt}, 
even within the same formalism~\cite{Qin:2007rn}. 

\subsection{In-Medium Spectral Functions below and above $T_c$}
\label{sec_spec-fct}
In the mid 1990's, dilepton data from the CERN SPS triggered vigorous
theoretical activity in trying to assess modifications of vector-meson
properties in hot/dense (hadronic) matter. The focus has been on
the $\rho$ meson due to its prevalent role in dilepton emission, see,
{\em e.g.} Refs.~\cite{Rapp:1999ej,Alam:1999sc,GH03,Harada:2003jx,Brown:2003ee}
for recent reviews.
In Secs.~\ref{sec_had} and \ref{sec_drop},
we briefly summarize some of the main features and insights
that have emerged over the last $\sim 10$ years.

\subsubsection{Hadronic Many-Body Theory and Chiral Virial Expansion}
\label{sec_had}
Effective hadronic models for vector mesons should be compatible
with basic symmetry principles, most notably electromagnetic gauge 
invariance,
vector-current conservation, and chiral symmetry\footnote{In many
instances, little is known
about the chiral structure of baryonic and mesonic resonance
couplings,  especially if no pions are involved.}. 
In addition, it is essential that the underlying
effective vertices are carefully constrained by phenomenological
information such as hadronic and electromagnetic decay widths or
scattering data.  
Measurements which provide information on in-medium effects near nuclear matter
density are particularly valuable, {\em e.g.} photoabsorption
data on both nucleons and nuclei~\cite{Rapp:1997ei,Steele:1997tv}.

\begin{figure}[!t]
\begin{center}
\includegraphics[width=0.45\linewidth,angle=90]{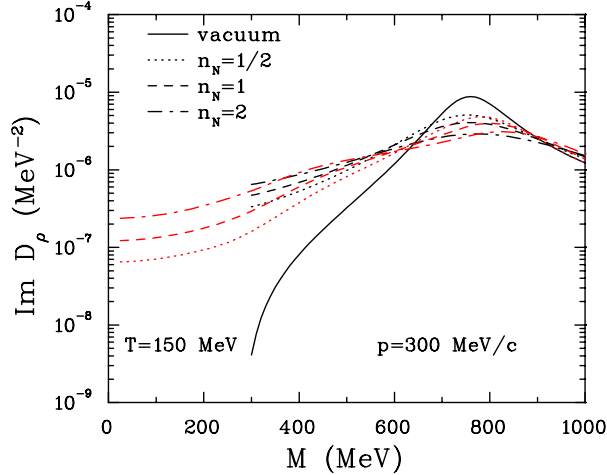}
\end{center}
\caption{Comparison of $\rho$ spectral functions in
hot hadronic matter within a many-body calculation~\cite{RW99}
(gray lines, extending to $M$=0) and an approach based on
imaginary parts of $\rho$-hadron scattering
amplitudes supplemented by dispersion relations to obtain the real
parts~\cite{Eletsky:2001bb} (black lines). Reasonable agreement
between the in-medium results is observed, especially for nucleon densities,
$n_N \le 1$ (in units of the nucleon saturation density, 0.16~fm$^{-3}$).}
\label{fig_Arhocomp}
\end{figure}

Most of the effective models with constraints built along the above
lines have reached a reasonable degree of agreement with the data, generically 
predicting substantial broadening in matter with little
mass shift\footnote{A simple explanation of this feature is
that imaginary parts of the in-medium self energies, which govern the
broadening, are negative definite (${\rm Im} \, \Sigma<0$) and therefore
strictly sum up whereas real parts, which induce mass shifts,
change sign around a resonance. Real parts therefore tend to cancel
if the system is characterized by a rich excitation spectrum, as is
the case for a hadronic resonance gas.}, characteristic of calculations
both in cold nuclear
matter~\cite{Klingl:1997kf,Rapp:1997fs,Post01,Lutz:2001mi,Cabrera02}
and in hot and dense matter~\cite{RW99,Eletsky:2001bb}, see 
Fig.~\ref{fig_Arhocomp} for two examples. Effective models also
suggest that the effects of the baryonic component of the medium
dominate over those from the mesonic one at comparable density,
consistent with findings in large-$N_c$ QCD where
meson-meson interactions are suppressed relative to meson-baryon
ones).

\begin{figure}[!t]
\hspace{-5cm}
\includegraphics[width=0.27\linewidth,height=0.25\linewidth]{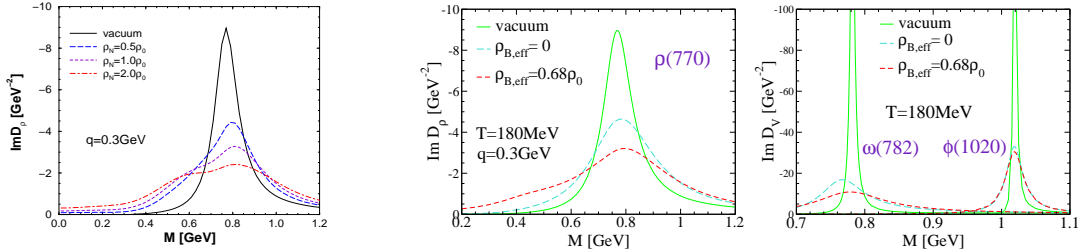}
\hspace{6cm}
\includegraphics[width=0.31\linewidth]{ArhoTr180.eps}
\includegraphics[width=0.31\linewidth]{Aom-phTr180.eps}
\vspace*{-5mm}
\caption{In-medium vector meson spectral functions within hadronic
many-body theory.
Left:
the $\rho$ in cold nuclear matter~\cite{Urban:1998eg}
at several densities proportional to the saturation density, 
$\varrho_0 = 0.16$~fm$^{-3}$.
Center:
$\rho$ behavior at RHIC conditions close to $T_c$
with and without baryon effects~\cite{Rapp:2000pe}.
Right:
same as the center panel but for $\omega$ and $\phi$~\cite{Rapp:2000pe}.
}
\label{fig:Vspec}
\end{figure}

The broadening of the spectral functions, amounting to a total width
of $\sim$500~MeV at nuclear matter saturation density and typically
accompanied by a slight {\em upward} mass shift (left panel of
Fig.~\ref{fig:Vspec}), is in fair agreement with constraints from
QCD sum rules~\cite{Leupold:1997dg}, recall Fig.~\ref{fig_qcdsr}.
When extrapolated to temperatures and densities close to the
expected chiral transition, an almost complete ``melting" of the
$\rho$-resonance structure emerges. This is not only true for the
net-baryon rich regime at SPS energies and below, but also in the
central rapidity region at collider energies where the baryon
chemical potential is small (center panel of
Fig.~\ref{fig:Vspec}). 
At the experimentally extracted chemical freezeout temperature, {\em e.g.}
$T_{\rm ch}\simeq180$\,MeV at RHIC, an appreciable density of
baryon-antibaryon, $B \overline B$, pairs is thermally excited
\cite{Rapp:2000pe} and mesons interact with both baryons and antibaryons.
In addition, the notion of chemical freeze-out implies that
baryon-antibaryon annihilation in the subsequent hadronic evolution
is suppressed~\cite{Rapp:2000gy}. 
Thus, the relevant quantity for medium effects on vector mesons 
is the sum of the $B$ and $\bar B$ densities
which, close to $T_{\rm ch}$, is quite comparable at $\mu_B =0$ and
$\mu_B=250$~MeV\footnote{Experimentally, the total baryon rapidity
density, $dN_{B+\bar B}(y=0)/dy$, is indeed very comparable at
the maximum SPS ($\sqrt{s_{_{NN}}} =17.3$~GeV) and RHIC ($\sqrt{s_{_{NN}}}
=200$~GeV) energies.  The total hadron rapidity density (mostly due to pions) 
is a factor of $\sim 2$ larger at RHIC, implying an accordingly lower
total baryon density at the transition.  However, most of the
pertinent medium effects on the $\rho$ spectral function build up at
densities at or below $\varrho_0$, see the left panel of
Fig.~\ref{fig:Vspec}.}. The baryon-density effects on the $\rho$ are
most pronounced at masses below $\sim 0.5$~GeV instead of at and above the
free $\rho$ mass (compare the long-dashed and short-dashed lines in
the center panel of Fig.~\ref{fig:Vspec}). While the
$\phi$ appears to be less sensitive to the baryonic component
of the medium, this conclusion may be altered once 
a better understanding of recent photon- and proton-nucleus
$\phi$ production data has been
achieved~\cite{Ishikawa05,Muto05}.

The $\rho$ and possible $\omega$ ``melting" has interesting
implications that deserve further theoretical investigations:
\begin{itemize}
\item[(i)] The very short mean-free path of the $\rho$ (and other
hadrons) close to $T_c$ is suggestive of hadronic liquid 
formation~\cite{Voskresensky:2004ux}.  Thus from a hadron structure objective, 
the matter properties in the phase
transition region may change rather smoothly from a hadronic liquid to the
sQGP liquid.
\item[(ii)] The hadronic in-medium EM correlator, based on a ``melted"
$\rho$, is surprisingly similar in shape and magnitude to the
QGP correlator evaluated in HTL pQCD~\cite{Braaten:1990wp} at
{\em all} masses~\cite{RW99,Rapp:1999if}.  This is also suggestive
of a rather continuous transition from HG to QGP close to $T_c$,
even on the level of spectral functions. The approximate coincidence
of the bottom-up and top-down extrapolated hadronic and pQCD 
calculations, while not understood theoretically, enhances our
confidence in applying these emission rates to fireball and hydrodynamic 
evolution across $T_c$, necessary for the space-time
integrated thermal dilepton spectra, rendering the dilepton spectra 
rather insensitive to the exact value of 
$T_c$~\cite{vanHees:2007th}.
\end{itemize}

Similar conclusions about the $\rho$ also emerge from the
chiral virial approach~\cite{Steele:1997tv} where medium
effects on the vacuum vector correlator are evaluated within a pion-
and nucleon-density expansion coupled with vacuum $V \pi$ and
$V N$ scattering amplitudes constrained by chiral symmetry.  The
$\rho$ peak is quenched, though not broadened, and its
low-mass shoulder is substantially enhanced, predominantly due to baryon 
effects in the heat bath~\cite{Steele:1999hf}. This
agreement, at least at low and moderate densities and temperatures, is
a consequence of the constraints imposed on the underlying
hadronic interactions.

\subsubsection{Dropping Mass}
\label{sec_drop}
Models involving dropping vector-meson masses~\cite{Brown:1991kk}
have recently been revisited within the so-called vector
manifestation of chiral
symmetry~\cite{Harada:2003jx,Sasaki05,Hidaka05}. Using the Hidden
Local Symmetry (HLS) framework, where the $\rho$ mass
is generated via a Higgs mechanism, an alternative representation of
the chiral group has been proposed in which the chiral partner of the
pion is the longitudinal $\rho$ meson, the so-called
``vector manifestation" of chiral symmetry instead of the conventional 
realization where the $\sigma$ is the chiral partner
of the pion. The HLS approach results in a satisfactory vacuum phenomenology.
A renormalization group analysis with hadronic loop effects reveals a
fixed point with a vanishing vector coupling constant. When applied to
the second order finite temperature chiral phase transition, matching 
the vector and axial vector correlators to the operator product expansion
(space-like $q^2$) requires that bare $\rho$
mass vanish at the critical point, becoming degenerate with the pion mass,
which persists when
carried on-shell due to the fixed-point nature of the transition. 
The vector dominance model (VDM), which works
well in the vacuum, is perhaps violated at finite temperature~\cite{Hidaka05}, 
suppressing $\pi\pi$ annihilation to
dileptons via an intermediate $\rho$, replaced by direct
annihilation via intermediate photons.  The violation of vector dominance
could make the observation of a dropping $\rho$ mass difficult
in dilepton spectra, perhaps reconciling the new NA60
data~\cite{Damjanovic:2005ni,Arnaldi:2006jq} with a dropping-mass
scenario.  Ref.~\cite{Harada:2006hu} calculates
the dilepton rates in the vector manifestation
scenario. Despite violation of the VDM, the rates clearly exhibit
a $\rho$ peak with dropping mass, at least up to
$T\sim 0.85 T_c$. We emphasize that the matching
procedure is only valid sufficiently close to $T_c$ and that
``ordinary" hadronic medium effects become
dominant at lower temperatures. In the low-temperature limit, modification
of the $\rho$ mass to leading order in temperature, ${\mathcal{O}}(T^2)$,
is at variance with chiral symmetry. Thus, the notion of a ``flash 
temperature" has been introduced~\cite{Brown:2003ee,Brown:2004qi}, 
below which the ``intrinsic" temperature dependencies of the parameters 
in the Lagrangian are void.     

At the quark-antiquark level, interactions in the vector channel are 
believed to be rather weak since the $\rho$
mass is nearly twice the constituent quark mass.  Finite
temperature effects resulting in enhanced interactions and an
accordingly reduced $\rho$ mass are not
easily conceived~\cite{Momchil97}. (See Ref.~\cite{Brown04}
for an alternative view.) It would also be interesting to work out
how the presence of hadronic many-body effects (especially baryons),
as discussed in the previous section, affect the matching procedure
and the resulting (axial) vector spectral functions.

\subsubsection{Resonances in the sQGP}
\label{sec_reso}
Another interesting development perhaps related to EM measurements 
are the conjectured hadronic bound states in the
(s)QGP~\cite{Shuryak:2003ty,Mannarelli:2005pz}.
Detecting signatures of vector states above $T_c$ in the
dilepton spectrum hinges on whether their mass is sufficiently large,
$M_V(T\ge T_c) > 1$\,GeV. As elaborated after Eq.~(\ref{Piem}),
QGP radiation can only compete with or dominate
contributions from the longer-lived and larger-volume hadronic phase
at large masses, especially if the resonance structure depends on temperature.
The lQCD spectral functions and dilepton rates shown in
Figs.~\ref{fig_spectral} and \ref{fig:rate-lat}, respectively,
indeed indicate resonances
with masses in the $M\simeq 2~{\rm GeV} \simeq 10 T_c$ regime, roughly
scaling with temperature. The existence of these states
may be understood~\cite{Shuryak:2003ty} to be due to heavy-quark 
quasiparticles bound by a rather strong color Coulomb-type
attraction, also present in heavy quarkonium states.  In this case, heavy-quark
symmetry implies approximate degeneracy of vector (``$\rho$" or $J/\psi$)
and pseudoscalar (``$\pi$" or $\eta_c$) states. The connection
between these resonances and the pion mass dropping to near zero when 
approaching $T_c$ from above has been addressed in Ref.~\cite{Brown04}.

\begin{figure}[!t]
\begin{minipage}{0.47\linewidth}
\begin{center}
\includegraphics[width=0.8\linewidth,angle=-0]{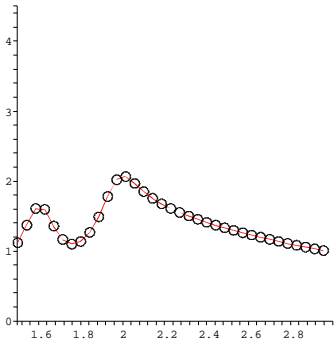}
\end{center}
\vspace{-0.2cm}
\caption{Ratio of dilepton spectra at RHIC from a resonance-enhanced 
QGP rate
to perturbative $q\bar q$ annihilation~\cite{Casalderrey-Solana:2004dc}
obtained from the QGP and mixed phases of an expanding thermal
fireball~\cite{Rapp:2000pe,Rapp:2002mm}.
The first peak is  the vector mass close to $T_c$ while the second is in
the vicinity of the ``zero-binding" line at $T\simeq 1.5-2T_c$. The
$x$-axis is in units of the quark-quasiparticle mass, $M_q\simeq1$~GeV.
The quark-quasiparticle width is assumed to be
0.1~GeV~\cite{Mannarelli:2005pz}.}
\label{fig:dlqgp}
\end{minipage}
\hspace{0.5cm}
\begin{minipage}{0.47\linewidth}
\begin{center}
\includegraphics[width=1.01\linewidth,angle=-0]{dl3rhic017-200.eps}
\end{center}
\vspace{-0.2cm}
\caption{Excitation function of low-mass dilepton spectra for
central Au+Au collisions
from top SPS to top RHIC energy. The upper (lower) set of four curves
corresponds to thermal radiation from hadronic matter (QGP) using
in-medium $\rho$, $\omega$ and $\phi$ spectral
functions~\cite{RW99,Rapp:2002mm} (HTL $q\bar q$
annihilation~\cite{Braaten:1990wp}),
assuming a critical temperature between 175\,MeV at $\sqrt{s_{_{NN}}}=17$ GeV
and 180\,MeV at $\sqrt{s_{_{NN}}} > 100$ GeV.}
\label{fig:excit}
\end{minipage}
\end{figure}

Quantitative signatures of vector resonances above $T_c$
in the dilepton spectrum have been evaluated in
Ref.~\cite{Casalderrey-Solana:2004dc}.
Convoluting the temperature-dependent resonance decays with an expanding
fireball at RHIC~\cite{Rapp:2000pe,Rapp:2002mm} predicts an enhancement 
over baseline pQCD emission scenario ($q\bar q$ annihilation) by about a
factor of $\sim 2$, see Fig.~\ref{fig:dlqgp}.
This result is sensitive to the vector-meson width
which in turn is governed by the width of the quark quasiparticles. 
Smaller widths lead to narrower peaks and thus a stronger
enhancement over the pQCD spectrum. 
The quasiparticle width is expected to be
$\sim 0.2$~GeV based on self-consistent solutions of a $q\bar q$
scattering equation~\cite{Mannarelli:2005pz} using input interaction 
potentials from finite temperature lQCD or Nambu-Jona-Lasinio
four-quark interactions~\cite{Kitazawa:2005mp}. The corresponding
elastic scattering rates of $\sim 1$/(fm/$c$) are suggestive of the
short thermalization times deduced from hydrodynamic analyses of
elliptic flow measurements and could therefore provide a link to
the early thermalization puzzle at RHIC.

\subsection{Low-Mass Dilepton Spectra}
\label{sec_lmdilep}
To illustrate the predictions of medium-modified vector mesons
within effective models, and in particular to investigate the
importance of the baryonic component of the medium, we summarize
a pertinent excitation function of low-mass
dilepton spectra in central Au+Au collisions in Fig.~\ref{fig:excit}. 
Thermal dilepton rates
in the QGP~\cite{Braaten:1990wp} and HG~\cite{RW99,Rapp:2000pe} phases
are convoluted with isentropic fireball evolution similar to
that underlying Figs.~\ref{fig:qgprad} and \ref{fig:dlqgp},
assuming a chemically equilibrated QGP which converts into a
chemically equilibrated HG at ($\mu_N^c$, $T_c$) values compatible
with: (i) thermal models for hadron production in central $AA$
collisions~\cite{Braun-Munzinger:2003zd} and (ii) a total entropy
that reproduces available particle
multiplicity data~\cite{Back:2004je,Adams:2005dq,Adcox:2004mh}. The
hadronic evolution subsequent to chemical freeze-out is augmented by
effective chemical potentials for hadrons that are stable under
strong interactions ({\em e.g.} $\pi$, $K$, $\eta$, baryons and
antibaryons). This is mandatory to maintain the observed chemical
composition until thermal freeze-out~\cite{Rapp:2002fc} and, in
particular, implies sizable {\em total} ($B+\overline B$) baryon
densities in the later stages of the hadronic evolution (recall
Sec.~\ref{sec_had}).  Surprisingly, there is
no large change in either shape or magnitude of the (hadronic)
dilepton-spectrum excitation function for $20 \leq \sqrt{s_{_{NN}}} \leq
200$~GeV. The
main reasons for this outcome are:
\begin{itemize}
\item[(i)]
Despite the large range in baryochemical potentials
($\mu_B^c = 25-250$~MeV at chemical freezeout with $T_c =175-180$~MeV) 
and thus in net baryon density,
the prevalent baryon-induced medium effects are comparable once
the $B + \overline B$ density is properly accounted for.
\item[(ii)]
The lifetime of the hadronic (and mixed) phase changes little
since the larger volume expansion at higher energies is essentially
compensated by an increase in radial flow inherited from the QGP phase.
\end{itemize}
This scenario should  be contrasted with one where the
medium effects are sensitive to the {\em net} baryon density such as
simple dropping mass parameterizations,
$m_V^*/m_V=(1-C\varrho_B/\varrho_0)\times (1-(T/T_c)^2)^{1/n}$.
In this case,
a stronger variation of the excitation function is anticipated,
with {\em weaker} effects at higher collision energy.
QGP emission increases appreciably with $\sqrt{s}$ but remains
subdominant ($\le$20\%) in the low-mass region at all energies
if no significant $q_t$ cut is applied, as may be expected for larger 
initial temperatures and longer QGP lifetimes.

At masses above $\sim $1\,GeV, the hadronic EM~spectral function is
dominated by four pion and higher contributions, encompassing annihilation
reactions such as $\rho \rho$, $\pi \omega$, and $\pi a_1$.
While these contributions are not included in the hadronic matter calculation
shown in Fig.~\ref{fig:excit}, they may become significant at 
$M \geq 0.9$~GeV where a similar enhancement could be
related to effects of partial chiral symmetry
restoration~\cite{vanHees:2006ng,vanHees:2006iv}. We will return
to this issue in Sec.~\ref{sec_mix}.

Finally, a few remarks on models of the space-time evolution are in order
since these provide crucial input on the thermodynamic
parameters for the equilibrium EM~emission rates. Hydrodynamic
models, if applicable, are the approach of choice since they are formulated 
using the same variables as the thermal emission rates.  The present RHIC data
suggest that ideal hydrodynamics gives a good approximation of
the first $\sim 5$~fm/$c$ after thermalization, encompassing the QGP,
``mixed" phases and possibly the early hot+dense hadronic liquid
phase for $T \geq 150$\,MeV. For lower temperatures, however, viscosity
effects are expected to become
significant~\cite{Hirano:2005wx}. It is presently not clear how finite viscosity
affects calculations of dilepton and photon emission.  The
underlying uncertainties must be scrutinized, especially since
the low-mass and low-momentum spectra receive significant contributions
from later stages.  One can either
implement viscosity effects in hydrodynamics to retain the
notion of thermodynamic variables or switch to transport
theory~\cite{Bass:2000ib,Teaney:2000cw,Lin:2004en}. In the
latter option, it is nontrivial to properly implement
broad resonances~\cite{Cassing:1999mh}. Alternatively, local temperatures 
and (baryon) densities could be extracted from
transport simulations and convoluted with the equilibrium
EM~emission rates. The relative agreement of this method with
viscous hydrodynamics could provide an estimate of the uncertainties
in the integrated EM~spectra and used to better calibrate
fireball models, which are suitable
parametrizations of microscopic evolution. Such comparisons
will become particularly relevant if less penetrating
probes, {\em e.g.} $\pi\pi$ or $\pi\gamma$ invariant-mass spectra, are 
calculated, as
discussed later.  Transport-based approaches are mandatory for thermal 
freezeout to
quantitatively account for finite sizes, lifetimes and mean-free paths.
In the late 1990s,  the agreement of
hydrodynamic~\cite{Huovinen:1998ze}, transport~\cite{Cassing:1997jz}
and fireball models~\cite{RW99,Renk:2002md} with CERES low-mass dileptons
\cite{Agakichiev:2005ai} at the SPS has been reasonable,
albeit with somewhat limited theoretical and experimental precision.

\subsection{Chiral Symmetry Restoration}
\label{sec_csr}
\subsubsection{Direct and Indirect Approach}
\begin{figure}[!t]
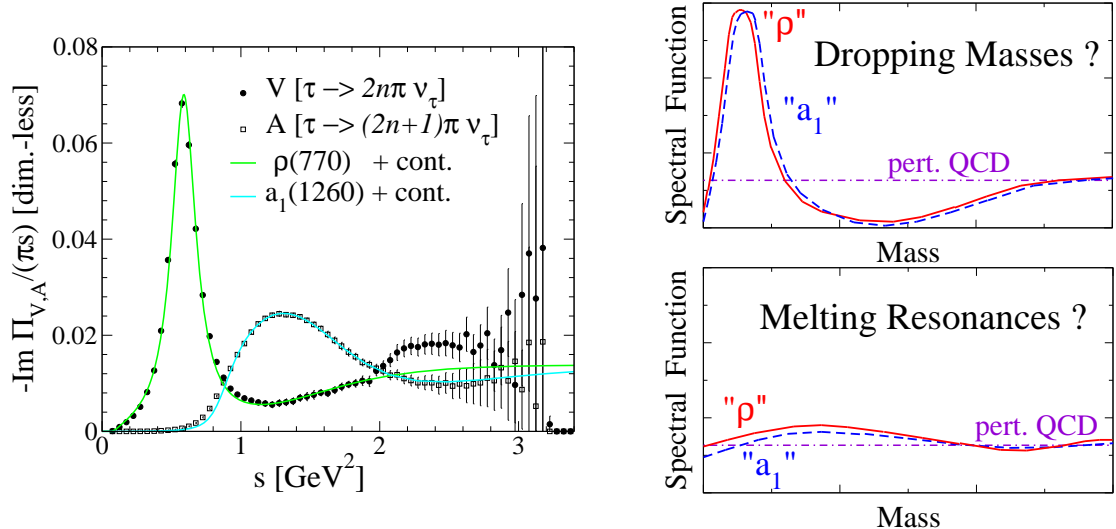

\begin{minipage}{7.5cm}
\epsfig{file=VAfit3.eps,width=7.5cm}
\end{minipage}
\hspace{1cm}
\begin{minipage}{7cm}
\epsfig{file=DM2-scenario.eps,width=6cm}
\epsfig{file=RW2-scenario.eps,width=6cm}
\end{minipage}
\caption{Left: isovector, vector and axial vector spectral 
functions in the vacuum
as measured in hadronic $\tau$ decays~\cite{aleph98} with model fits using
$\rho$ and $a_1$ resonances plus a perturbative continuum~\cite{Rapp02}.
Right: two schematic scenarios of chiral symmetry restoration
in hot and dense matter.}
\label{fig:VA}
\end{figure}

We now address the question of how, in principle, in-medium effects
detected in dilepton spectra can be used to draw conclusions about 
chiral symmetry restoration ($\chi$SR). An
unambiguous consequence of the $\chi$SR is that isovector, vector
and axial vector correlation functions, which are very different in
the vacuum, become degenerate at and beyond the chiral transition.  The
question is {\it how} this happens, see Fig.~\ref{fig:VA} for an
illustration. We reiterate that the effects of chiral symmetry
breaking are concentrated at low masses since, already in vacuum, the
correlators become degenerate in the pQCD regime.  Therefore, symmetry breaking 
constitutes an inherently nonperturbative phenomenon which, 
ideally, can be addressed
with input from experiment, theoretical models and lattice QCD 
computations.

A direct way to search for $\chi$SR is {\em measurement} of the
in-medium axial vector spectral function in connection with model
comparisons, as also done for dileptons. 
Measurements of the $\pi^\pm\gamma$
invariant-mass spectra were suggested \cite{Rapp:2003ar} as a means of 
searching for $\chi$SR. This was partly motivated by similar 
measurements of $\pi^+\pi^-$ invariant-mass spectra in heavy-ion
collisions which indicated $\rho$ modifications in the late stages of 
peripheral 200~GeV Au+Au
collisions at RHIC~\cite{Adams:2003cc}. Absorption effects on the
outgoing pions limit the $\pi\pi$ information to rather
dilute stages while $\pi\gamma$ spectra may probe somewhat 
deeper into the fireball. However, emission from later collision
stages is advantageous since a narrower density and temperature window
is probed so that the convolution over the
space-time history becomes less of an issue. Experimentally the
challenges are the rather low rates\footnote{Even though the radiative decay
branching ratio of the $a_1$ is small, the (not so small)
absolute decay width, $\Gamma_{a_1\to\pi\gamma}\simeq0.7$~MeV, is 
the relevant quantity for thermal radiation.} and the rather
broad structure of the $a_1(1260)$ resonance ($\sim 0.4$~GeV in
vacuum), making it susceptible to distortions due to background
subtractions.  Simulations will be shown in Sec.~\ref{sec:a1}.

In addition to the direct experimental approach to the axial vector
channel, we now formulate a well-defined theoretical procedure based
on chiral hadronic models to connect experimental information on
the vector correlator (dileptons) to (first-principle) information
on $\chi$SR from lattice QCD. 
The in-medium versions of the chiral sum rules, Eqs.~(\ref{wsr1med}),
(\ref{wsr2med}) and (\ref{wsr3med}), critical to the calculation, are
obtained as follows:
\begin{itemize}
\item[(1)] First, calculate the vector ($V$) and axial vector ($A$) spectral
      functions as a function of temperature and density 
      in a chirally invariant model, including as many of the 
      constraints as possible (see~Sec.~\ref{sec_had}).
\item[(2)] Then insert the spectral functions into the Weinberg sum rules 
to evaluate the
     temperature dependence of the pion decay constant and four-quark
     condensate and compare to lattice QCD results.  Note that
     $f_\pi(T)$ and $\langle (\bar qq)^2\rangle (T)$ are presumably
     more easily evaluated in lQCD than spectral functions
     since lQCD is primarily applicable to the finite temperature axis,
     {\em i.e.}, at $\mu_q$=0, the closest relation between lQCD and 
heavy-ion experiments
     is realized in the central rapidity regions at RHIC and the LHC.
\item[(3)] Finally, perform detailed comparisons of the in-medium
effects on the
     vector correlator with dilepton data as a function of centrality, 
$\sqrt{s_{_{NN}}}$, mass and $q_t$-spectra.  The comparison requires 
additional input
     from realistic expansion models ({\em e.g.} hydrodynamical and transport
     simulations), which have been thoroughly tested against 
the large body of hadronic
     observables.
\end{itemize}
The three different energy moments of $\rho_V-\rho_A$
probed by the chiral sum rules, provide detailed constraints
on the energy dependence of the in-medium spectral functions.
In addition, each in-medium chiral sum is valid for a given three-momentum,
providing further kinematic information.
Therefore, if a chiral hadronic model complies with both theoretical
(2) and experimental (3) tests, a tight
connection between lattice QCD and data has been established, producing
explicit evidence for chiral symmetry restoration
without a direct measurement of the axial vector correlator.
In the absence of (unquenched) lattice data for (low-mass) spectral
functions for at least the next ten years, a systematic approach 
involving effective models is the {\em only} way to 
interpret data in terms of $\chi SR$.  Experimental
guidance is crucial for progress in understanding the underlying
nonperturbative physics.

\subsubsection{LMR-IMR Transition: Chiral Mixing}
\label{sec_mix}
In a low-temperature pion gas, the expectation values of vector
and axial vector correlators can be evaluated model-independently
based on chiral reduction formulae in connection with a low-density
expansion. The leading medium effect has first been derived in 
Ref.~\cite{DEI90} in the chiral limit ($m_\pi=m_{u,d}=0$) and 
amounts to chiral correlator mixing,
\begin{eqnarray}
\Pi_V(q) &=& (1-\epsilon)\, \Pi_V^{\rm vac}(q)
             + \epsilon \, \Pi_A^{\rm vac}(q)
\nonumber\\
\Pi_A(q) &=& (1-\epsilon)\, \Pi_A^{\rm vac}(q)
             + \epsilon \, \Pi_V^{\rm vac}(q) \ ,
\label{mix}
\end{eqnarray}
where $\epsilon=T^2/6f_\pi^2$ encodes the thermal pion density 
(smaller for $m_\pi>0$).  Interactions of the vector current with
pions from the heat bath quench the vacuum vector
correlator\footnote{The shape of the correlator is unaffected.} and its 
admixture of the axial vector correlator induced by
$V+\pi\to A$ and $A+\pi\to V$ processes.  The axial vector current is 
analogously affected. Full mixing corresponds to $\epsilon =1/2$,
implying degenerate correlators.
\begin{figure}[!t]
\begin{minipage}{0.47\linewidth}
\epsfig{file=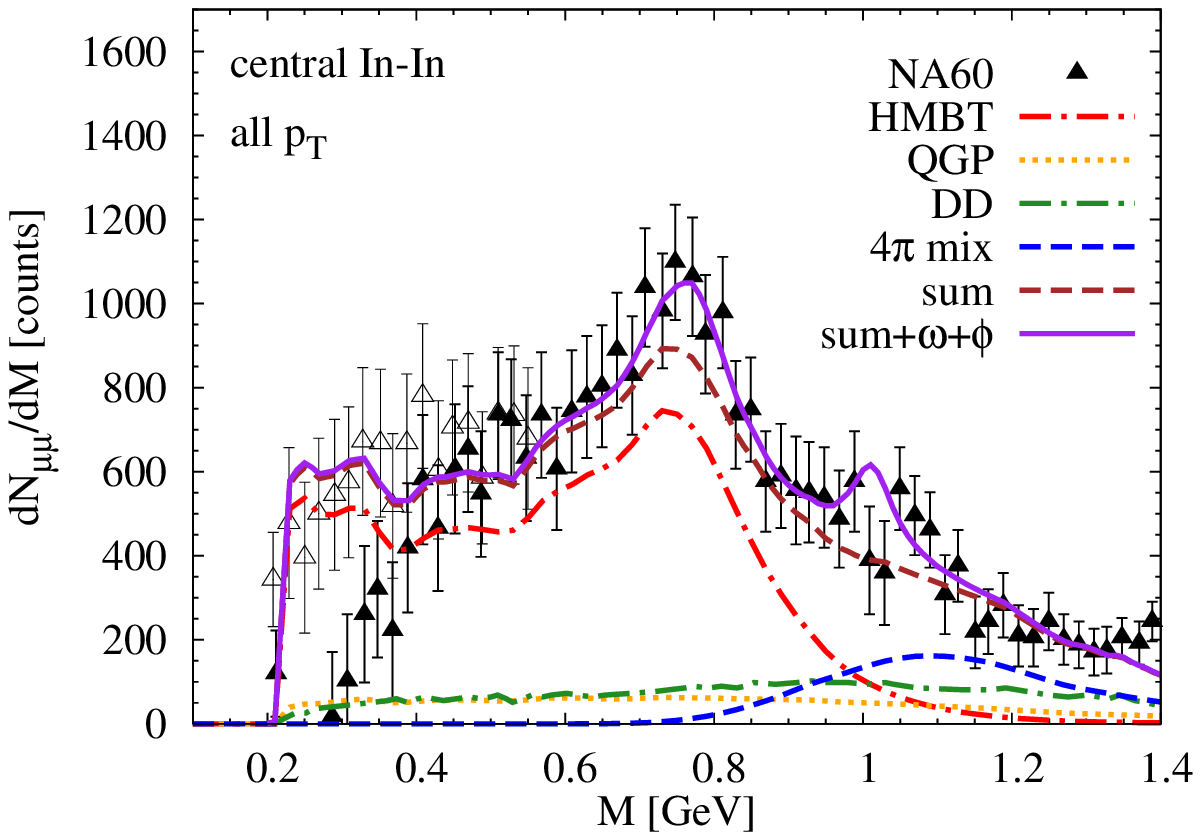,width=\linewidth}
\end{minipage}
\hspace{0.4cm}
\begin{minipage}{0.47\linewidth}
\vspace{-0.2cm}
\epsfig{file=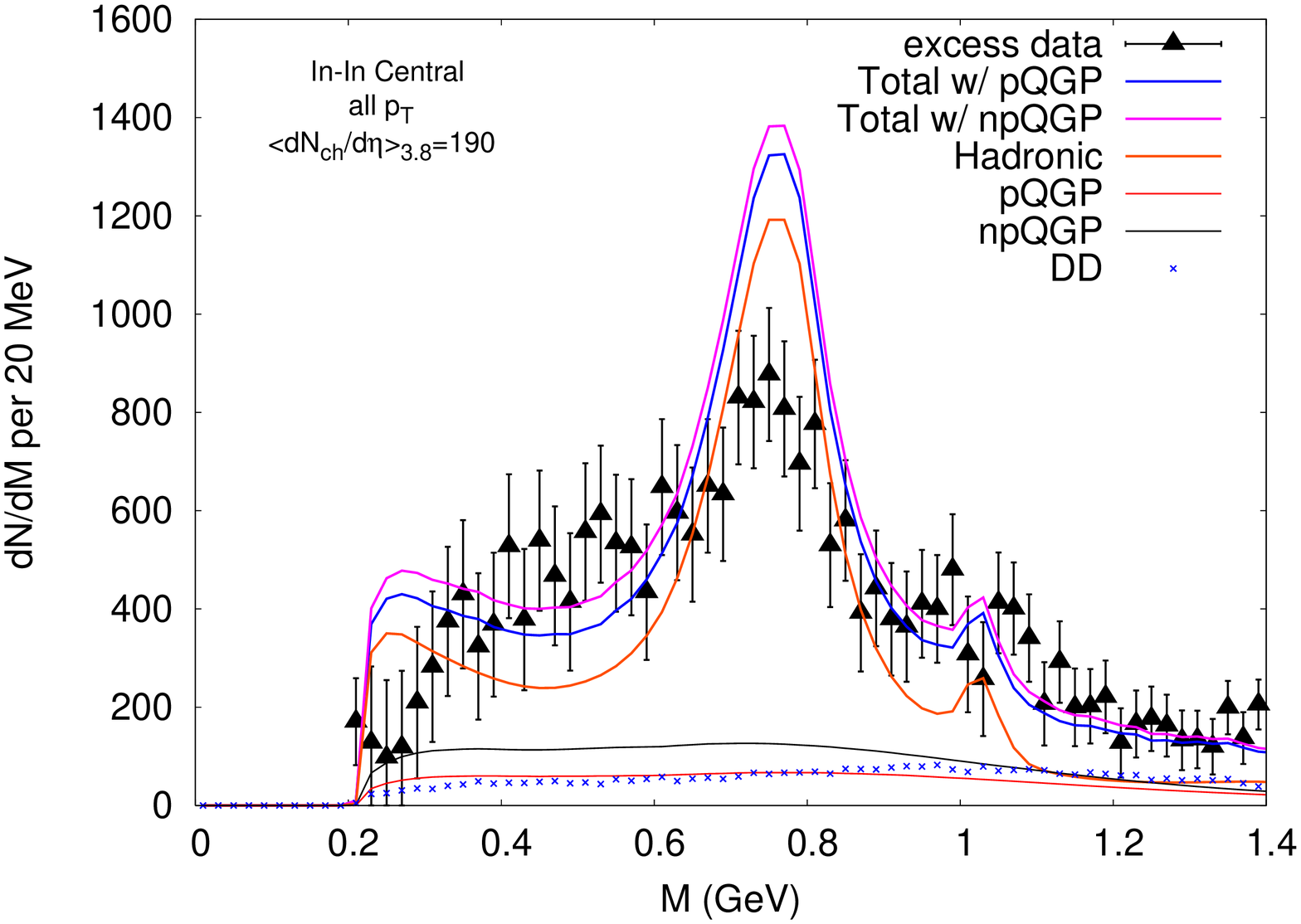,width=1.02\linewidth}
\end{minipage}
\caption{Comparison of NA60 dimuon data~\cite{Damjanovic:2005ni,Arnaldi:2006jq}
in central In (158~$A$GeV) + In collisions with calculations.
Left: an expanding fireball
calculation~\cite{vanHees:2006ng,vanHees:2006iv}
for QGP and hadronic emission.  The hadronic emission is based on  
in-medium $\rho$~\cite{RW99},
$\omega$~\cite{Rapp:2000pe} and $\phi$ spectral functions
as well as four-pion contributions with an
enhancement from chiral mixing according to Eq.~(\ref{mix}).
Right: hydrodymanic convolution~\cite{Dusling:2006yv} of
the dilepton rates following from the chiral virial
expansion~\cite{Steele:1997tv,Steele:1999hf}.}
\label{fig:mix}
\end{figure}
Broadening and possible mass shifts of the $\rho$
are thus due to higher-order effects in both $T$ and $\varrho_N$.
The mixing has the interesting feature of filling in 
the ``dip" in the $s =1-2$\,GeV$^2$ region of the vacuum vector 
correlator\footnote{Full mixing in this regime
leads to degenerate $V$ and $A$ correlators that closely coincide
with the pQCD $q\bar q$ continuum level, interpreted
as lowering the ``duality scale" from $s \simeq 2.5$~GeV$^2$ in
the vacuum to about 1~GeV$^2$ for full mixing.  It is further tempting 
to interpret the $\rho$ melting 
found in hadronic
many-body theory (recall Sec.~\ref{sec_had}) as 
lowering the duality scale as $s\to 0$, implying chiral
restoration~\cite{Rapp:1999if}.}
(see the left panel of Fig.~\ref{fig:VA}).  Dilepton enhancement 
by up to a factor of 2 over the vacuum vector spectral
function in this mass region 
is therefore a signature of the approach to
chiral restoration, via $\pi a_1$ annihilation (four-pion
contributions not present in the vacuum EM~correlator)
or QGP emission.
While the earlier SPS dilepton data did not have the necessary
precision for the required quantitative analysis (recall
Fig.~\ref{fig_spsdilep}),  it may be feasible with the new
NA60 data~\cite{Damjanovic:2005ni,Arnaldi:2006jq} shown in the left panel
of Fig.~\ref{fig:mix}.  The blue dashed curve, a theoretical
upper estimate~\cite{vanHees:2006iv}, employs Eq.~(\ref{mix}) with
$\epsilon(T)=\frac{1}{2} n_\pi(T)/n_\pi(T_c)$ where $n_\pi(T)$ is
the pion density at $T\le T_c$, including pion chemical
potentials below $T_c=175$~MeV, and removing the $a_1\to\pi\rho$ 
decay, included in the $\rho$ in-medium spectral 
function. On one hand, it is gratifying to see that
this calculation properly accounts for the excess spectrum in the
relevant mixing regime.  On the other hand,
the data are still reasonably well described without mixing.  To be sensitive
to the mixing effect, both the data and the theory require an accuracy of
at least 20\%.
The chiral virial approach, when folded over hydrodynamic evolution,
as shown in the right panel of Fig.~\ref{fig:mix}, also describes
the region $M \geq 1$\,GeV well~\cite{Dusling:2006yv}.  The free EM~correlator 
with mixing effects is a key ingredient,
resulting in an enhancement consistent with maximal mixing as the upper 
limit and the free EM correlator as the lower limit.  Although the $\rho$ peak
is quenched (Sec.~\ref{sec_had}), the abscence of $\rho$ resonance 
broadening results in a $\sim 40$\% overestimate of the yield
around the free $\rho$ mass.  The enhancement below the $\rho$ mass
is again accounted for, with important baryon contributions.
In both approaches underlying Fig.~\ref{fig:mix}, the QGP 
yield is small.
The conclusions of Ref.~\cite{vanHees:2006iv} on $\rho$ broadening
and the importance of baryon-driven medium effects in the context of the
NA60 data have been confirmed in Ref.~\cite{Ruppert:2007cr} which, however,
attributes the bulk of the enhancement above 
the $\rho$ mass to QGP radiation.
The relative QGP to four-pion yields in the IMR is essentially determined by 
the choice of $T_c$ in the fireball evolution~\cite{vanHees:2007th}: QGP
dominates the IMR if $T_c = 160$ MeV while hadronic contributions are
dominant if $T_c = 175$ MeV. Irrespective
of whether the source is of QGP or hadronic origin, the IMR 
enhancement is associated with matter at temperatures close to $T_c$.
The robustness of this conclusion is again a consequence of 
``parton-hadron duality" in the underlying emission rates.  

\begin{figure}[!t]
\begin{minipage}{0.31\linewidth}
\epsfig{file=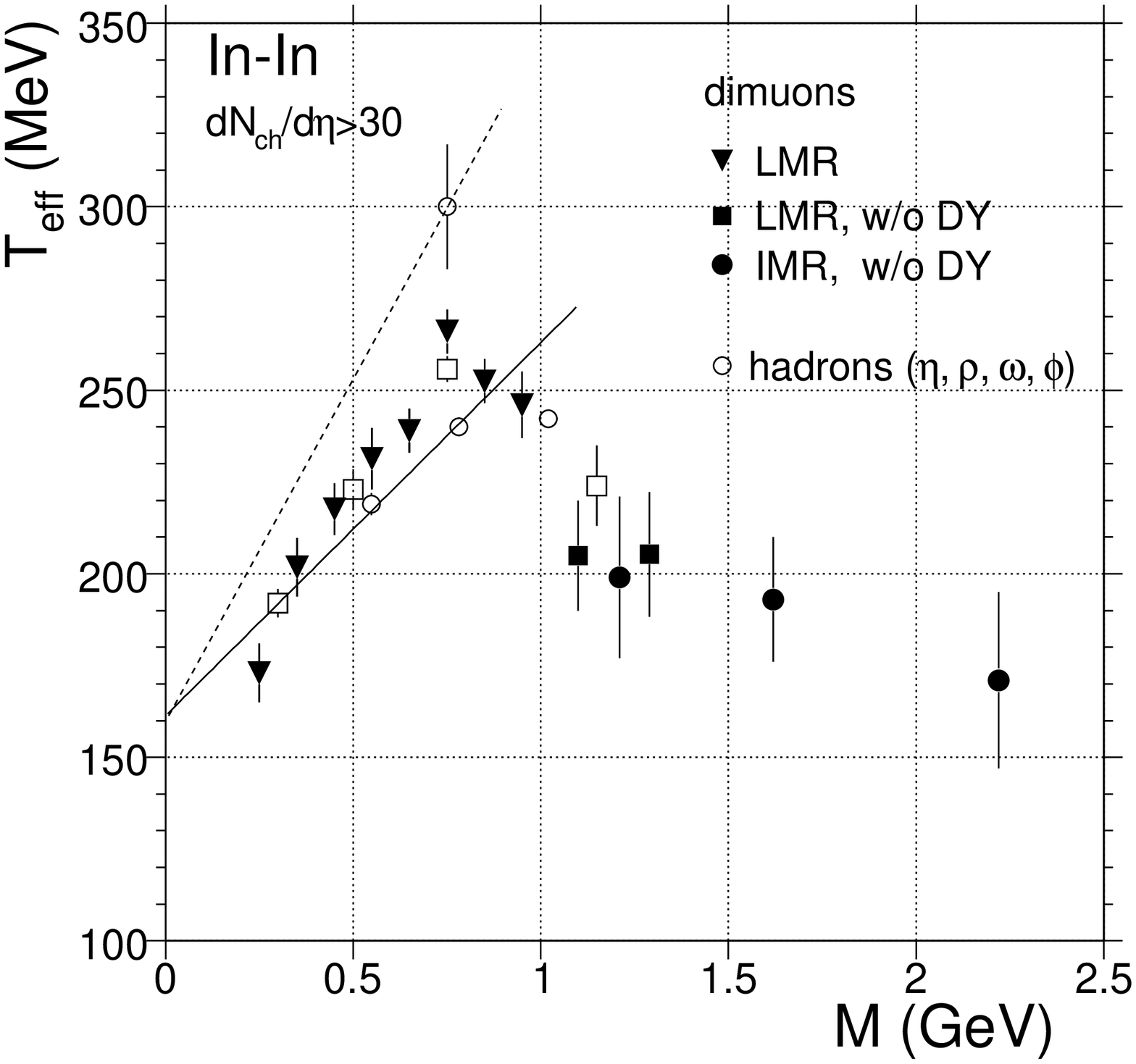,width=\linewidth}
\end{minipage}
\hspace{-0.1cm}
\begin{minipage}{0.33\linewidth}
\epsfig{file=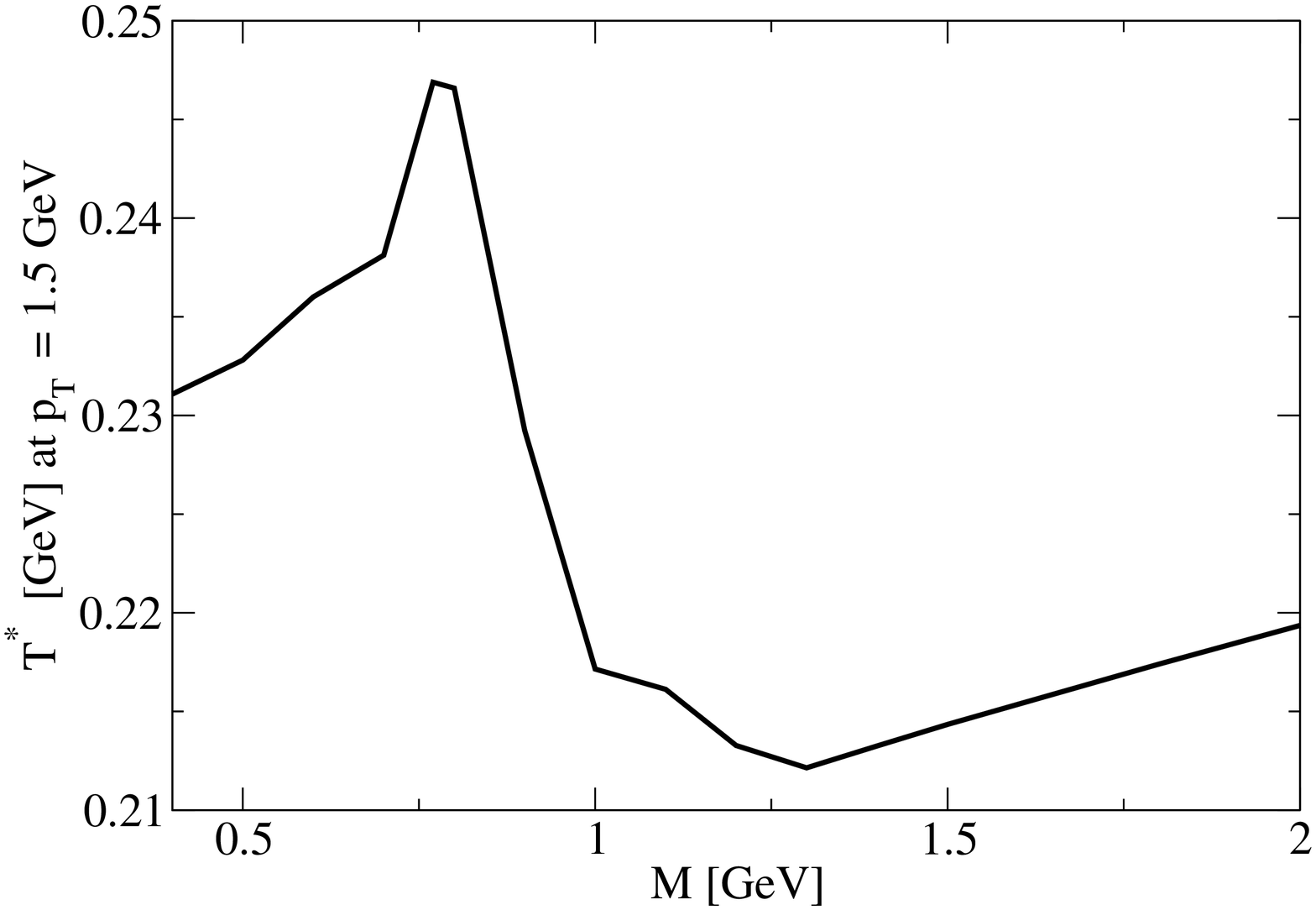,height=0.8\linewidth,width=\linewidth}
\end{minipage}
\hspace{-0.1cm}
\begin{minipage}{0.33\linewidth}
\epsfig{file=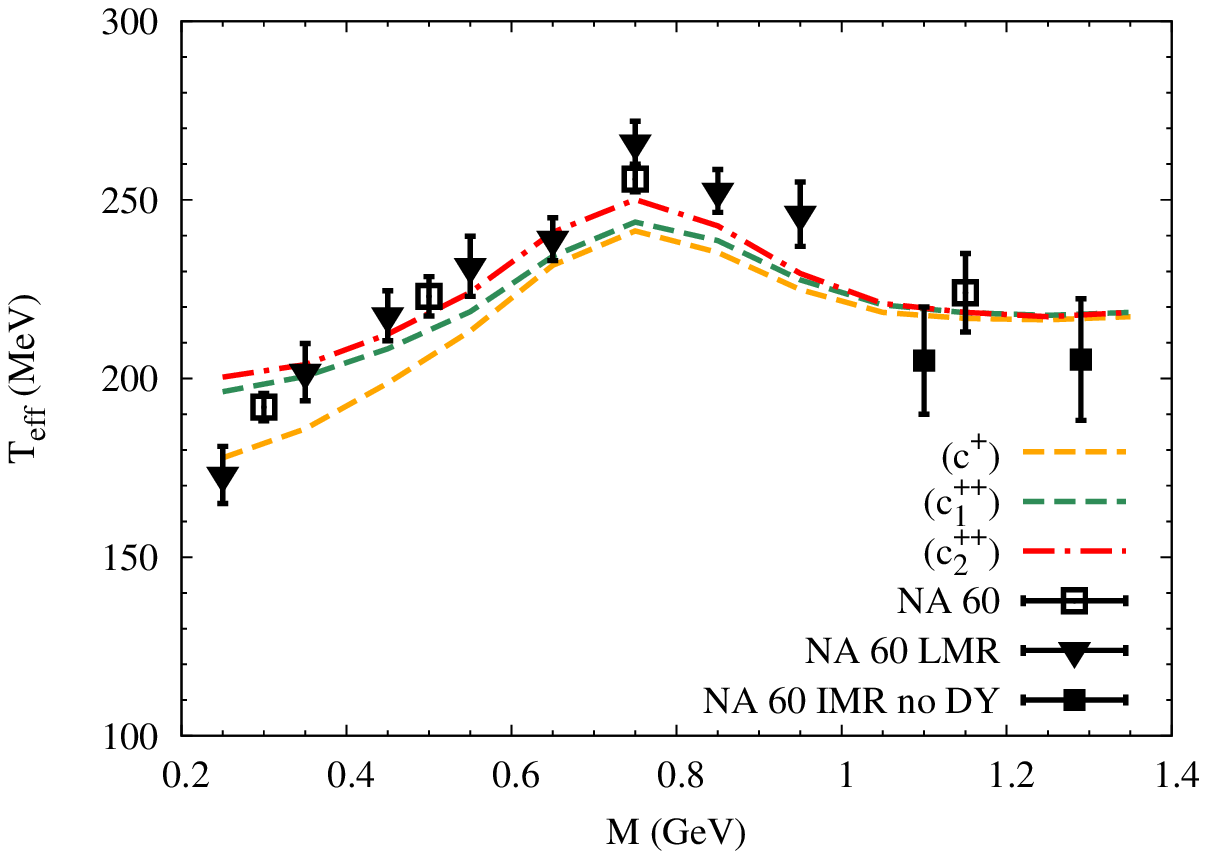,height=0.82\linewidth,width=1.04\linewidth}
\end{minipage}
\caption{Effective slope parameters as a function of dimuon invariant 
mass for excess radiation in In (158~$A$GeV) + In collisions at the SPS.
Note the different vertical and horizontal scales on each plot. 
Left: experimental results with $dN_{\rm ch}/dy>30$ summed
over all centralities and fit over $0.4 <q_t< 1.$ 8 
GeV/$c$~\cite{Arnaldi:2007ru}.
Center: fireball calculations of semicentral collisions
based on thermal emission plus free 
$\rho$ decays at freezeout using the local slope at
$q_t=1.5$~GeV~\cite{Renk:2006qr}.  Right:
fireball calculations of semicentral collisions based on thermal and
non-thermal sources fit over $1.0 <q_t< 1.8$ GeV.
The lower dashed line is a baseline calculations while the 
upper dashed and dot-dashed lines also include $t$-channel
meson exchange in $\pi\rho\to\pi\mu\mu$ 
reactions~\cite{vanHees:2007th}.}
\label{fig:slopes}
\end{figure}
Additional information on the nature of the emitting source might be 
obtained from quantitative analysis of dilepton $q_t$ spectra,
including elliptic flow~\cite{Chatterjee:2007xk}.  Such analysis
has become possible with the NA60 data~\cite{Arnaldi:2007ru}. 
Effective inverse slope parameters, $T_{\rm eff}$, have been extracted 
from dilepton excess spectra in the momentum range $0.4 <q_t< 1.8$~GeV,
obtained from
inclusive In+In collisions at the SPS with $dN_{\rm ch}/dy>30$.
The resulting $T_{\rm eff}$ are shown as a function of dilepton invariant
mass on the left side of Fig.~\ref{fig:slopes}.  The extracted values of 
$T_{\rm eff}$ appear quite large relative to the hadronic slopes at 
freezeout, requiring a surprisingly strong collective flow for the
moderate system size at SPS energies. 
The maximum $T_{\rm eff}(M)$, in the free $\rho$ mass region,
is indicative of $\rho$ decays in the late stages of the collision where
the line shape is expected to approach the vacuum shape while the
effective temperature, $T_{\rm eff}\simeq T_{\rm fo} + m\bar{v}^2$, is 
primarily due to the blue shift imprinted by the collective 
expansion velocity, $\bar v$.
Note that the slope of the excess radiation at $M=m_{\rho,\omega}$
is larger than that of the $\omega$ (lower circle at $M= m_{\rho,\omega}$,
suggesting
that the $\rho$ freezes out significantly later than the $\eta$,
$\omega$ and $\phi$ (the upper open circle at $M=m_{\rho,\omega}$, estimated
for $\rho$ decays around freezeout).  
In the IMR, $T_{\rm eff}$ is significantly reduced,
in agreement with thermal emission
early in the evolution, from temperatures close to $T_c$ where the flow has
not yet developed much, implying a small blue shift.
Quantitative theoretical analyses~\cite{vanHees:2007th,Renk:2006qr} are 
rather involved, see the center and right panels of Fig.~\ref{fig:slopes}, 
especially for momenta above $q_t\simeq1-1.5$~GeV where nonthermal sources 
are expected to become significant~\cite{Rapp:2007zz}, e.g., 
Drell-Yan dileptons or primordial $\rho$ decays.
 
It would be very valuable to obtain similar information
for Au+Au collisions at RHIC.  The thermal yield is
expected dominate further, contributing at higher momenta, 
due to the larger system size and collision 
energy\footnote{The multiplicity, $dN_{\rm ch}/dy$ in Au+Au collisions at 
RHIC is about a factor of four larger than that of In+In collisions
at the SPS.}.  Unfortunately, the increased contribution from correlated 
charm decays is sensitive not only to the total charm cross section but also 
to the charm momentum spectra and thus to charm thermalization (at low and
intermediate $p_T$) and energy loss (at high $p_T$). However, 
open heavy flavor spectra are interesting independent probes 
of the medium and their measurement therefore constitutes one of the 
central goals of RHIC-II, see the heavy flavor part of this report.
It would also be very illuminating to check whether the much 
increased {\em partonic} 
collectivity at RHIC reflects itself by an increased
$T_{\rm eff}$ in the IMR\footnote{We thank Nu Xu for an interesting
discussion on this point.}. 

\subsection{Electromagnetic Signatures of the Color Glass Condensate}
\label{sec_cgc}


Saturation physics has been applied to the description of RHIC data
quite successfully,
from hadron multiplicities and the phenomenon of limiting fragmentation in
Au+Au and d+Au collisions to the produced hadron
transverse-momentum spectra in d+Au collisions at mid- and forward
rapidity (for recent reviews and an extensive list of references,
see~\cite{Iancu:2003}).
Nevertheless, in order to establish gluon saturation as the
dominant physics responsible for these phenomena at RHIC and beyond,
and to rule out other
phenomenological scenarios, one needs to consider further tests of the
CGC formalism such as the predictions of saturation physics
for electromagnetic processes. In this section we outline electromagnetic
signatures of the CGC at RHIC. Specifically, photon and dilepton
production in d+Au collisions are considered. These processes have an
advantage over hadronic processes in that they do not interact
strongly after they are produced and the nonperturbative hadronization process
is absent. Furthermore, photon and dilepton production can shed light
on the validity of the recombination model approach to hadron production which
are also capable of fitting the available data, albeit with a few assumptions.
{\it Since saturation physics predicts similar suppression patterns for
photon, dilepton and hadron production in
d+Au collisions, an experimental confirmation of this generic prediction
would be a major step in establishing saturation physics at RHIC}
and in ruling out recombination as the physics
of hadron production in d+Au collisions.

\subsubsection{Dilepton and Photon Production}
\begin{figure}[!t]
\begin{center}
\epsfxsize=7cm
\leavevmode
\hbox{\epsffile{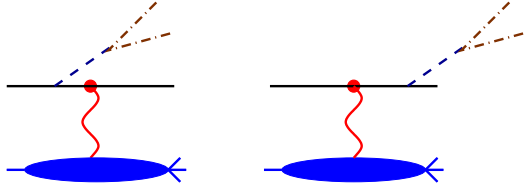}}
\end{center}
\caption{Dilepton production in quark-nucleus scattering.}
\label{fig:dilep}
\end{figure}

We consider the dilepton-production cross section in quark-nucleus
scattering~\cite{fgjjm}
\begin{eqnarray}
q(p) + A \to q(q) + l^+ (k_1) +l^- (k_2) + X \ ,
\end{eqnarray}
shown in Fig.~\ref{fig:dilep}, where $k_1$ and $k_2$ are the momenta of the
two leptons.
Photon production can be obtained by taking the photon virtuality
(dilepton invariant-mass) to zero. The differential cross section at
fixed impact parameter, $b_t$, is given by
\begin{eqnarray}
&&\!\!{d\sigma^{q\,A\rightarrow q\,l^+l^-\,X}
\over d^2 b_t \,d^2 k_t \,d\ln M^2 \, dz}=
{{2 e_q^2 \, \alpha^2} \over{3\pi}} \int
{{d^2 q}\over{(2\pi)^4}} \sigma_{\rm dipole}^F (l_t,b_t,x_A)\nonumber\\
&&
\left\{
\left[{{1+(1-z)^2}\over{z}}\right]
{z^2 l_t^2 \over [k_t^2+M^2(1-z)] [(k_t -z \, l_t)^2 + M^2 (1-z)]}
\right.\nonumber\\ &&\left.  - z(1-z)\,M^2\left[{1\over
[k_t^2+M^2(1-z)]} - {1\over [(k_t - z \, l_t)^2+M^2(1-z)]}\right]^2
\right\}
\label{eq:cs_dilep}
\end{eqnarray}
where $z$ is the fraction of the incoming quark light cone energy
carried away by the (virtual) photon while $x_A$ is the Bjorken $x$ probed in
the target nucleus.  All quark masses
are ignored, $M^2$ is the dilepton invariant mass squared 
with $l_t = q_t + k_t$ and $k_t$ is the transverse momentum of the lepton pair.

Eq.~(\ref{eq:cs_dilep}) is the standard expression obtained if 
propagation of the incident quark through the nucleus is assumed to be
eikonal, {\it i.e.},
the transverse momentum transferred to the nucleus by the incident quark
is much less than its longitudinal momentum~\cite{boris}. 
All the information on the degrees of freedom in the target is contained 
in the dipole cross section,
$\sigma_{\rm dipole}^F (b_t, l_t, x_A)$. Gluon saturation physics comes 
in via the dipole cross section, which is 
determined by the evolution equations of the Color Glass
Condensate. Given an initial condition for the dipole cross section, the JIMWLK
equations in the Color Glass Condensate formalism determine the dependence of
the dipole cross section on the collision energy (or alternatively, $x_A$) and
$l_t$. This is the
main difference between the CGC formalism and the other
approaches \cite{boris}: while the CGC formalism can predict the
$x_A$ and $l_t$ dependence of the dipole cross section, other approaches cannot
and must instead motivate a suitable form from phenomenological considerations.
We note that to obtain the invariant cross section for photon production,
$d\sigma^{q\,A\rightarrow q\,\gamma\,X}/ d^2 b_t \,d^2 k_t \, dz$,
from Eq.~(\ref{eq:cs_dilep}), the mass is set to zero, $M=0$, and the
dilepton vertex factor, $\alpha / 3\pi\, M^2$, is removed.

Since the dipole cross section is also the main ingredient in the
hadron production cross sections in deuteron (proton)-nucleus collisions,
a similar nuclear modification factor, $R_{dA}$, is predicted
for dilepton production and hadron production. Parametrically, one expects
the nuclear modification factor, $R_{pA}$, to scale with the nucleon
number $A$ like
$R_{pA} \sim A^{ -{\gamma_0 \over 3}}$
where $1 - \gamma_0\simeq 0.628$ is the BFKL anomalous dimension. It should
be emphasized that in QCD, small-$x$ evolution (BFKL) is the only way to
generate leading-twist shadowing if partonic degrees of freedom 
(quarks and gluons) are used.

\begin{figure}[!t]
\begin{center}
\epsfxsize=7cm
\leavevmode
\hbox{\epsffile{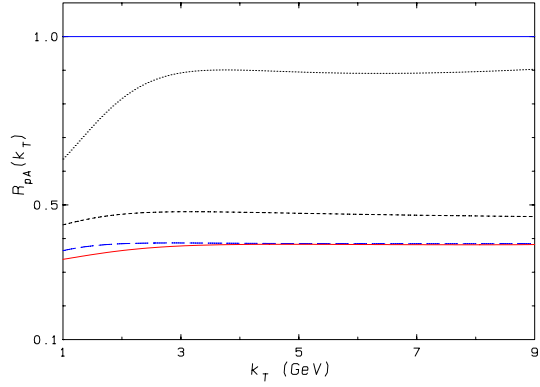}}
\end{center}
\caption{Nuclear modification factor, $R_{dA}$, of dilepton production in 
quark-nucleus scattering \cite{bms}.  The dotted, short-dashed and long-dashed
curves are for $M = 2$ GeV and $y = 0.5$, 1.5 and 3 respectively.  The solid
curve is for $M=4$ GeV and $y=3$.}
\label{fig:bms}
\end{figure}

Dilepton production has an additional knob to turn
in order to change the kinematics and probe QCD dynamics in different
settings, the dilepton invariant mass.  
The nuclear modification factor, calculated using a
saturation inspired model of the dipole cross section \cite{bms}, is shown
in Fig.~\ref{fig:bms}.

Since the dilepton production rates are small due to the 
electromagnetic coupling,
we consider the invariant-mass dependence of the 
transverse-momentum integrated cross section. The integration over
Eq.~(\ref{eq:cs_dilep}) can be done analytically to obtain
\begin{eqnarray}
z {d\sigma^{q\,A\rightarrow q\,l^+l^-\,X}
\over d^2 b \,d M^2 \, dz}=&&
{{\alpha^2}\over{3\pi^2}}\,
 {1-z \over z^2} \int \, dr_T^2 \,
\sigma_{\rm dipole}^F (x_g, \underline{b}, r_T)
\nonumber\\
&&
\nonumber \\
&& \!\!\!\!\!\!\!\!\!\!\!\!\!\!\!\!\!\!\!\!\!\!\!\!\!\!\!\!\!\!\!\!\!
\Bigg[[1 + (1-z)^2] \,K_1^2[{\sqrt{1-z}\over z} M r_T] +
2 (1-z) \, K_0^2[{\sqrt{1-z}\over z} M r_T]\Bigg]
\label{eq:cs_dilep_M} \ .
\end{eqnarray}
To relate this to proton (deuteron)-nucleus scattering,  
Eq.~(\ref{eq:cs_dilep_M}) must be convoluted with the quark (and antiquark) 
distributions, $q(x,M^2)$ ($\overline q(x,M^2)$) in a proton or deuteron.  
This can be written
in terms of the proton (deuteron) structure function $F_2$ \cite{boris}, 
\begin{eqnarray}
{d\sigma^{p\,A\rightarrow l^+l^-\,X}
\over d^2b \,d M^2\, dx_F}= &&
{{\alpha^2}\over{6\pi^2}}{1 \over x_q + x_g}\,
\int_{x_q}^1 \,dz \,\int dr_T^2 \, {1-z \over z^2} \, F_2^{p} (x_q/z) \,
\sigma_{\rm dipole}^F (x_g,\underline{b}, r_T)
\nonumber\\
&&
\nonumber \\
&& \!\!\!\!\!\!\!\!\!\!
\Bigg[[1 + (1-z)^2] \,K_1^2[{\sqrt{1-z}\over z} M r_T] +
2 (1-z) \, K_0^2[{\sqrt{1-z}\over z} M r_T]\Bigg]
\label{eq:dpA}
\end{eqnarray}
where
\begin{eqnarray}
x_q & = & {1\over 2} \bigg[\sqrt{x_F^2 + {4M^2 \over s}} + x_F\bigg ] \,\, ,
\nonumber \\
x_g & = &  {1\over 2} \bigg[\sqrt{x_F^2 + {4M^2 \over s}} - x_F\bigg ] \, \, ,
\nonumber \\
x_F & \equiv & {M\over \sqrt{s}}[e^y - e^{-y}] \, \, ,
\end{eqnarray}
and 
\[ F_2^{p}\equiv \sum_f e_f^2 x \, [q_f(x,M^2) + \overline{q}_f (x,M^2)]\] 
is the proton structure function.

\begin{figure}[!t]
\begin{minipage}{0.47\linewidth}
\epsfig{file=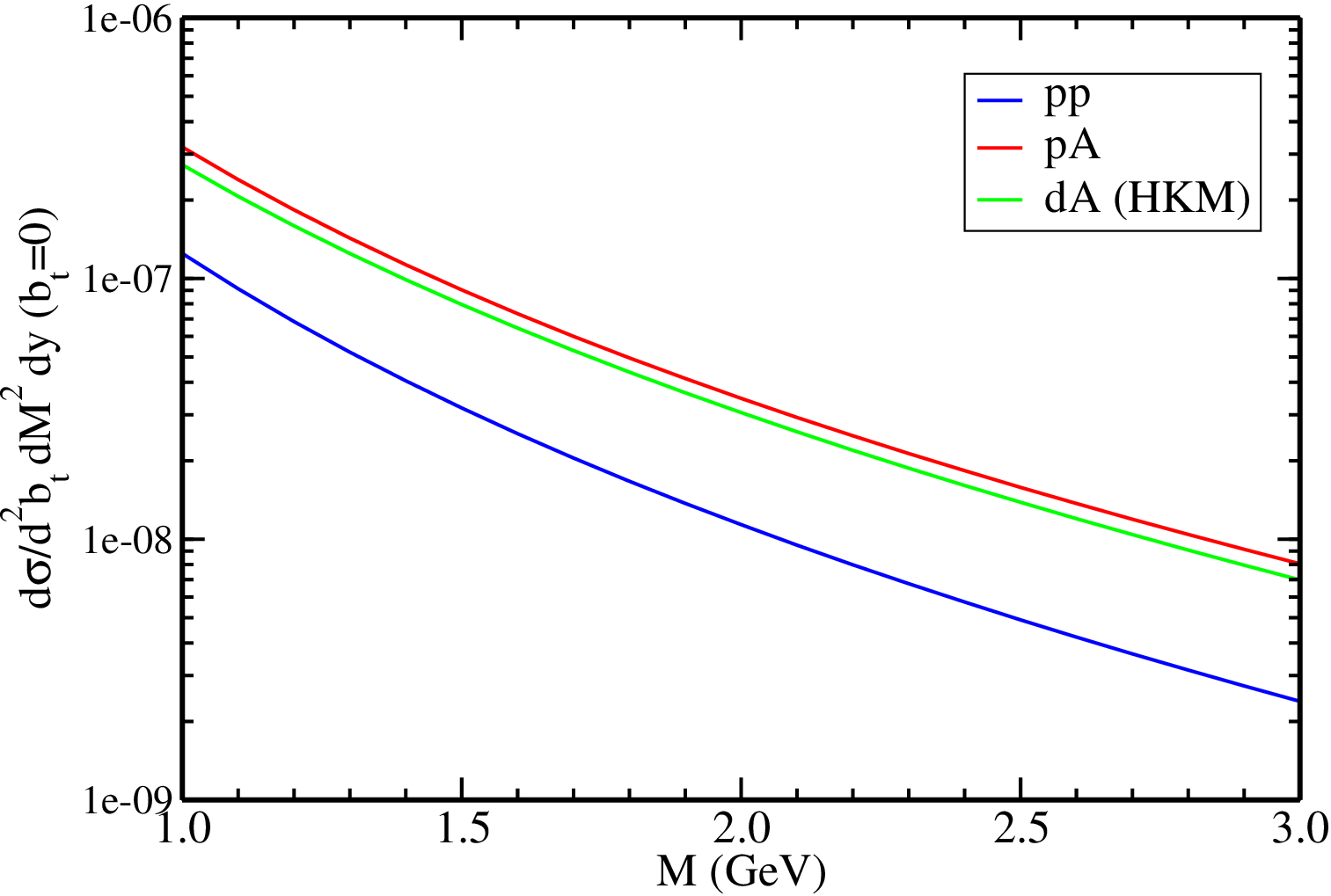,width=\linewidth}
\caption{Dilepton production cross section at $y=2.2$.}
\label{fig:cs_rhic}
\end{minipage}
\hspace{0.5cm}
\begin{minipage}{0.47\linewidth}
\epsfig{file=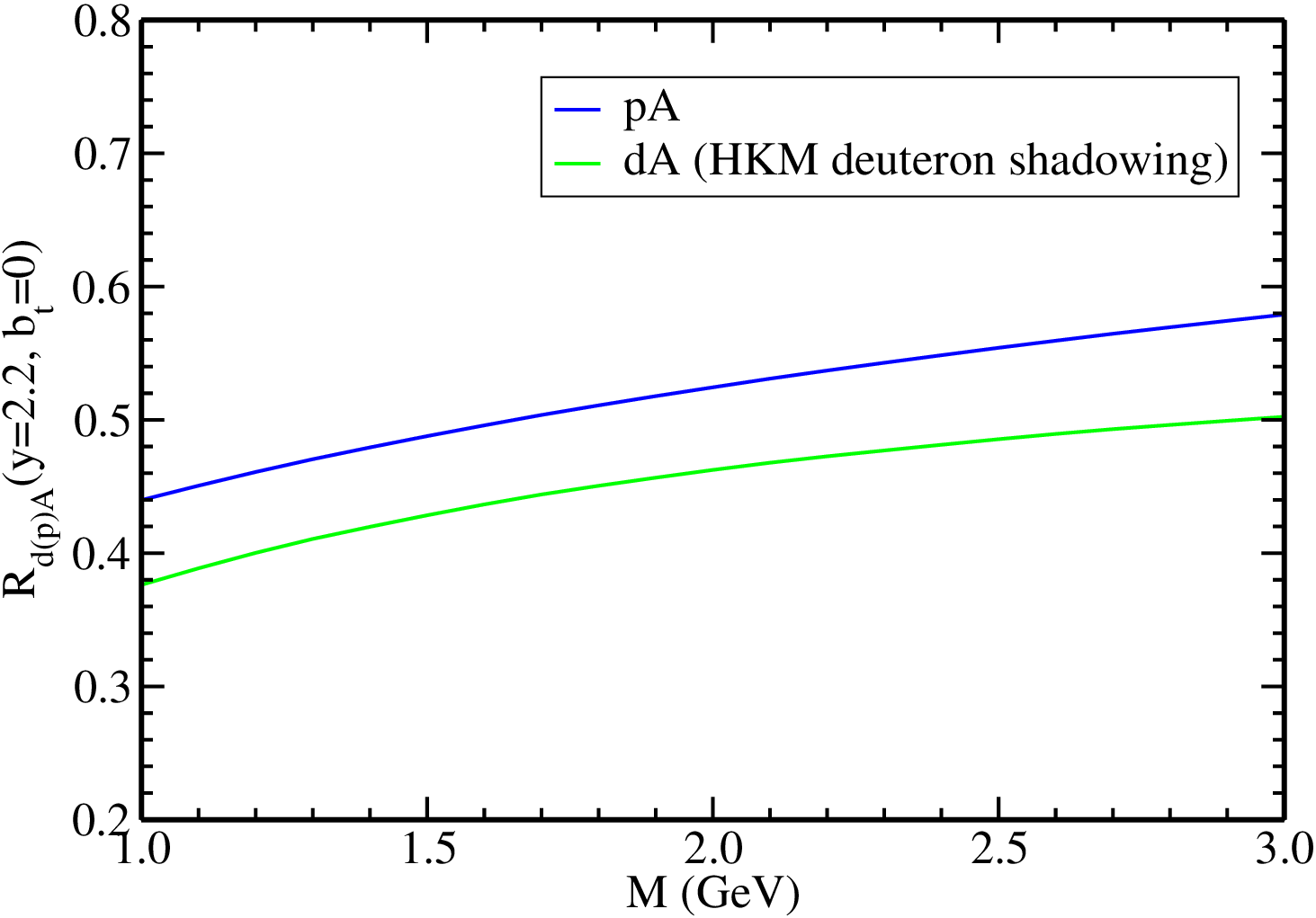,width=0.97\linewidth}
\caption{The dilepton nuclear modification factor in $pA$ and d$A$ collisions
at $y=2.2$.}
\label{fig:R_rhic}
\end{minipage}
\end{figure}

The invariant dilepton production cross section is shown in 
Fig.~\ref{fig:cs_rhic} for proton-proton,
proton-nucleus and deuteron-nucleus collisions as a function of dilepton
invariant mass, while Fig.~\ref{fig:R_rhic} shows the nuclear
modification factor for
both $pA$ and d$A$ collisions at RHIC. The calculations are for fixed rapidity,
$y=2.2$, and the most central
collisions~\cite{jjm}.  The HKM
parameterization of the deuteron wave function
is used for the deuteron projectile, a few percent effect.

Photon production and photon and hadron
correlation functions have been calculated using the same formalism
\cite{jjm}.  The
hadron-photon cross section is a very sensitive probe of the dipole profile, 
the main ingredient of single-particle production cross section in the
CGC formalism. Therefore, experimental studies of electromagnetic probes
in deuteron-nucleus collisions at RHIC can shed light both
on the dynamics of gluon saturation and on the role of saturation physics
in the observed suppression of the
hadron spectra in the forward rapidity region at RHIC.
This measurement can also clarify the role of hadron recombination models at
RHIC, at least in the forward rapidity region, since recombination effects 
will not be present in electromagnetic final states.  Therefore, observation
of dilepton or photon suppression would be strong evidence for
the CGC.

It is worth noting that, since the saturation scale of the proton (deuteron) 
is very
small at mid- and forward rapidity at RHIC, CGC predictions
for proton-proton collisions will have large uncertainties. 
Indeed, as recently shown~\cite{aaj}, particle production in proton-proton
collisions cannot be reliably calculated at RHIC, since the saturation
momentum is small, while d+Au calculations 
are under much better quantitative
control. Therefore, in any saturation-inspired calculation of the 
nuclear modification factor, the $pp$ cross section in the denominator 
should be understood to be a fit to the data while the
d+Au cross section can be calculated
using saturation physics and used to determine the physics
dominant in the forward-rapidity region at RHIC. This, in turn,
will have significant ramifications for the LHC.

\section{OBSERVABLES: STATE OF THE ART}
\label{sec_obs}
\subsection{Low-Mass Dileptons}
\label{sec:na60}
\begin{figure}[!t]
\begin{minipage}{0.400\linewidth}
\includegraphics[width=\linewidth]{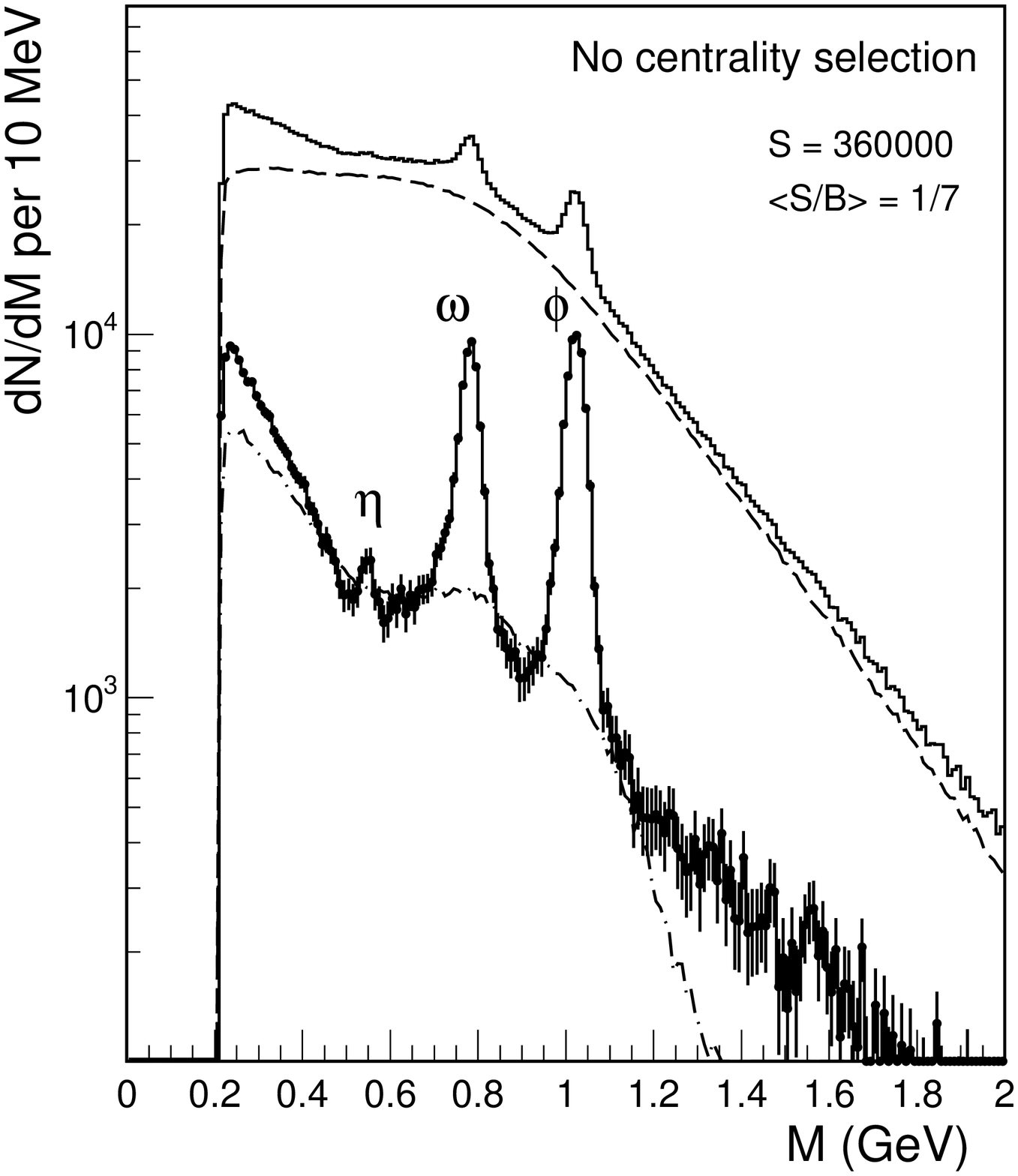}
\vspace*{-10mm}
\caption{\label{fig:lmspectrum} Dimuon invariant-mass spectra
  from NA60~\cite{Damjanovic:2005ni,Arnaldi:2006jq}.  The raw
  data (upper histogram) show clear $\omega$ and $\phi$ peaks. In
  the background-subtracted spectrum, (lower histogram) the
  $\eta\to\mu\mu$ decay is recognizable.}
\end{minipage}
\hspace{\fill}
\begin{minipage}{0.525\linewidth}
\vspace{0.4cm}
\includegraphics[width=\linewidth]{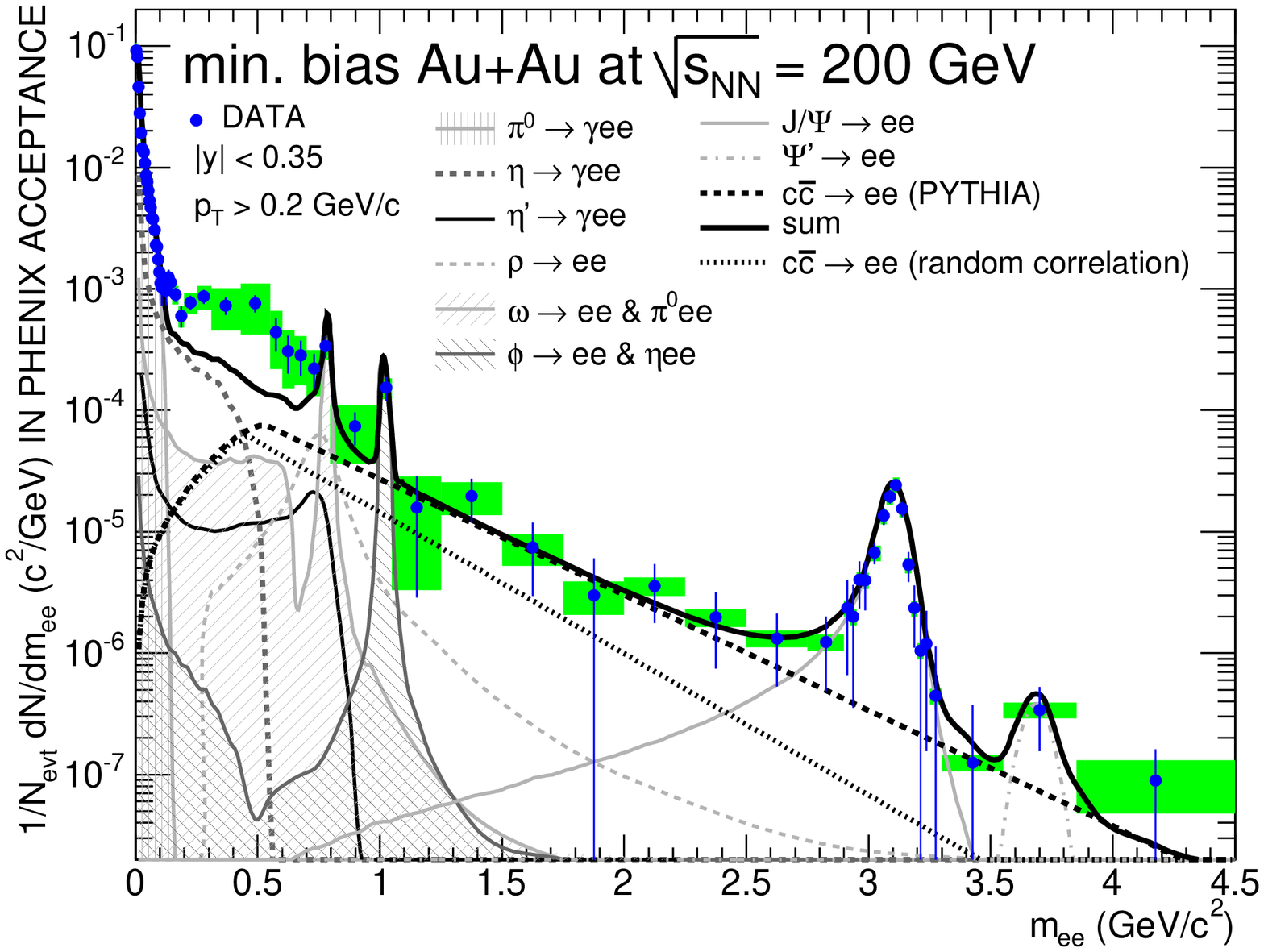}
\vspace*{-10mm}
\caption{Dilepton mass spectrum from PHENIX~\cite{ppg075}
compared to a cocktail of hadronic sources and charm decays.}
\label{fig:Pre_cocktail}
\end{minipage}
\end{figure}

The $\rho$ meson, with its short lifetime (1.3~fm/$c$ in vacuum),
large dilepton decay width ($\Gamma_{ee}=7$\,keV)
and prevalent coupling to $\pi\pi$ (and possibly $q\bar{q}$)
annihilation has long been identified as the most promising probe
of in-medium
modifications of hadron properties close to the QGP phase boundary.
Changes in $\rho$ mass or width were anticipated as precursors of
the chiral transition more than two decades ago~\cite{Pisarski}.
Dielectron mass spectra from
CERES~\cite{Agakichiev:2005ai,Adamova:2003xx} over the
last ten years have shown a significant (factor of $2-3$) excess over
known hadronic sources in the mass region below the
free $\rho$ mass -- a result which spurred vigorous
theoretical activity.  However, the CERES experimental uncertainties, including
limited statistics and mass resolution, could neither confirm nor refute 
scenarios with medium effects\footnote{A significant improvement in mass 
resolution was achieved with the CERES Time Projection Chamber
upgrade~\cite{Miskowiec:2005dn}.}.
Recently, the NA60 experiment at the SPS, a significant upgrade
of NA50 including high-precision tracking close to the
vertex in a high magnetic field, measured dimuon mass spectra in
158~$A$GeV In+In collisions with an unprecedented $\sim 20$~MeV mass
resolution and high statistics in both the low and intermediate mass 
regions~\cite{Damjanovic:2005ni,Arnaldi:2006jq}. The results
in the light vector-meson region are shown in Fig.~\ref{fig:lmspectrum}.
After subtracting the ``hadron decay cocktail'' with $\eta$, $\omega$, and
$\phi$ decays, the excess spectrum has a clear albeit rather broad peak around
the nominal $\rho$ mass, as shown in Fig.~\ref{fig:mix} and compared
to theoretical predictions using in-medium $\rho$ spectral functions
with either a dropping mass~\cite{Brown:1995qt} or a strong broadening,
as calculated in hadronic many-body theory~\cite{RW99}. The excellent
resolution and statistics, {\it made possible by detector upgrades and high
luminosity}, can now clearly distinguish between these two
approaches and be considered as a benchmark for future dilepton
measurements.  (How well other approaches can explain the measured data remains
to be seen.)  Measurements of similar precision - made possible by
RHIC-II - will be essential to further improve our understanding
and elevate it from the qualitative to the quantitative level, 
necessary for drawing conclusions about the nature of chiral symmetry
restoration.

Comparable quality measurements of the $\rho$ spectral function do
not yet exist at RHIC.  PHENIX is measuring dielectrons in the central
arm.  The mass spectrum in minimum bias 200 GeV Au+Au 
collisions is shown in Fig.~\ref{fig:Pre_cocktail} along with the
hadronic cocktail and the background from charm.  While the $J/\psi$
measurement is very clean, the resolution in the low mass region is
not adequate to test in-medium modifications of vector mesons.  Such
precision is, however, expected once the hadron-blind detector, one of
the major PHENIX upgrades becomes operational (see Sec.~\ref{sec:hbd}).

STAR measured the $\pi^+\pi^-$
invariant-mass spectra~\cite{Adams:2003cc} and found indications that the 
$\rho$ mass is both $p_T$ and multiplicity-dependent in $pp$, d+Au and
peripheral Au+Au collisions.
At low transverse momenta, $0.5<p_T<1.5$ GeV/$c$, the $\rho$ mass peak
is $3-8$\% lower.  Whether this is caused by a $\rho$ mass
shift~\cite{Brown:2003xx,Broniowski:2003ax} or other effects
({\em e.g.} Bose-Einstein correlations, thermal phase space in connection
with resonance broadening, underlying ``$\sigma$" decays,
{\em etc.}~\cite{Rapp:2003ar,Pratt:2003vb}) still needs to be clarified.
The measurement was insensitive to
possible changes in the $\rho$ width~\cite{Fachini:2004jx} which requires 
an improved determination of the background sources
including the ``physics background"\footnote{In a different energy
  range, KEK (E325) measured a drop of the $\rho$ mass whereas JLAB
  (CLAS) reported broadening without any drop of the $\rho$ mass.}.

The combination of leptonic and hadronic $\rho$ decays is a valuable tool for
disentangling in-medium modifications due to 
the hot and dense phases from the more dilute phases.  The leptonic decays 
probe the entire collision history while the hadronic decays probe the fireball
surface and the late stages.

\subsection{The $\phi$(1020): Hadronic {\em vs} Leptonic Decays and $v_2$}
\label{sec:phi1020}
%
\begin{figure}[!t]
\begin{minipage}[t]{0.45\linewidth}
\includegraphics[width=\linewidth]{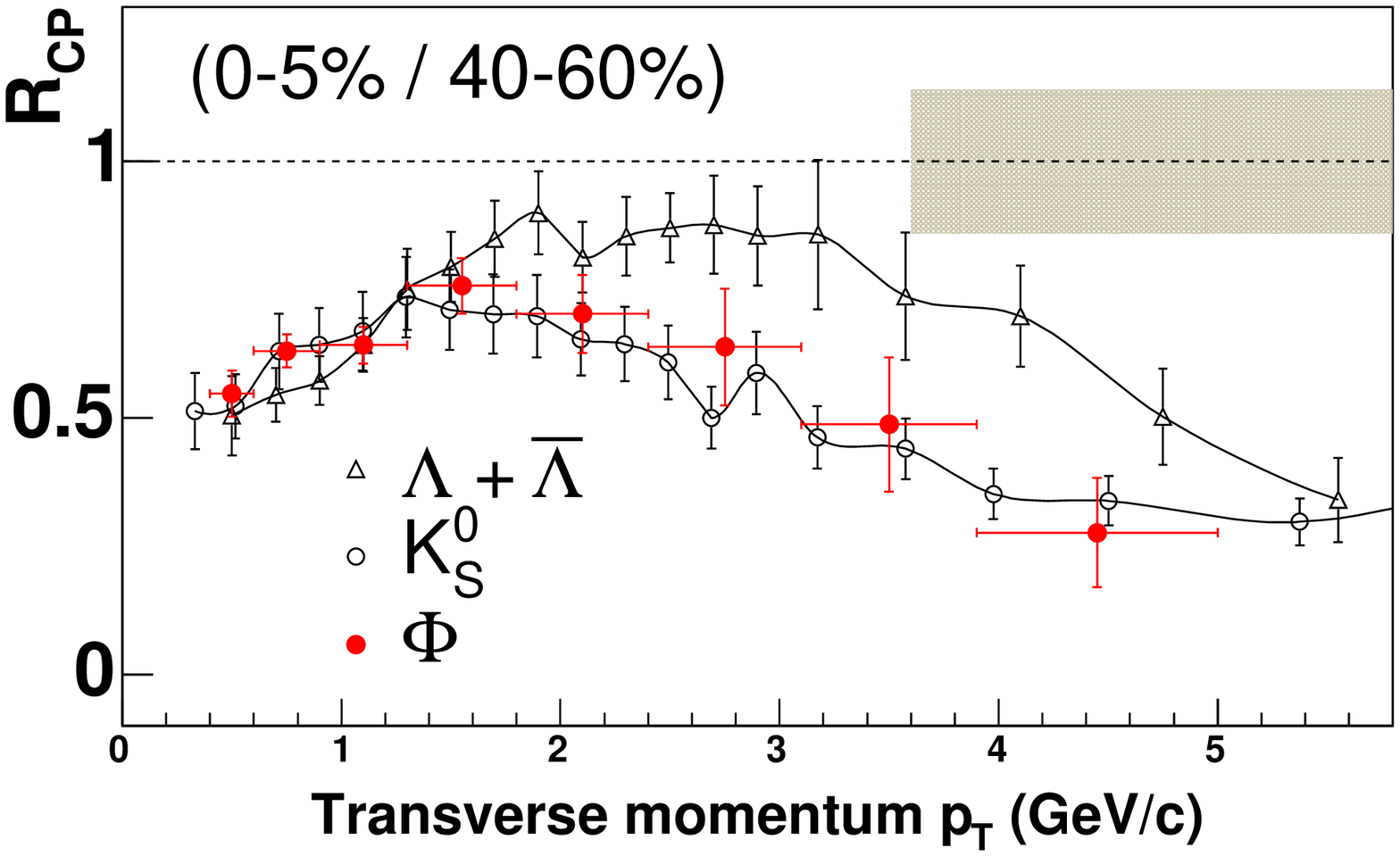}
\vspace*{-10mm}
\caption{\label{fig:AuAu_Phi} The nuclear modification factor
  $R_{CP}$ (ratio of yields in central and peripheral collisions,
  divided by the respective number of binary $NN$ collisions)
  for $\phi$, $K_s^0$ and $\Lambda$ in 200\,GeV Au+Au collisions~\cite{Cai}.}
\end{minipage}
\hspace{\fill}
\begin{minipage}[t]{0.45\linewidth}
\includegraphics[width=\linewidth]{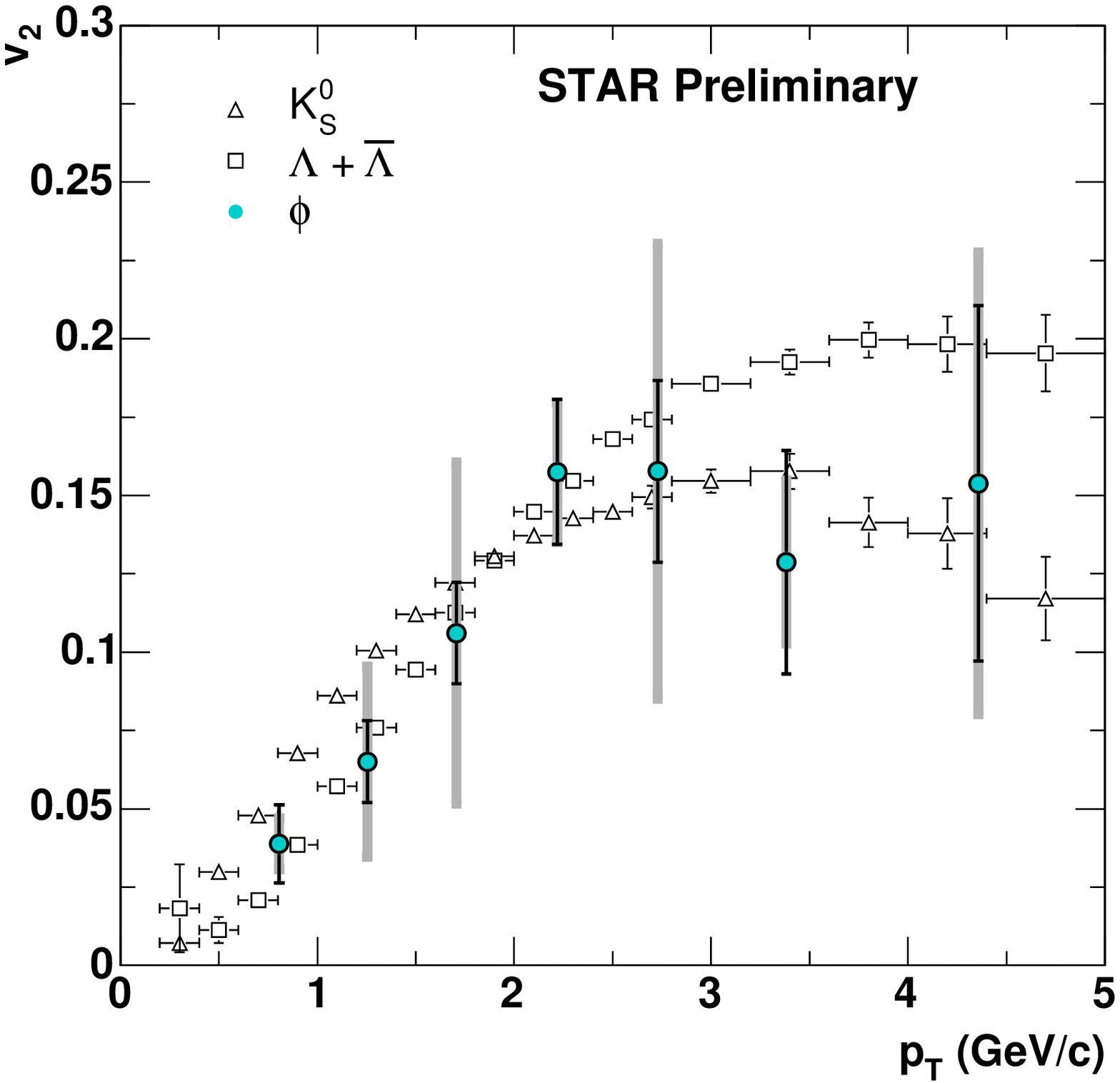}
\vspace*{-10mm}
\caption{\label{fig:Phi_V2} Elliptic flow as a function of $p_T$
  for $\phi$, $K_s^0$ and $\Lambda$ in 200\,GeV Au+Au collisions~\cite{Cai}.}
\end{minipage}
\end{figure}
Since the $\phi$ mass is barely twice the kaon mass,
$m_\phi-2m_K \simeq 30$\,MeV, a study of hadronic,
$\phi \rightarrow K \overline K$, relative to leptonic, $\phi \rightarrow e^+
e^-$, decays has long been suggested as a
sensitive test of a dropping $\phi$ mass or braodening of the $\phi$ width, 
possibly related to partial chiral symmetry restoration.
If the mass drops, the dominant $\phi \rightarrow K \overline K$ channel is 
quickly kinematically suppressed and the ratio of production
cross sections
of the hadronic, $B(\phi \rightarrow K \overline K) =49.2$\%, and leptonic,
$B(\phi \rightarrow e^+ e^-) = 3 \times 10^{-4}$, channels,
$\sigma(\phi \rightarrow K \overline K)/\sigma(\phi \rightarrow e^+e^-)$,
should change dramatically.  Similar
arguments may apply if a reduced kaon mass or inelastic scatterings
such as $\phi+\pi\to K+K^*$ lead to a strong increase of the
$\phi$ width.  In this case, $\phi$ regeneration has to
be accounted for as well~\cite{phiinmedium}.  
Recent preliminary
results from PHENIX~\cite{Kozlov} suggest that,
when normalized by the branching ratio,
$dN(\phi \rightarrow e^+e^-)/dy$ is larger and the $m_T$ slopes are
steeper in $AA$ relative to $pp$ collisions.  However, the experimental
uncertainties prevent definite conclusions.  The main problem is the small
signal-to-background ratio (S/B) in the dielectron channel {\em which can only
be cured by better background rejection}.  However, the
$\phi$ yields from $\phi \rightarrow K \overline K$ extracted by 
STAR~\cite{Cai} are systematically higher than those measured
by PHENIX in the same decay channel.  Careful analysis shows
that $p_T$ spectra measured by the two experiments are not 
inconsistent~\cite{rafelskiphi}.  

A difference was observed between 
$\phi \rightarrow K \overline K$ measured by NA49 and $\phi
\rightarrow ee$ measured by NA50. Recent CERES~\cite{Adamova:2005jr}
results for both decay channels in Pb+Au collisions 
and measurements of $\phi \rightarrow \mu^+ \mu^-$ decays in In+In collisions
from NA60~\cite{NA60QM05} are consistent with data
from NA49.  Thus the difference was not due to the observed decay channel
but the method of observation. This experience emphasizes both
the importance of powerful detector upgrades and the ability to
measure different decay channels in the same apparatus over a wide momentum 
range.

Since the $\phi$ meson mass is comparable to some
baryon masses,
it can also play a major role in distinguishing between the mass and
species dependence of quantities such as the nuclear
modification factor, $R_{CP}$, and elliptic flow.
Recent results from STAR~\cite{Cai} in 200 GeV Au+Au collisions
show that both $R_{CP}$ (Fig.~\ref{fig:AuAu_Phi})
and $v_2$ (Fig.~\ref{fig:Phi_V2}) are consistent with
parton recombination and collective flow at the partonic level, as well as
with the PHENIX data~\cite{PHENIXphi}.

\subsection{Direct Photon Spectra}
\label{sec:photonspectra}

\begin{figure}[!t]
\begin{minipage}{0.470\linewidth}
\includegraphics[width=\linewidth]{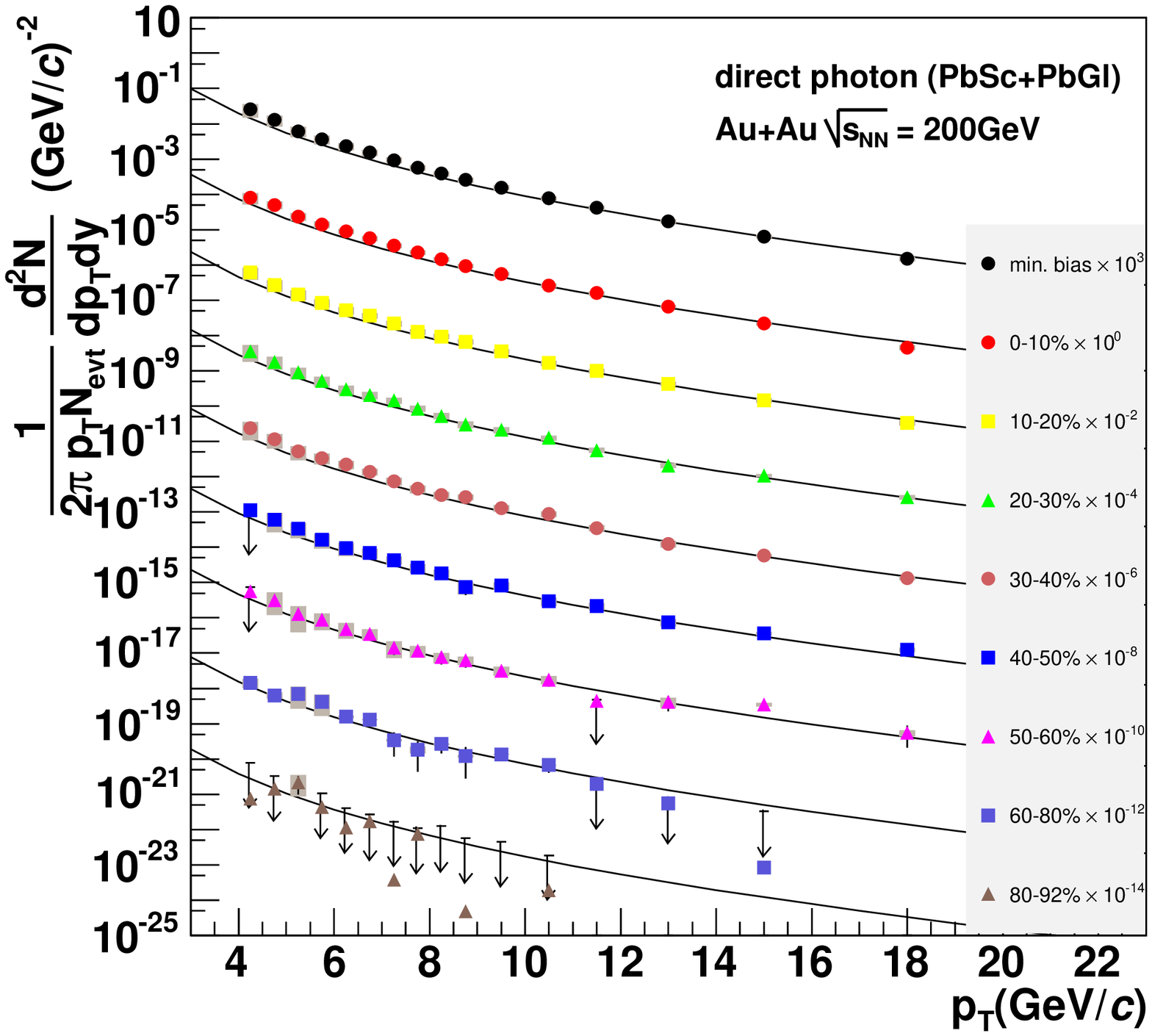}
\vspace*{-10mm}
\caption{Preliminary PHENIX direct photon spectra in 200 GeV Au+Au collisions,
  for all centralities and minimum bias collisions.  
  The curves are NLO pQCD calculations with the scale set equal to $p_T$.  }
\label{fig:cdirphoton}
\end{minipage}
\hspace{\fill}
\begin{minipage}{0.470\linewidth}
\vspace{0.4cm}
\includegraphics[width=\linewidth]{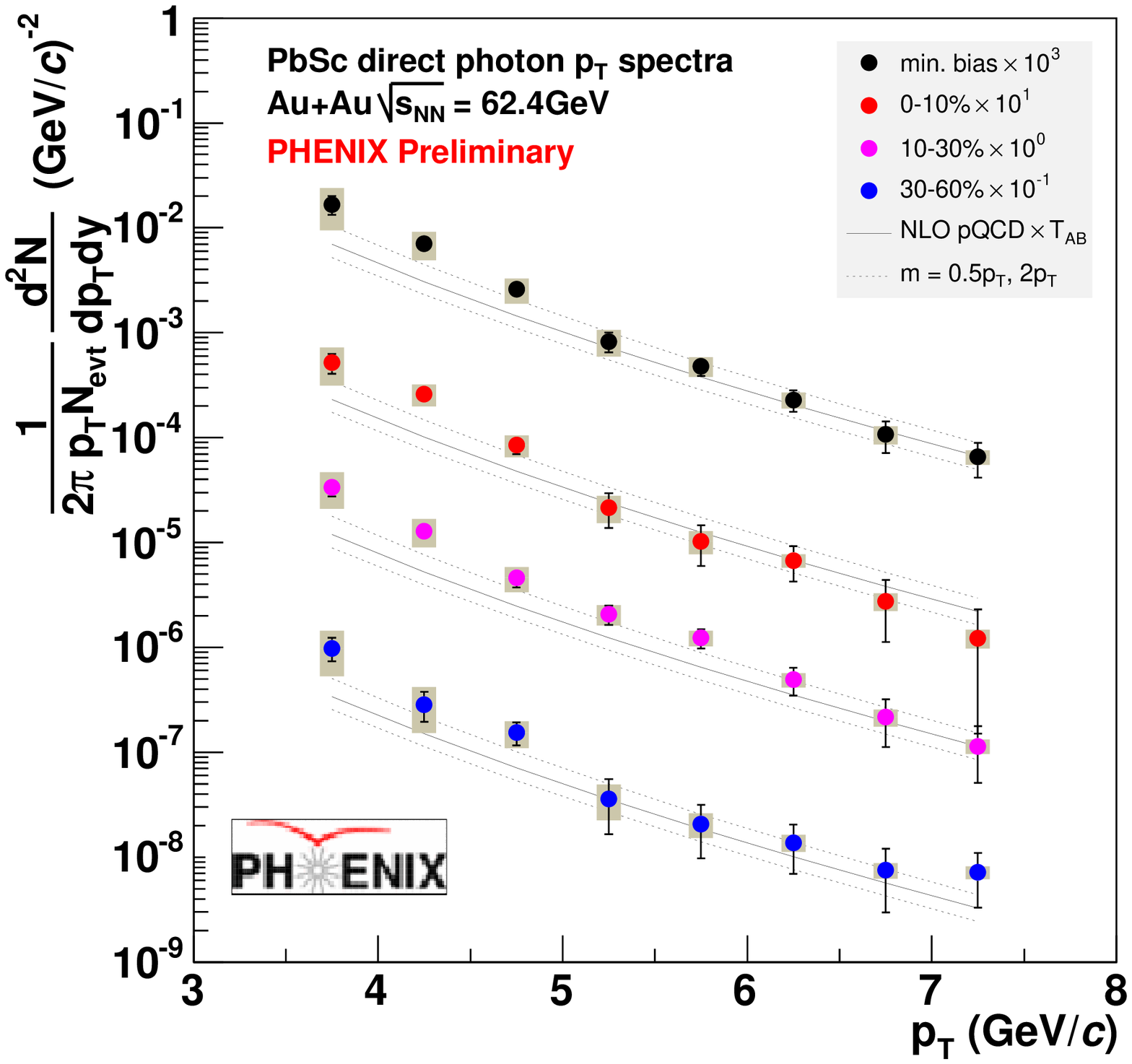}
\vspace*{-10mm}
\caption{Preliminary PHENIX direct photon spectra in 62 GeV Au+Au collisions
  at different centralities compared to NLO pQCD predictions at scales of
  $0.5p_T$, $p_T$, and $2.0p_T$.}
\label{fig:DirectPhoton62}
\end{minipage}
\end{figure}

The latest direct photon spectra from Au+Au collisions over the full 
centrality range are compared to NLO pQCD calculations at 200 GeV 
(Fig.~\ref{fig:cdirphoton}) and 62GeV (Fig.~\ref{fig:DirectPhoton62}).
While the data seem to favor a perturbative scale of $0.5p_T$,
particularly at lower transverse momenta, they are consistent with a scale
equal to $p_T$ at higher 
momenta\footnote{A similar trend is observed in the $\pi^0$ 
spectra~\cite{ppg063}.}.  
One explanation could be the onset of additional photon sources such as 
photons from jet-medium interactions at intermediate $p_T$ 
(see Fig.~\ref{fig_jet-qgp}).  Before drawing any conclusions, high quality 
$pp$ data have to determine the level of agreement of the data with the 
NLO pQCD predictions, see Sec~\ref{sec:photonraa}.
Unfortunately there are different competing sources of the excess relative
to baseline pQCD calculations proposed - all with their respective
theoretical uncertainties.  Studying inclusive cross sections is
insufficient for disentangling these sources.  A
possible solution, involving much larger data samples than currently
available, is proposed in Sec.~\ref{sec:photon_elliptic}.

\subsection{Direct Photon $R_{AA}$}
\label{sec:photonraa}

\begin{figure}[!t]
\begin{minipage}{0.470\linewidth}
\includegraphics[width=\linewidth]{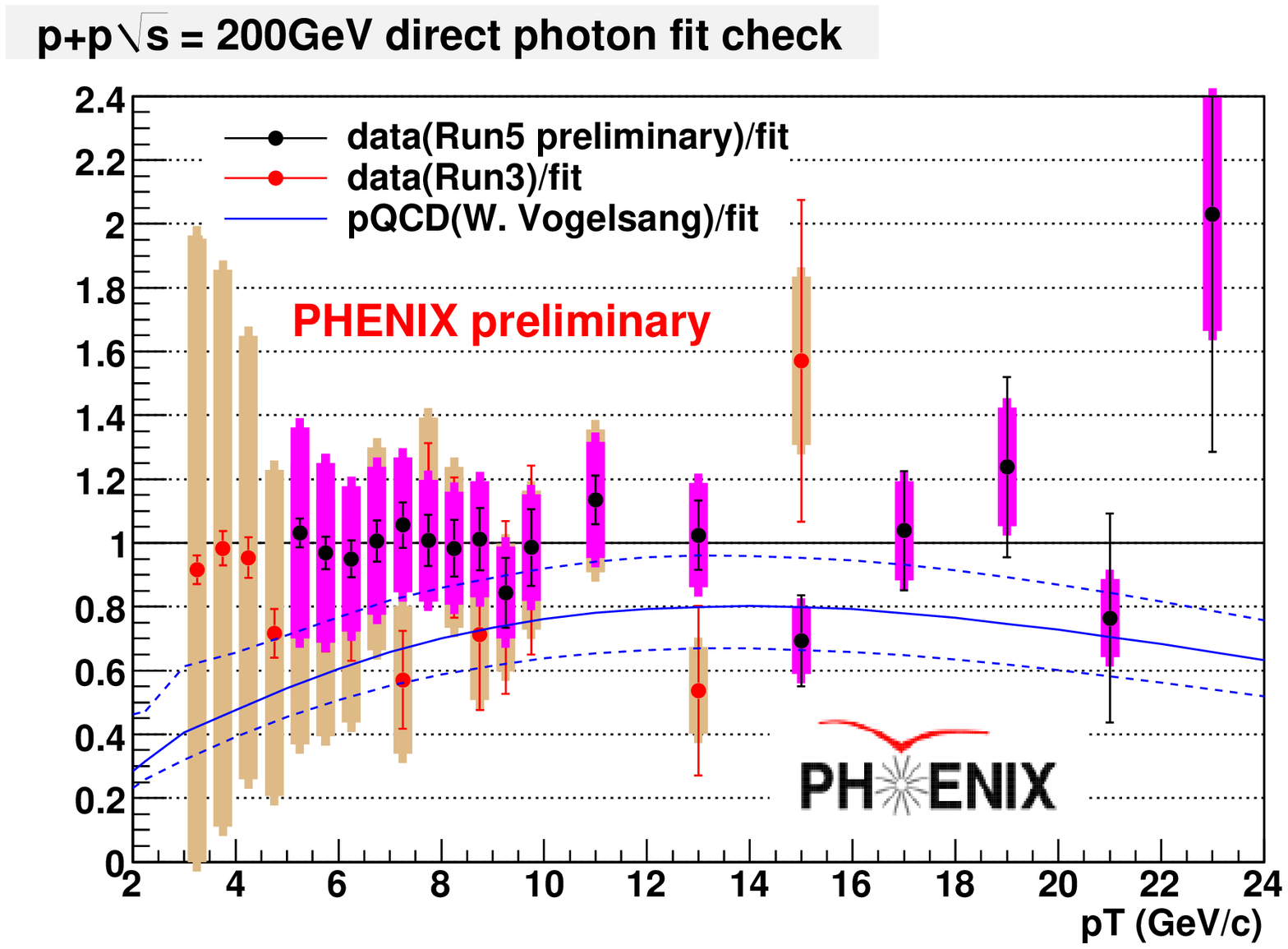}
\vspace*{-10mm}
\caption{Direct photons in 200 GeV $pp$ collisions from different PHENIX runs
  divided by a fit to the preliminary Run-5 results.  
  The black circles show preliminary Run-5 results to illustrate the
  quality of the fit while the red circles are published Run-3 data.
  The blue curves are NLO pQCD calculations.
  }
\label{fig:cppdirgamma}
\end{minipage}
\hspace{\fill}
\begin{minipage}{0.470\linewidth}
\vspace{0.4cm}
  \includegraphics[width=\linewidth]{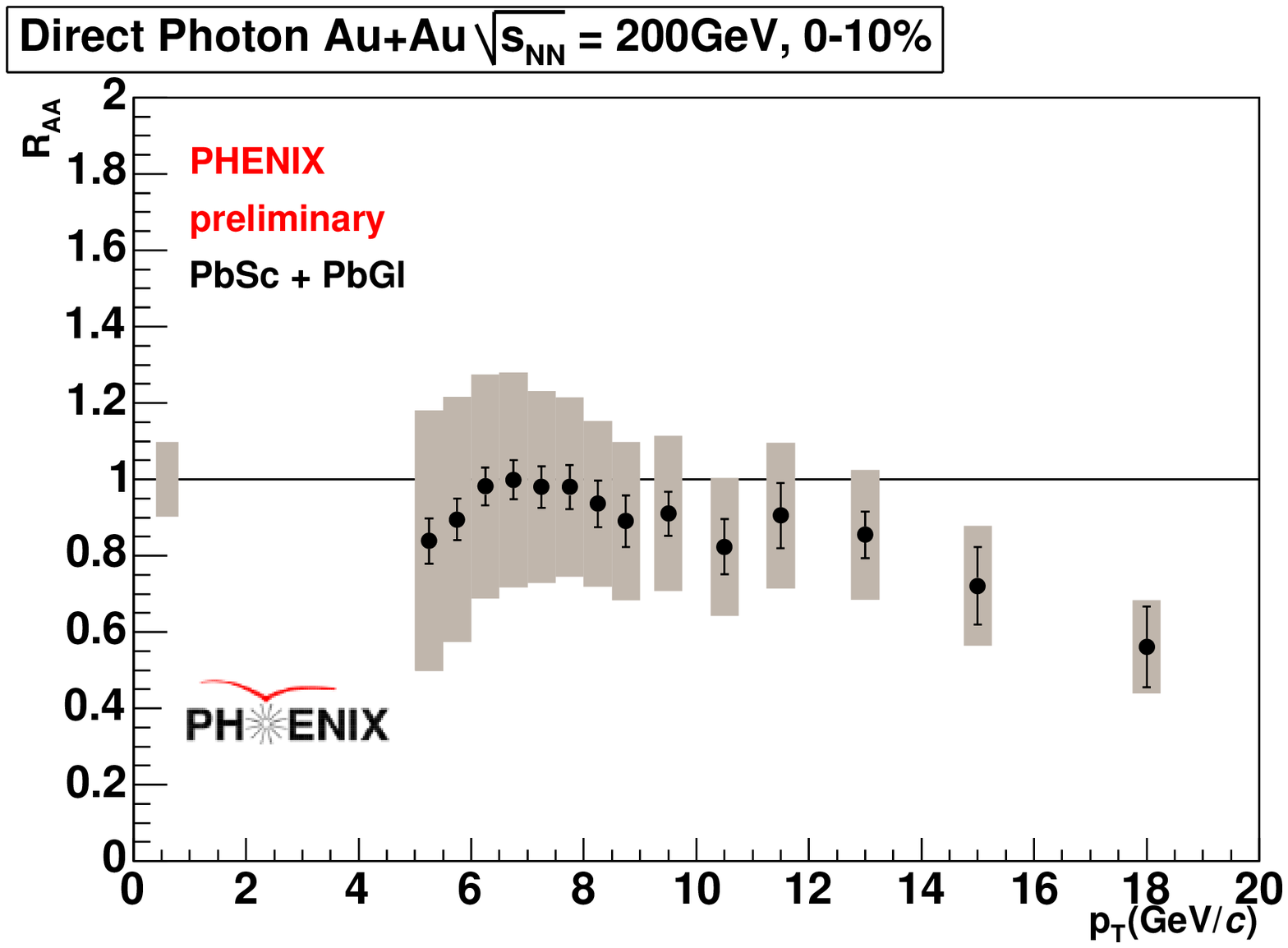}
\vspace*{-10mm}
\caption{Preliminary PHENIX direct photon $R_{AA}$
  in the most central (0-10\%) 200 GeV Au+Au collisions, fit to the
  $pp$ direct photon spectra measured in the same experiment.}
\label{fig:raa_qm06_cent0010}
\end{minipage}
\end{figure}

The first RHIC measurement of the photon nuclear modification factor,
shown in Fig.~\ref{fig:ppg042}, suggests that
the direct photon $R_{AA}$ is neither enhanced
nor suppressed, as opposed to the observed strong hadron suppression
shown on the right side of Fig.~\ref{fig:ppg042}.
The $\pi^0$ suppression observed in the
very first RHIC Au+Au run but absent in d+Au collisions was interpreted as 
a final-state effect.  Both the $\pi^0$ and direct photon $R_{AA}$ 
interpretations assume that, at high $p_T$ 
where hard scattering should dominate, scaling the
$pp$ cross sections by the nuclear overlap function 
$T_{AB}$\footnote{Colloquially, but quite misleadingly often called
``scaling with the number of binary collisions'' or  $N_{\rm coll}$ scaling}
is a sensible estimate of the expected $AA$ yields {\it in the
absence of a medium}.  Jets are quenched relative to this expectation.
Strictly speaking, only the direct photon $R_{AA}$
{\it proves} that this expectation is reasonable: photons produced by
hard scattering should leave the medium unchanged and their yield should thus
scale with $T_{AB}$.  This assumption
has indeed been confirmed within $\sim 20-30$\% uncertainty by
RHIC Run-2 Au+Au data~\cite{ppg042} 
and by the {\it integrated} $R_{AA}$ with $p_T>6$ GeV/$c$ where $R_{AA}$
was calculated using an NLO pQCD calculation since $pp$ 
direct photon data were not yet available.

\begin{figure}
\includegraphics[width=0.7\linewidth]{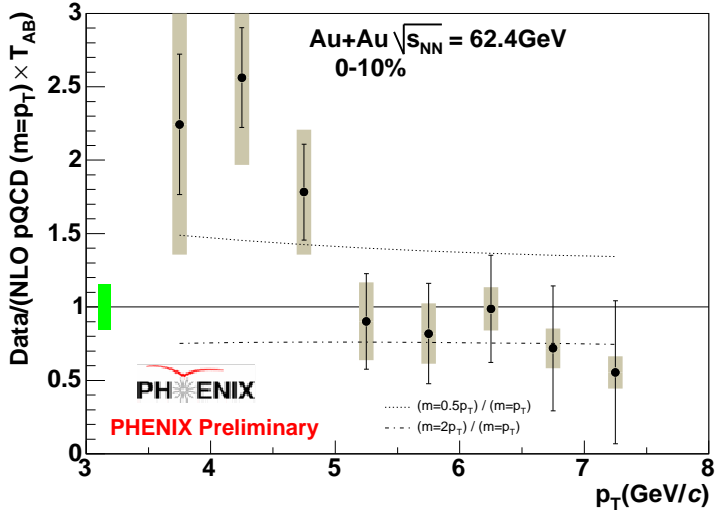}
\hspace*{3.5cm}
\caption{Preliminary PHENIX direct photon $R_{AA}$
  in the most central (0-10\%) 62 GeV Au+Au collisions, using an
  NLO pQCD calculation with scales of $0.5p_T$, $p_T$, and $2.0p_T$.} 
\label{fig:RAA62}
\end{figure}

The much larger Au+Au data set from Run-4, the two different
collision energies ($\sqrt{s_{NN}}= 200$ and 62 GeV) and measurement of
$pp$ reference data in the same experiment with similar systematic errors made
it possible to study the evolution of $R_{AA}$ with $p_T$ and
collision centrality in much greater detail.  
The fit to the $pp$ data differs both in shape and magnitude from NLO 
pQCD \cite{Isobe:2006qm}.  The data are higher,
particularly at the lower and upper ends of the $5<p_T<23$ GeV/$c$ range,
see Fig.~\ref{fig:cppdirgamma}.
The $R_{AA}$ obtained from a fit to the $pp$ data  is shown in
Fig.~\ref{fig:raa_qm06_cent0010} for the most
central 200 GeV Au+Au collisions.  There is an apparent suppression at 
very high $p_T$.   At least part of the suppression
has been predicted due to the ``isospin effect''~\cite{Arleo:2006jh},
a natural consequence of the different quark content 
($\sigma \propto \Sigma e_q^2$) of
protons and neutrons\footnote{This implies that while $pp$ data
are the proper reference for the hadron $R_{AA}$, in the photon case
the proper mixture of $\sigma_{pp}$, $\sigma_{pn}$, and $\sigma_{nn}$ must be
used.  For example, in minimum bias collisions
$\sigma_{AA}/N_{\rm coll} = (1/A^2)(Z^2\sigma_{pp} +
2Z(A-Z)\sigma_{pn} + (A-Z)^2\sigma_{nn})$.}.  Hence the suppression in the
photon $R_{AA}$ is referred to as ``apparent''.  Other mechanisms such as 
modifications of the parton densities and fragmentation functions
which enhance or deplete high $p_T$ direct photon yields
in heavy-ion collisions have also been proposed.  Finally, 
distinguishing direct photons from $\pi^0$ decay photons
producing an overlapping shower in the PHENIX calorimeter becomes
increasingly difficult at $p_T>14$ GeV/$c$.

\begin{figure}[!t]
\begin{minipage}{0.470\linewidth}
  \includegraphics[width=\linewidth]{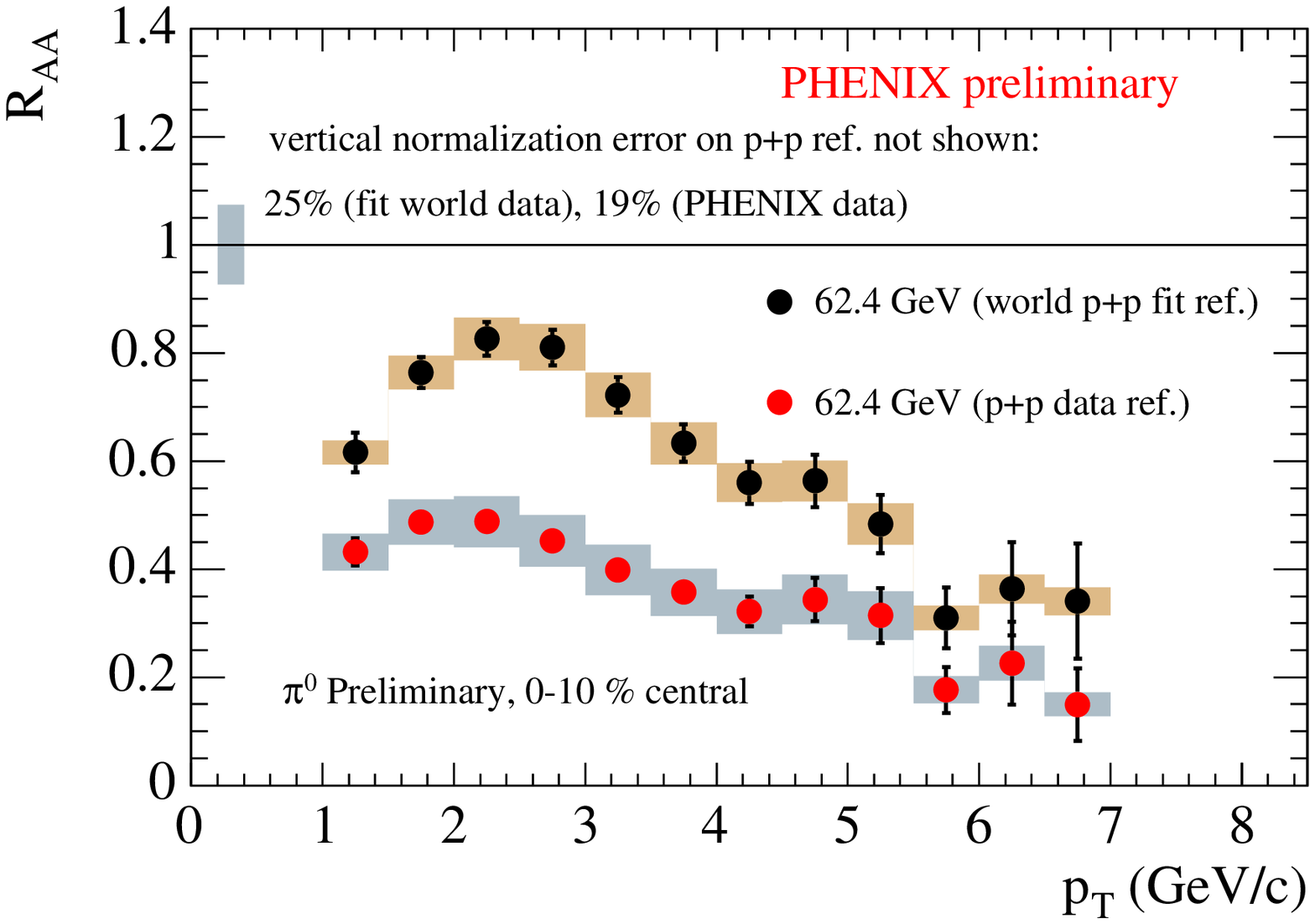}
\vspace*{-10mm}
\caption{Preliminary PHENIX $\pi^0$ $R_{AA}$ in central 62 GeV
  Au+Au collisions using a fit to the world average of previous $pp$ 
  data (black circles) and a PHENIX $pp$ measurement (red circles).}
\label{fig:pi0_62_plot1}
\end{minipage}
\hspace{\fill}
\begin{minipage}{0.470\linewidth}
\vspace{0.4cm}
\includegraphics[width=\linewidth]{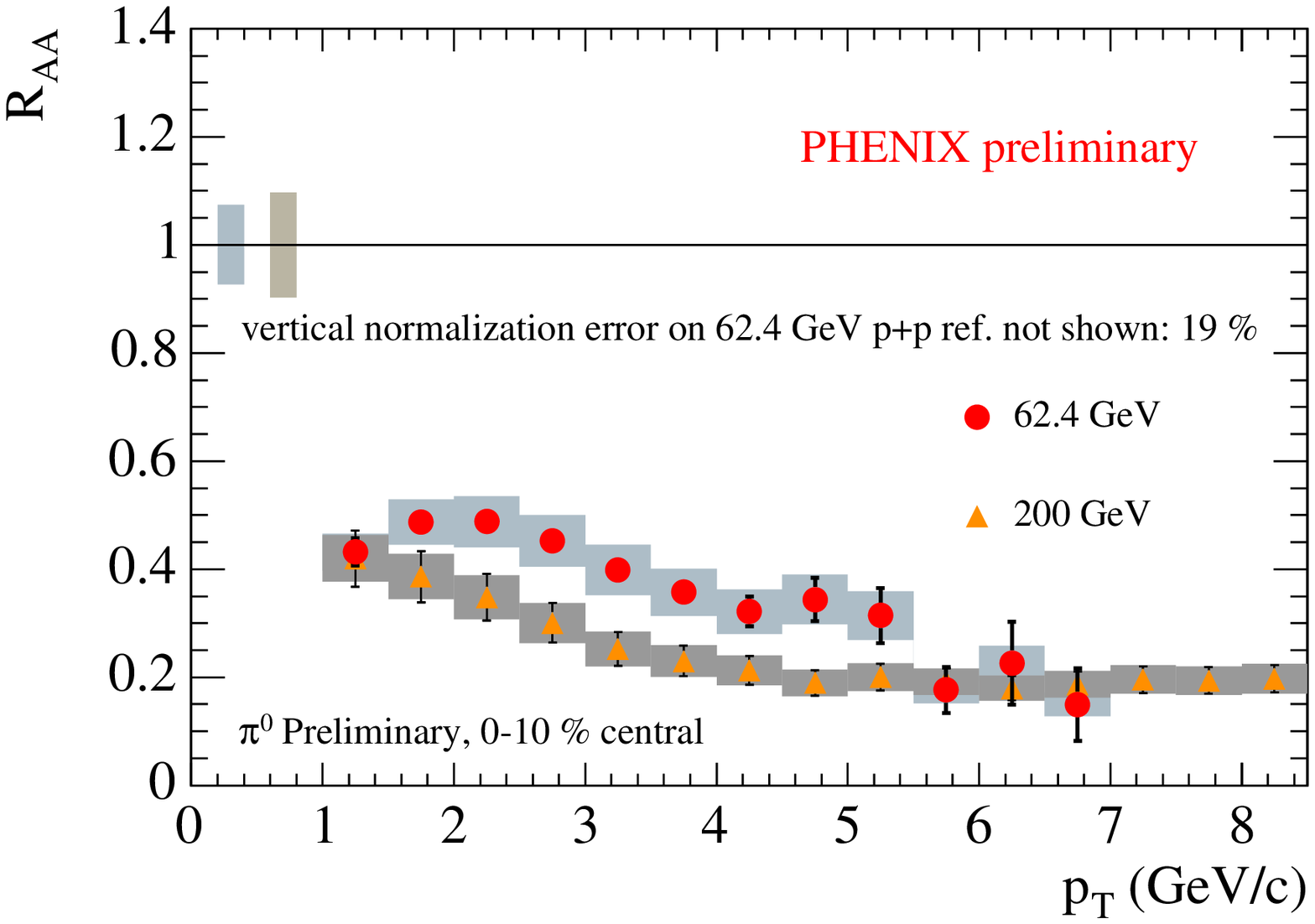}
\vspace*{-10mm}
\caption{Preliminary PHENIX $\pi^0$ $R_{AA}$ in central 62 GeV
(red circles) and 200 GeV (orange triangles) Au+Au collisions.  Both
ratios use PHENIX $pp$ measurements.}
\label{fig:pi0_62_plot3}
\end{minipage}
\end{figure}

However, RHIC's ability to provide collisions over a wide energy range
may allow some means of separation.  If the suppression at 200 GeV, as shown in
Fig.~\ref{fig:raa_qm06_cent0010}, is indeed only an isospin effect, it
should scale with $x_T=2p_T/\sqrt{s}$.  Therefore, in lower energy
collisions it should manifest itself at lower $p_T$ where the 
experimental difficulties due to overlapping showers are absent.  
PHENIX measured a similar photon $R_{AA}$ in 62 GeV Au+Au collisions 
using NLO pQCD calculations for $pp$ (preliminary data from PHENIX on direct
photon production in $pp$ at 62GeV will be available soon.)
The results for
the 10\% most central collisions is shown in Fig.~\ref{fig:RAA62}.  While
within errors it is consistent with unity, using measured
data in the ratio may change this conclusion, as was the case at 200 GeV.

Although not an electromagnetic probe {\it per se}, the $\pi^0$
$R_{AA}$ in Figs.~\ref{fig:pi0_62_plot1} and~\ref{fig:pi0_62_plot3}
illustrates the importance of measuring the reference in the
same experiment with similar systematic errors. 
Fig.~\ref{fig:pi0_62_plot1} compares the previous and the current preliminary
$\pi^0$ suppression in 62 GeV Au+Au collisions.
The earlier result used a fit
to the ISR $pp$ data (upper, black circles) whereas
the new result uses the $pp$ cross sections now measured in the same
experiment (PHENIX).  
The two results in Fig.~\ref{fig:pi0_62_plot1} would clearly lead
to very different physics conclusions.
Fig.~\ref{fig:pi0_62_plot3} shows the new 62 GeV
result compared to the $R_{AA}$ obtained in 200 GeV collisions. 


\subsection{Photon Azimuthal Asymmetries (Elliptic Flow)}
\label{sec:photon_elliptic}
\begin{figure}
\hspace{3.5cm}
\includegraphics[width=0.5\linewidth]{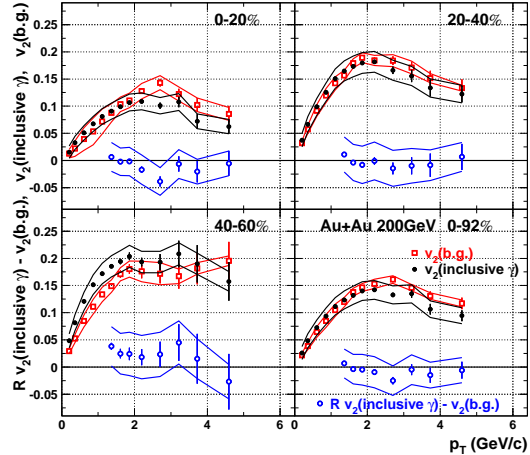}
\caption{The measured $v_2$ of inclusive photons
($v_2^{{\rm inc}~\gamma}$,
solid circles) and expected photon $v_2$ from hadronic decays
($v_2^{\rm bg}$, open squares)~\protect\cite{Adler:2005rg}.
The subtracted quantity, $R v_2^{{\rm inc}~\gamma} - v_2^{\rm bg}$
is plotted at the bottom of each panel (open circles).  Here
$R=(N_{{\rm dir}~\gamma}+N_{\rm bg})/N_{\rm bg}$
corresponds to the product $v_2^{{\rm dir}~\gamma} (R-1)$).
Four centralities are shown, including the minimum bias result.
}
\label{fig:ppg046_fig2}
\end{figure}

Fig.~\ref{fig:ppg042} showed that the {\it overall} azimuthally-integrated
high-$p_T$ photon yields scale with $N_{\rm coll}$
and are well described by pQCD
within current experimental and theoretical uncertainties.
However, this global agreement may mask more subtle effects.
It is even possible that the agreement is only accidental,
due to cancellations of processes that enhance and
quench the photon yield.  An important step toward clarifying the level of
agreement is the study of azimuthal asymmetries in the photon distributions,
specifically their elliptic flow, $v_2$.
If photons from the initial hard scattering do not interact with the medium, 
their $v_2$ is expected to be zero. However, initial hard scattering is
not the only source of photons in $AA$ collisions.
They may also originate from jet-thermal parton interactions
or from Brehmsstrahlung off a quark.  These photons are expected
to exhibit a negative
$v_2$~\cite{Turbide:2006bz,Chatterjee:2005de} (since more
material is traversed out-of-plane, the major axis in
coordinate space) with a strong $p_T$-dependence.
However, photons from thermal radiation should reflect the
dynamical evolution of the hot and dense matter, thus carrying
a positive $v_2$.

The first measurement of photon elliptic flow at RHIC is shown in
Fig.~\ref{fig:ppg046_fig2} \cite{Adler:2005rg}.  The
measurement is quite delicate due to the large background from
$\pi^0$ decay-photons that inherit the $v_2$ of the parent $\pi^0$.
While the measured $v_2$ of inclusive photons is consistent with
the $v_2$ of hadronic decay photons, {\it i.e.} a zero net direct
photon flow, the errors are appreciable and the
direct-to-inclusive photon ratio is very small at low $p_T$.
The quality of the current data is insufficient to verify the
predictions~\cite{Turbide:2006bz,Chatterjee:2005de}, including the sign
of the net flow.  Much higher statistics can help remedy the situation,
at least at higher $p_T$.  Although the net direct photon flow is
predicted to decrease, the statistical errors will become
smaller and, equally important, the direct-to-inclusive photon ratio
will increase dramatically.  But even at high $p_T$ the net flow
will be a competition between processes with positive and negative $v_2$.
New analysis techniques may be able to statistically disentangle
isolated and non-isolated direct photons in heavy-ion collisions.
Jet-photon conversions primarily produce isolated
photons~\cite{Turbide:2006bz} with negative $v_2$. The magnitude of this
flow depends strongly both on $p_T$ and the jet energy-loss mechanism
in heavy-ion collisions. Therefore, a measurement of
the isolated photon $v_2$ may provide an independent constraint on
energy-loss models.

\subsection{Electron $R_{AA}$ and Flow}
\label{sec:charm}
The recent measurement of ``non-photonic'' single-electron spectra, 
associated with decays of open heavy-flavor hadrons, at
RHIC lead to two unexpected and very important results: (i) the
nuclear modification factor, $R_{AA}$, shows a strong suppression in
central Au+Au collisions~\cite{phenix-raa,ppg066,star-raa}, 
comparable to
the suppression observed for pions; (ii) the elliptic flow, $v_2$,
is significant at low $p_T$, up to 10\%.  Although such strong
suppresion was predicted a decade ago~\cite{Shuryak:1997}, it was later
argued that the suppression would be mild due to
the ``dead cone effect'' for heavy quarks.  Measurements from both STAR
and PHENIX have, in fact, shown strong charm suppression and significant
charm flow (early thermalization) in 200 GeV Au+Au collisions:
observations not {\it simultaneously} explained by current theories.
These issues primarily concern charm and bottom physics and are
discussed in great detail in the heavy flavor part of this Report.
However, they may be significant for ``classic''
electromagnetic probes as well since heavy flavor suppression
actually aids measurements of intermediate mass continuum dilepton
radiation by reducing the combinatorial
background from charm\footnote{This background is irreducible for the PHENIX 
HBD.} in
this regime, see Fig.~\ref{fig:hbd_PHENIX}.

\subsection{Direct Thermal Photons via Low-Mass Dielectrons}
\label{sec:internalconv}

\begin{figure}[!t]
\begin{minipage}{0.48\linewidth}
\includegraphics[width=\linewidth]{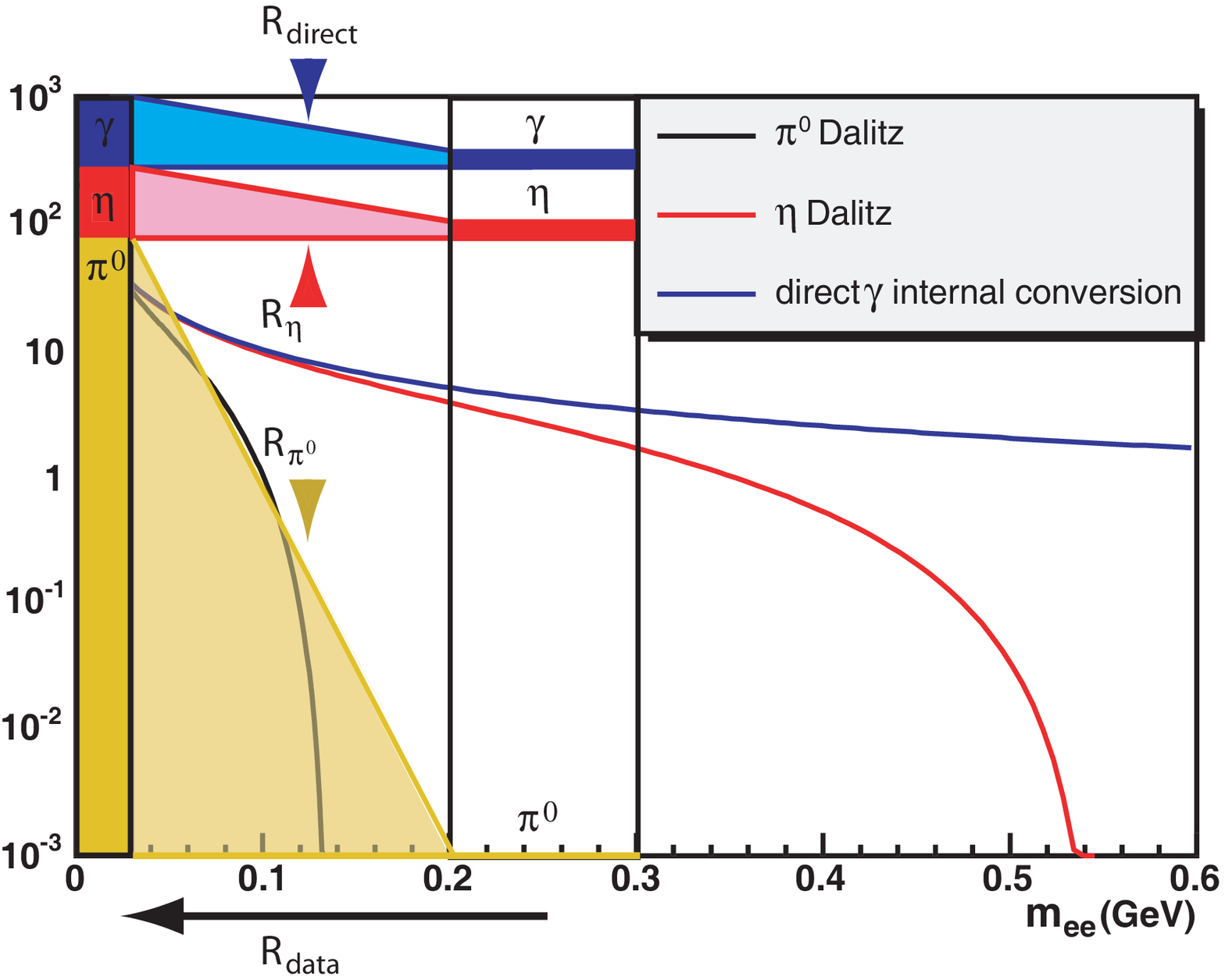}
\end{minipage}
\hspace{0.6cm}
\begin{minipage}{0.49\linewidth}
\includegraphics[width=\linewidth]{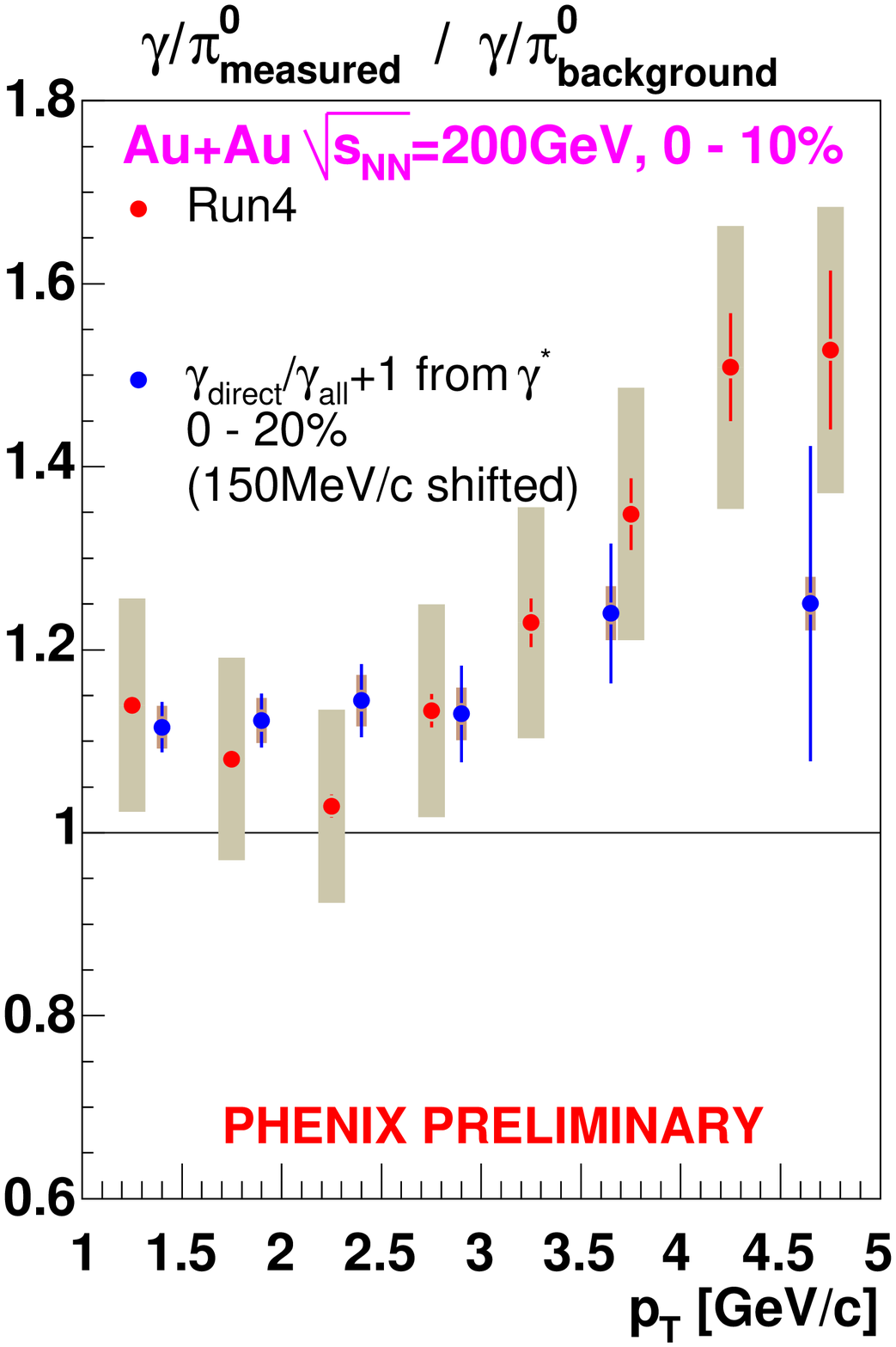}
\end{minipage}
\caption{Left: illustration of the direct photon
measurement via low-mass dileptons~\protect\cite{Bathe:2005nz}.
Right: preliminary results
on low- and intermediate-$p_T$ direct photons,
the ``photon excess ratio'',
with the traditional calorimeter method (red circles) and
low mass electron pairs (blue circles).}
\label{fig:akiba_method}
\end{figure}

A very promising approach for measuring low-$p_T$ direct photons
is to utilize low-mass electron pairs from ``internal conversions'',
as was first applied in heavy-ion collisions by
PHENIX~\cite{Bathe:2005nz}.
The basic idea is that any process
producing a real photon can also produce a very low mass virtual 
photon~\cite{KrollWada} which subsequently decays into an $e^+e^-$ pair.
This direct photon signal competes, of course, with dielectrons
from $\pi^0$, $\eta$, {\em etc}  Dalitz decays.
The dielectron rates and mass distributions are described both
for low-mass direct photons and Dalitz decays by the
Kroll-Wada formula~\cite{KrollWada},
\begin{equation}
  \frac{1}{N} \frac{dN_{ee}}{dm_{ee}} = \frac{2\alpha}{3\pi}
   \sqrt{1 - \frac{4m_e^2}{m_{ee}^2}} \bigg(1 + \frac{2m_e^2}{m_{ee}^2}\bigg)
   \frac{1}{m_{ee}} |F(m_{ee}^2)|^2 \bigg(1 - \frac{m_{ee}^2}{M^2}\bigg)^3 \; ,
 \label{eq:KW}
\end{equation}
where the form factor, $F$, is unity for real photons.
Note that the phase space for Dalitz decays is limited by the
mass of the parent meson, $m_{ee} < M_{\pi^0,\eta,\omega}$,
while for direct photons it is not, $m_{ee}\sim p_T$.
Therefore, the measurement becomes relatively clean for
$p_T > 1$ GeV/$c$, still in the low-$p_T$ realm where
``traditional'' calorimeter measurements have serious
difficulties.  The method is illustrated in the left
panel of Fig.~\ref{fig:akiba_method} while the resulting photon excess
ratios are shown on the right. 
The systematic errors are
much smaller for the new method\footnote{The errors are smaller, at least for
the direct photon excess {\it ratio}.  However, the absolute normalization 
and the direct photon cross sections are difficult to assess and therefore 
have been inferred using the calorimeter.}.  The
statistical errors become large for $p_T>4$ GeV/$c$ although $\sim 15$
times more data have been analyzed for dielectrons (solid red
circles) than for the calorimeter analysis (open blue
circles).  There is an important lesson here:
even signals such as thermal photon radiation, typically
not considered to be ``starving for statistics'', novel but
{\it promising new analyses techniques require large luminosity increases.}

\begin{figure}[!t]
\centering{
\includegraphics[width=0.8\linewidth]{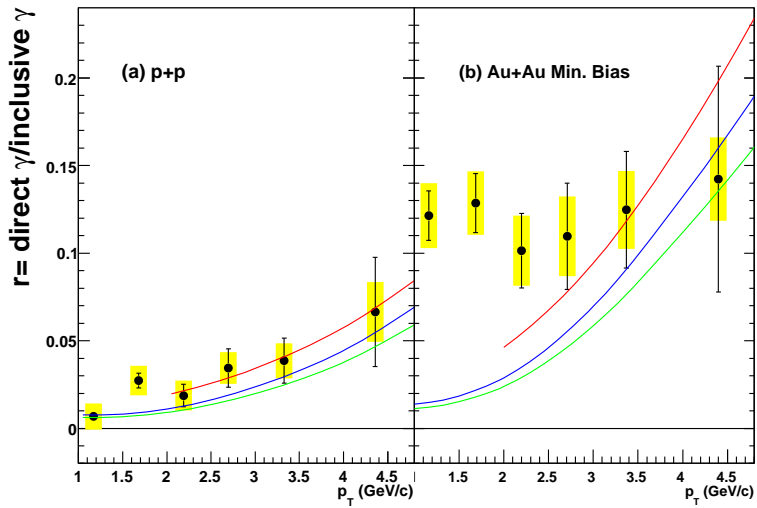}
}
\caption[]{The fraction of the direct photon component as a function
  of $p_T$ in 200GeV
  $pp$ (left) and Au+Au (right) collisions.  The error bars and the
  error band are the statistical and systematic uncertainties,
  respectively.  The curves are from NLO pQCD calculations.
}
\label{fig:ppg086_fig3}
\end{figure}

Before drawing any conclusions about the thermal nature of the
apparent direct photon signal at low $p_T$ the baseline in $pp$ 
has to be established.  This is shown on the left panel of
Fig.~\ref{fig:ppg086_fig3} (final results from PHENIX): the signal is 
in good agreement with NLO pQCD predictions.  In contrast, in
Au+Au (right panel) there is clearly an additional source at low
$p_T$. 

The conversion approach is also appealing from the theoretical point
of view since the framework outlined in Sect.~\ref{sec_frame}
accommodates a comprehensive treatment of real and virtual photons,
as represented by Eqs.~(\ref{Rphot}) and (\ref{Rdilep}). However,
additional theoretical work is needed for a more complete assessment of
possible very-low mass dielectron sources~\cite{Liu:2006zy}.

\section{THE FUTURE}
\label{sec_future}
\subsection{Detector upgrades}

The original designs of both PHENIX and STAR anticipated
later upgrades to enhance their physics capabilities.
In fact ``upgrading'' as a response to spectacular rises in luminosity
and physics insights from data already taken was and remains an almost
continuous process from the very start of operations.  Few
can be tied uniquely to RHIC-II with the exception of upgrades
facilitating the handling of high luminosities and data rates.
Nonetheless, several major projects should be discussed here 
because of their magnitude and their
impact on the future capabilities of RHIC-II, see Table~\ref{tab:axel_table}.
These projects are in different stages of development.
Some are in the early R\&D phase while others are ready to be
installed soon.  Here we concentrate on
upgrades that are particularly important for
electromagnetic probes.  The STAR Time-of-Flight (TOF) detector and
Heavy Flavor Tracker (HFT) improve the electron identification
and the displaced vertex measurement by rejection of electrons from
charm, bottom and Dalitz decays as well as conversion electrons.
The PHENIX Hadron Blind Detector (HBD) will reject Dalitz pairs for
light vector meson measurements.  The Silicon Vertex Detector
(SVTX) will measure displaced vertices to identify electrons from open
charm.  The Nose-Cone Calorimeter (NCC) will measure photons and
$\pi^0$'s at forward rapidities.  
A new high-resolution sampling calorimeter is proposed
for STAR to measure direct photon correlations.
In the following, we will discuss these systems in more detail.

\subsubsection{STAR Time-of-Flight and Heavy Flavor Tracker}
\label{sec_star}
STAR electron identification is made possible by a combination of
energy loss, $dE/dx$, by charged particles due to ionization
of the time-projection chamber (TPC) gas and a velocity
measurement with the TOF system~\cite{starepid}.  The
relativistic rise of the electron $dE/dx$ separates electrons from
hadrons except at the crossovers with pions at
$\sim 0.2$ GeV/$c$, kaons at $\sim 0.6$ GeV/$c$, protons at
$\sim 1.1$ GeV/$c$ and
deuterons at $\sim 1.5$ GeV/$c$. A time-of-flight measurement, with the
requirement that $|1-\beta|< 0.03$, eliminates slow hadrons and cleans
up the crossovers, resulting in clean electron
identification~\cite{starepid}.

In addition to direct measurements of open-charm hadrons via $K\pi$
decays, the STAR HFT~\cite{starHFT} will be a powerful means of discriminating
primordial electrons from
background electrons in the measurement of electromagnetic probes.

As discussed earlier, direct photon and lepton production is rare
and overwhelmed by photons and leptons from electromagnetic decays of
hadrons and $\gamma$ conversions to electrons, $\gamma
\rightarrow e^+e^-$.  The $\gamma$ conversions, to a large extent, 
occur in the detector material.  The HFT detector will reduce
the background electrons and positrons from these $\gamma$
conversions.  By requiring hits in the HFT, electrons from photon
conversion outside the HFT, i.e., in the Silicon Strip Detector
(the upgraded STAR Silicon Vertex Tracker) and the TPC inner field
cage, are rejected. To estimate the signal-to-background ratio in the
vector-meson measurements, STAR adopted a reasonably conservative approach
and assume that the HFT can reject $\gamma$ conversions by a factor of
10~\cite{starHFT}. Although the configuration of HFT has evolved 
significantly, the simulation of background rejection is not expected to 
change. Another electron background arises from
semileptonic decays of heavy quarks.  If
heavy-quark spectra are extrapolated from $pp$ collisions, the
dominant dilepton source in central Au+Au collisions at intermediate
mass~\cite{Rapp:2002mm} is due to semileptonic $c\overline{c}$ decays. With
the large charm yield at RHIC, the latter are comparable to the
yield from $\gamma$ conversions and $\pi^0$ and $\eta$ Dalitz decays
after HFT rejection. Detailed simulations show
that the HFT is capable of rejecting $\sim 75$\% of $e^+e^-$ pairs from 
$D^0$ decay while preserving 50\% of the direct $e^+e^-$ pairs.

The large reduction in electron background will enable STAR to observe
the electromagnetic signal from low-mass vector mesons and radiation
of intermediate-mass dileptons with a few hundred thousand central
Au+Au events.  The rejection of $\pi^0$ and $\eta$ Dalitz
decays by a factor of three (single track) can be achieved by measuring
both electrons of the pair, made possible by the large
acceptance of the STAR TPC. Future work on further rejection of such 
dilepton-background pairs with the inner tracker alone will be exploited. 
With the upgrades, STAR expects to detect 6K
$\phi$ and 22K $\omega$ decays in 200 million recorded central Au+Au
collisions.  These are to be compared with $\sim10$K for the $\phi$ and 
$\sim6$K for the $\omega$ in central In+In collisions,
presented by NA60~\cite{NA60QM05}.

\begin{figure}[!t]
\centering{
\includegraphics[width=0.6\linewidth]{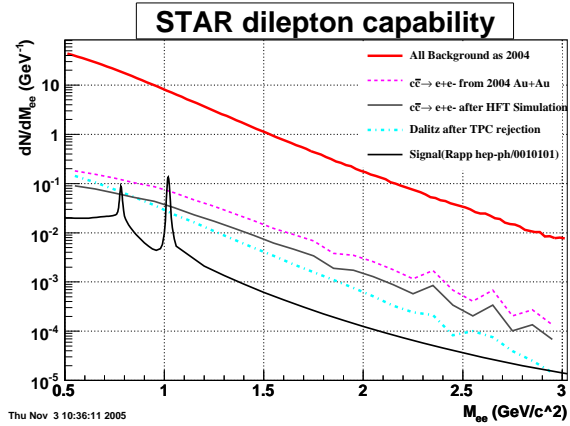}
}
\caption[]{Dielectron invariant mass distributions for central
200~GeV Au+Au collisions.  The solid black
curve is the prediction for thermal emission~\cite{Rapp:2002mm} in the STAR
acceptance.  The red curve at the top is the total dielectron
invariant-mass spectrum in the 2004 configuration assuming electron
PID from full TOF coverage.  The pink dashed line is the $e^+e^-$ pair
spectrum from semileptonic charm decays, the grey curve is the
charm $e^+e^-$ spectrum after HFT ${\rm DCA} <80$ $\mu$m, and the
dot-dashed line are from $\pi^0$ and $\eta$ Dalitz decays after
rejection by the TPC.
}
\label{fig:STAR_2}
\end{figure}
Fig.~\ref{fig:STAR_2} summarizes the dielectron background and signal
invariant-mass distributions.  The signals of
medium-modified vector mesons and thermal QGP radiation (black curve)
are from calculations~\cite{Rapp:2002mm} folded over the STAR
acceptance.  The uppermost (red) curve is the total dielectron
invariant-mass spectrum in the 2004 configuration, obtained from
the single-inclusive electron spectrum measured in 200~GeV Au+Au
collisions assuming electron PID from full TOF coverage.
The pink dashed line is the $e^+e^-$ spectrum from semileptonic
charm decays derived from the non-photonic single-electron spectra
measured in Au+Au collisions.  The gray curve is the charm $e^+e^-$
distribution after applying the HFT distance of closest approach cut,
${\rm DCA} <80$ $\mu$m.  The dot-dashed line is
$\pi^0$ and $\eta$ Dalitz decays after TPC rejection.
The net result is a signal-to-background ratio that, even in the
continuum regime, is $\leq 0.1$,
comparable to the central In+In NA60 measurements~\cite{Arnaldi:2006jq}.
The standard way of dealing with the residual background
is the mixed-event method, used by CERES and NA50/NA60
at the SPS and will be used in PHENIX and STAR.

With the proposed
Time-of-Flight, Heavy-Flavor Tracker and Data Acquisition System
(DAQ1000) upgrades, STAR will be able to take data at a rate of 1000~Hz
with very little dead time.  At the same time, the collision
vertex has to be limited to $\sim\pm 5$~cm due to the acceptance of the
HFT~\cite{starHFT}. The current data-taking
rate is 50~Hz with a collision diamond of $\sim\pm 50$~cm. Without a
machine upgrade, the average luminosity delivered to STAR is
$8\times10^{26}$ cm$^{-2}$s$^{-1}$ or 6~kHz Au+Au minbias nucleus
collision rate. Taking into account the factor of $3-5$ loss from the
vertex constraint and centrality binning ({\em e.g.} 10\%
most central), central-triggered Au+Au events can be recorded at a
rate of about 200~Hz with all the available luminosity, significantly
below the DAQ1000 capability. A factor of 10 luminosity upgrade will
enhance both the collider and detector. In peripheral collisions, the 
statistics will be lower due to lower event
multiplicities. However, triggering will be more effective even
for low-$p_T$ lepton pairs. Since the luminosity decreases quadratically
with decreasing beam energy, an efficient energy scan at RHIC
requires a luminosity upgrade, as discussed in more
detail in Sect.~\ref{sec:scan}.

\subsubsection{PHENIX Hadron Blind Detector and Silicon Vertex
  Detector}
\label{sec:hbd}
\begin{figure}[!t]
\hspace{3.5cm}
\includegraphics[width=0.5\linewidth]{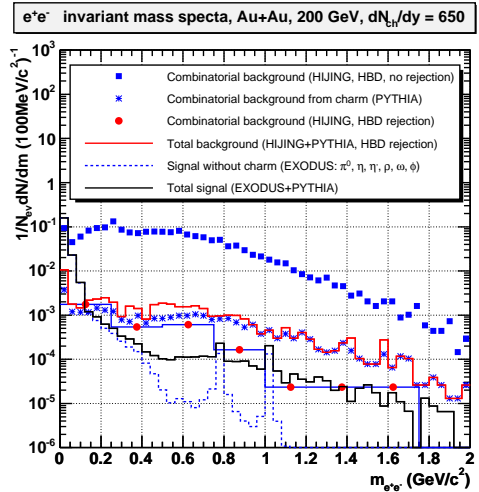}
\caption{
Combinatorial background for low-mass electron pairs compared to the
total signal from vector mesons and charm with and without the HBD.
Blue squares: total $e^+e^-$ combinatorial (HIJING), no rejection
from the HBD.  Red circles: combinatorial background after rejection
with HBD.  Blue stars: combinatorial background from charm alone
(PYTHIA), not rejected by the HBD, measured separately by the Silicon
Vertex Detector (SVTX).  Solid red line: total combinatorial background
after HBD rejection (HIJING+PYTHIA).  Dashed blue line: pure signal
from direct vector-meson and Dalitz decays after freezeout ``cocktail"
without the charm contribution (EXODUS).
Solid black line: total signal from vector mesons, Dalitz decays and
charm (EXODUS+PYTHIA).
}
\label{fig:hbd_PHENIX}
\end{figure}
As discussed earlier, the measurement of the low-mass electron pair
continuum to infer the in-medium modifications of the light vector-meson
($\rho$, $\omega$, and $\phi$) spectral functions is crucial for understanding
the fate of hadron masses and the approach to chiral symmetry
restoration in the hot,
dense matter created in heavy-ion collisions.  However, the $e^+e^-$
signal is overwhelmed by the combinatorics from $\pi^0$ Dalitz decays
and photon conversions.  The combinatorial background from open-charm decays
is also significant, see Fig.~\ref{fig:hbd_PHENIX}.  Fortunately, the typical 
opening angle of the $e^+e^-$ pair is very small, both
for $\pi^0$ Dalitz decays and photon conversions.  
Thus unless they are swept apart by
a magnetic field, a proximity cut on electron pairs is an effective
veto on Dalitz and conversion electrons.  The PHENIX magnet has an
inner and an outer coil which are typically powered by currents in the
same direction for maximum field to optimize tracking to the
highest possible $p_T$.  However, the current in the inner coil can
be reversed, making the magnetic field vanishingly small around
the collision region, allowing the
installation of a Hadron Blind Detector (HBD) to eliminate the
combinatorial background from Dalitz decays and conversions.  This detector,
described in Ref.~\cite{hbdnim}, is a windowless Cherenkov detector
operated with pure CF$_4$ in proximity focus configuration with a CsI
photocathode and a triple GEM detector with pad readout.  It is a
crucial upgrade for the exploration of the properties of the new matter
created. As illustrated in Fig.~\ref{fig:hbd_PHENIX} it will reduce
the combinatorial background and increase the signal/background ratio
from $\sim10^{-3}$ to $\sim10^{-1}$. The latter is again very comparable
to NA60 and the expected performance of STAR after upgrades,
described in the previous section.

The remaining background from
open charm will be separately measured in PHENIX by a Silicon Vertex
Detector (SVTX), which will measure the heavy-flavor
displaced vertex with a 40 $\mu$m resolution of the distance of closest
approach.  The SVTX resolution is driven by the $c\tau$ of 123 $\mu$m and 
462 $\mu$m for $D^0$ and $B^0$ decays, respectively.
The SVTX will have a central barrel and two endcap detectors,
covering both central and forward rapidities and providing
inner tracking with full azimuthal coverage and up to $|\eta|<2.4$.
This, in particular, will enable the
measurement of correlated $e\mu$ invariant-mass spectra and thus
provide a standalone determination of the correlated open heavy flavor 
component of the dilepton spectra.

\subsubsection{The PHENIX Nose-Cone Calorimeter}
\label{sec:ncc}
PHENIX can measure photons, $\pi^0$'s and $\eta$'s well at
midrapidity but the rapidity coverage of the electromagnetic
calorimeter in the central arm is limited
to $|y|<0.35$, making full jet reconstruction very difficult.
Also, several large $y$ measurements in d+Au
collisions suggest that low-$x$
gluons might be saturated so that the CGC describes
the results, including hadron suppression at large rapidity
relative to no suppression at $y=0$.
If the large $y$ suppression is indeed a consequence of initial-state gluon
saturation, photons should also be suppressed, see Sec.~\ref{sec_cgc}. 
Such a study is only feasible
with a calorimeter at large
rapidities.  The limited acceptance of the central arm
also makes crucial $\gamma-$jet measurements very difficult.

PHENIX thus proposed to replace the current copper magnet nosecones by
a calorimeter covering
$1 < |\eta| < 3$.  This Nose-Cone Calorimeter (NCC) is a
silicon-tungsten sandwich sampling calorimeter, longitudinally
segmented in three sections (two electromagnetic and one hadronic) 
and read out by $1.5\times1.5$ cm$^2$ Si pads.
In addition, two layers of 468 $\mu$m
pitch strip-pixels are added to achieve separation of direct photons
from high-energy $\pi^0$'s by shower-shape reconstruction.
The total depth is $\sim 35X_0$ radiation
and $1.3L_{\rm abs}$ nuclear absorption lengths.  The expected resolution
for electromagnetic showers is $23\%/\sqrt{E}$.
The longitudinal segmentation allows to distinguish 
between electromagnetic and hadronic showers.  The
coverage is sufficient to reconstruct the entire jet energy.

Therefore, jet physics and energy-loss
studies using both photon-tagged jets and leading $\pi^0$'s will be
possible far away from central rapidity with the NCC.  It will also
be very useful for studying $P$-state quarkonia via 
$\chi_c \rightarrow \gamma J/\psi$ and possibly
$\chi_b \rightarrow \gamma \Upsilon$ radiative decays together with the
muon spectrometer, giving information on the
different quarkonium dissociation temperatures and feed down
contributions to the $J/\psi$ and $\Upsilon$ 
yields~\cite{Satz05,Rapp:2006ii}.
The NCC is also an important part of the
spin program by measuring the
polarized gluon structure function at low $x$.

\subsection{New Measurements}


\subsubsection{Direct Photon Correlations (HBT)}
\label{sec:gamma_hbt}
The HBT correlation of photons from $\pi^{0}$ or $\eta$ decays is
exceedingly small because of the ``large'' distances,  $>10^7$ fm, at
which the decay occurs. Typical correlations from $\pi^{0}$ decay
photons are $\mu$eV/$c$ in relative photon four-momentum.
Therefore, any measurable HBT correlation will come from photons
directly emitted from the collision.

Photon HBT correlations encode space-time information about the system
emitting the photons. Many
calculations, as well as ``simple'' quantum mechanics, indicate that
by studying HBT in different $p_T$ regions,
emitting systems of different temperatures and thus at different
stages of evolution can be identified~\cite{Bass04}.  In
addition, interferometry can also be used to measure the yield of
direct photons~\cite{wa98-lowpt}.

While the effect of the $\pi^{0}$ decay background is to {\it dilute} the
HBT signal, the decay in itself does not generate a fake
correlation.  However, the residual correlation of decay
photons due to HBT $\pi^0$ correlations
{\it could} cause a fake HBT correlation.  When
the measured HBT parameters determined from charged pions are added
into a Monte Carlo calculation of the residual correlation, the change
is negligible compared to reasonable estimates of the direct photon
correlation assuming that the charged and neutral pion HBT parameters are 
the same~\cite{Sandweiss06}.

The dilution effect, on the other hand, is a serious
issue. Reasonable estimates of the visible HBT correlation indicate
that $\lambda$, the effective amplitude of the HBT correlation, is only
a few parts per thousand.  Consequently, great care must be taken in
designing experiments to measure the direct photon HBT. Corresponding
work on this problem is briefly summarized in the remainder of this
section, based on a more complete, technical paper~\cite{Sandweiss06}.

STAR studies showed that the
best way to measure the HBT correlation is to use one photon that
converted in a thin (10\% $X_0$) converter
placed inside the inner-field cage of the TPC and another (real)
photon detected in the barrel calorimeter.  It is thus possible to be
sensitive to photons with nearly identical directions.  However, 
in order to make useful
measurements of the HBT parameters, the energy resolution of
the electromagnetic calorimeter must be substantially improved over
the resolution of the present calorimeter, $\delta
E/E=15\%/\sqrt{E}$. Pertinent studies show that $\delta
E/E=3\%/\sqrt{E}$ is certainly sufficient and some preliminary Monte
Carlo studies indicate that even $5\%/\sqrt{E}$ may be adequate albeit near 
the limit of usefulness.  Work is in progress to
determine the precise requirements, along with studies of the ``Shashlyk'' 
design, a relatively cheap calorimeter
where alternating plates of radiator and scintillator are read out
by an array of wavelength-shifting fibers which run parallel to the
long axis of the calorimeter and penetrate the radiators and the
scintillators.  The technology is well established: a Shashlyk
calorimeter used in E-865 at the AGS achieved $8\%/\sqrt{E}$
resolution, and ten years ago 75\% of the PHENIX calorimeter was built
with a similar design and shows similar performance.  Subsequent
studies done for the KOPIO experiment have demonstrated that
a $3\%/\sqrt{E}$ resolution is achievable. The designers of this
calorimeter are now engaged in a design study of a Shashlyk
calorimeter for STAR.

\begin{figure}[!t]
\begin{center}
\includegraphics[width=0.6\linewidth]{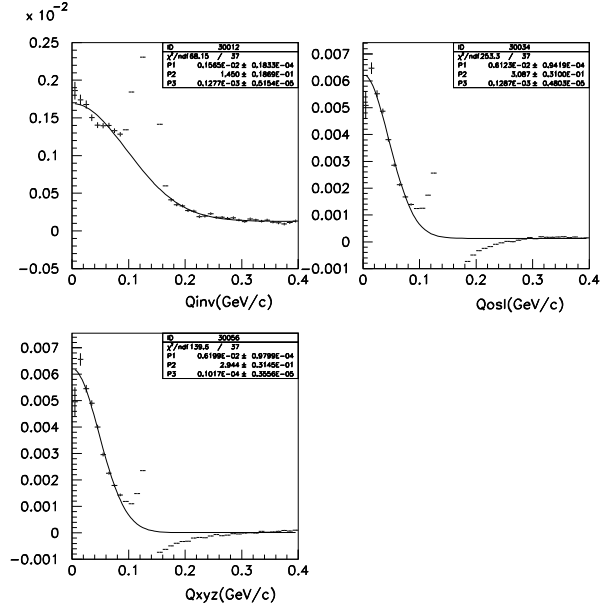}
\end{center}
\vspace{-0.2cm}
\caption{Simulated correlation functions using 40 million
events with full background, as described in the text.
Top left: $Q_{\rm inv}$, top right: $Q_{\rm osl}$, bottom left: $Q_{xyz}$.}
\label{fig_cor}
\end{figure}
In subsequent feasibility studies, a model of direct photon production
has been constructed which agrees with both existing measurements at
$p_{T}\geq 2.0$~GeV/$c$ and a recent calculation below that
value~\cite{d'Enterria:2005vz}. The model assumes Bjorken (boost
invariant) expansion, and thermal spectra from three regions of
temperature at different times in each rapidity slice.  The $\pi^{0}$
spectrum is taken from PHENIX data. The rapidity distribution is known from
the PHOBOS measurement, however, in the STAR range ($\eta=\pm 1.0$) it
is effectively constant. An analysis procedure for the calorimeter
has been devised which takes into account both the effects of isolation
cuts and of unavoidable unrecognized overlaps of
background photons.

STAR can measure the direct photon yields and HBT
parameters for $p_T\leq 600$~MeV/$c$ with about 40
million events.  Fig.~\ref{fig_cor} shows the simulated results for three
one-dimensional correlation functions:  $Q_{\rm inv}$ is the usual
invariant four-momentum difference; $Q_{\rm osl}$ is the four-momentum
difference in the usual out-side-long Bertsch-Pratt
system~\cite{Bertschxx}; and $Q_{xyz}$ is the four-momentum difference
in the $x$, $y$, and $z$ Cartesian system. The figure shows that, despite the
very low $\lambda$, quite good signals can be
obtained. The individual out, side, and long radii are functions of
the photon $k_T$, described in detail in a
forthcoming paper~\cite{Sandweiss06}. 
The number of events necessary to observe direct photon correlations for
$p_T$ between 1.0 and 2.0~GeV/$c$ is being studied, but it would clearly 
involve much larger data samples.  An interesting possibility is to use 
the high $E_T$ trigger to access the higher $p_T$ region since it
measures the system size at a different time~\cite{Bass04}.


\subsubsection{Spectral Distributions of the $a_1$ Meson}
\label{sec:a1}

\begin{figure}[!t]
\centering
\includegraphics[width=0.7\textwidth]{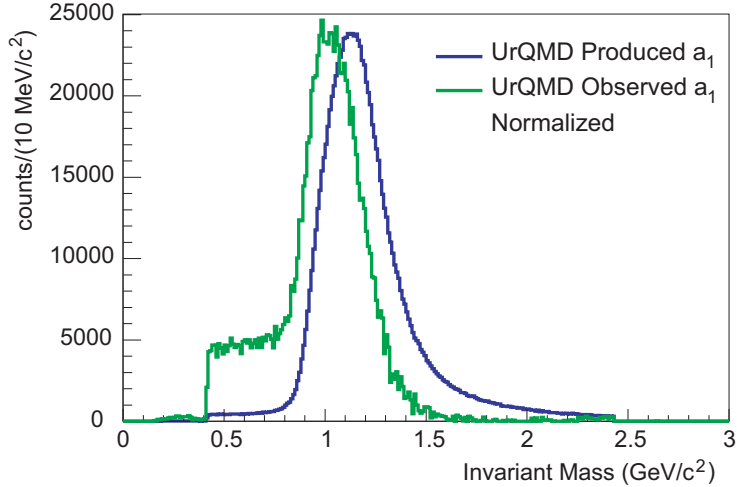}
\caption{The spectral shape of the initially-produced $a_1 \rightarrow 
\gamma \pi$ (blue line) compared to the spectral shape of the observed
$a_1 \rightarrow \gamma \pi$ which have passed through the medium without
being absorbed (green line),
normalized to the number of produced $a_1$.}\label{fig:a1}
\end{figure}

An experimental measurement of the iso-vector--axial vector ($a_1$)
spectral function is desirable because of its important role in
the search for chiral symmetry restoration.  It is related to
measurements of thermal low-mass dileptons, dominated by
the $\rho$ (iso-vector--vector correlator), the chiral partner
of the $a_1$, see~Sec.~\ref{sec_csr}.  It has been suggested
in~\cite{Rapp:2003ar}  
to make such a measurement via the $\pi^\pm\gamma$
invariant-mass spectra. Calculations of photons from $a_1$ decays
have concentrated on the inclusive photon
yield.  It is very difficult to isolate the $a_1$
contribution and determine the spectral shape
of the $a_1$ from inclusive measurements alone. This is the main 
motivation for measuring the
associated charged pion even though they experience strong
final-state interactions, including absorption. Therefore, the
in-medium spectral information will primarily pertain to the more
dilute stages of a heavy-ion reaction. However, the $a_1$ may
undergo significant modifications even at moderate temperatures and
densities~\cite{Rapp:2002pn}. Such effects have possibly been seen
in the $\rho \rightarrow \pi^+ \pi^-$~\cite{AdamsPRL92} decay,
which is ``penalized" by two pion absorption factors. 
Important information
about chiral symmetry restoration may thus be obtained
together with
the dilepton measurements of the $\rho$ meson.

We report on studies~\cite{Fachini06} of the
$a_1 \rightarrow \gamma \pi$ channel using transport model simulations,
i.e., minimum bias Au+Au UrQMD events~\cite{Vogel:2006rm} at
$\sqrt{s_{_{NN}}} = 200$ GeV. In UrQMD, the $a_1$ was introduced with
$B(a_1 \rightarrow \gamma \pi) = 0.1$,
$B(a_1 \rightarrow \rho \pi) = 0.9$, $M_{a_1}=1230$ MeV, and
$\Gamma_{a_1}^{\rm tot}=400$ MeV. Only 5$\%$ of the produced $a_1$'s
are not absorbed in the medium.  Of these unabsorbed $a_1$'s,
80$\%$ decay via $\gamma \pi$ and 20$\%$ via
$\rho \pi$. The spectral shape of $a_1 \rightarrow
\gamma \pi$ that are not absorbed (green line in Fig.
\ref{fig:a1}) is significantly different from the spectral shape
of the produced $a_1$'s (blue line in Fig. \ref{fig:a1}).  The peak is
shifted toward lower masses by about 200 MeV. Thus, the extraction
of medium effects seems feasible but requires a good understanding
of final-state absorption effects.

At RHIC-II, the $a_1 \rightarrow \gamma \pi$ decay can be measured
using the STAR full coverage TOF. In STAR, the efficiency of measuring
a conversion $\gamma$ is
5\%~\cite{AdamsPRC70}.  Therefore, 55M minimum bias Au+Au collisions
are necessary to measure a 3$\sigma$ signal (using HIJING
to estimate the background). If a Shashlyk
calorimeter measuring low-momentum $\gamma$'s could be used
in STAR, the $a_1$ efficiency could improve by a
factor of 10.

\subsection{High Statistics and Energy/Species Scan}
\label{sec:scan}

\begin{figure}[!t]
\begin{minipage}{0.48\linewidth}
\includegraphics[width=\linewidth]{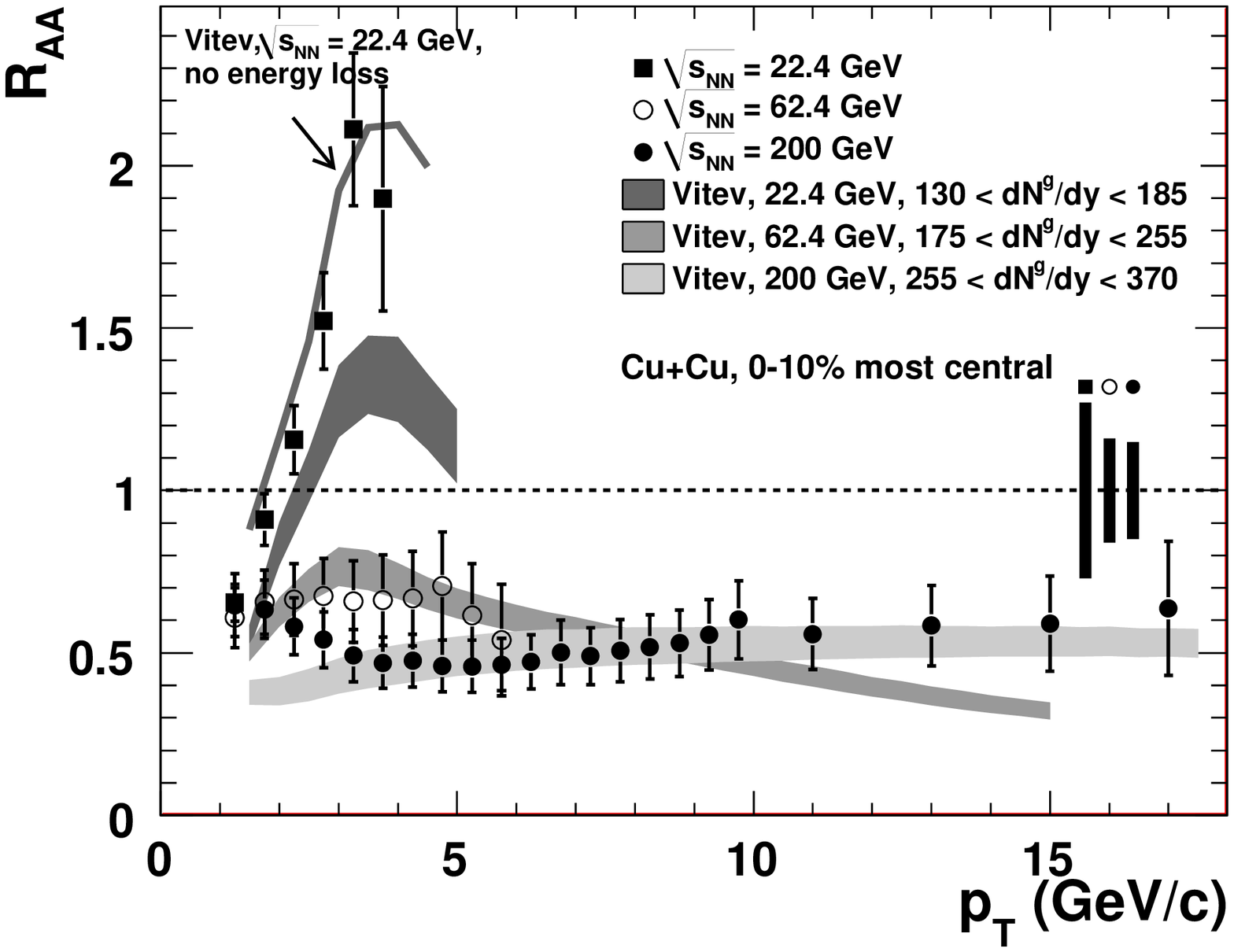}
\end{minipage}
\hspace{0.6cm}
\begin{minipage}{0.49\linewidth}
\includegraphics[width=\linewidth]{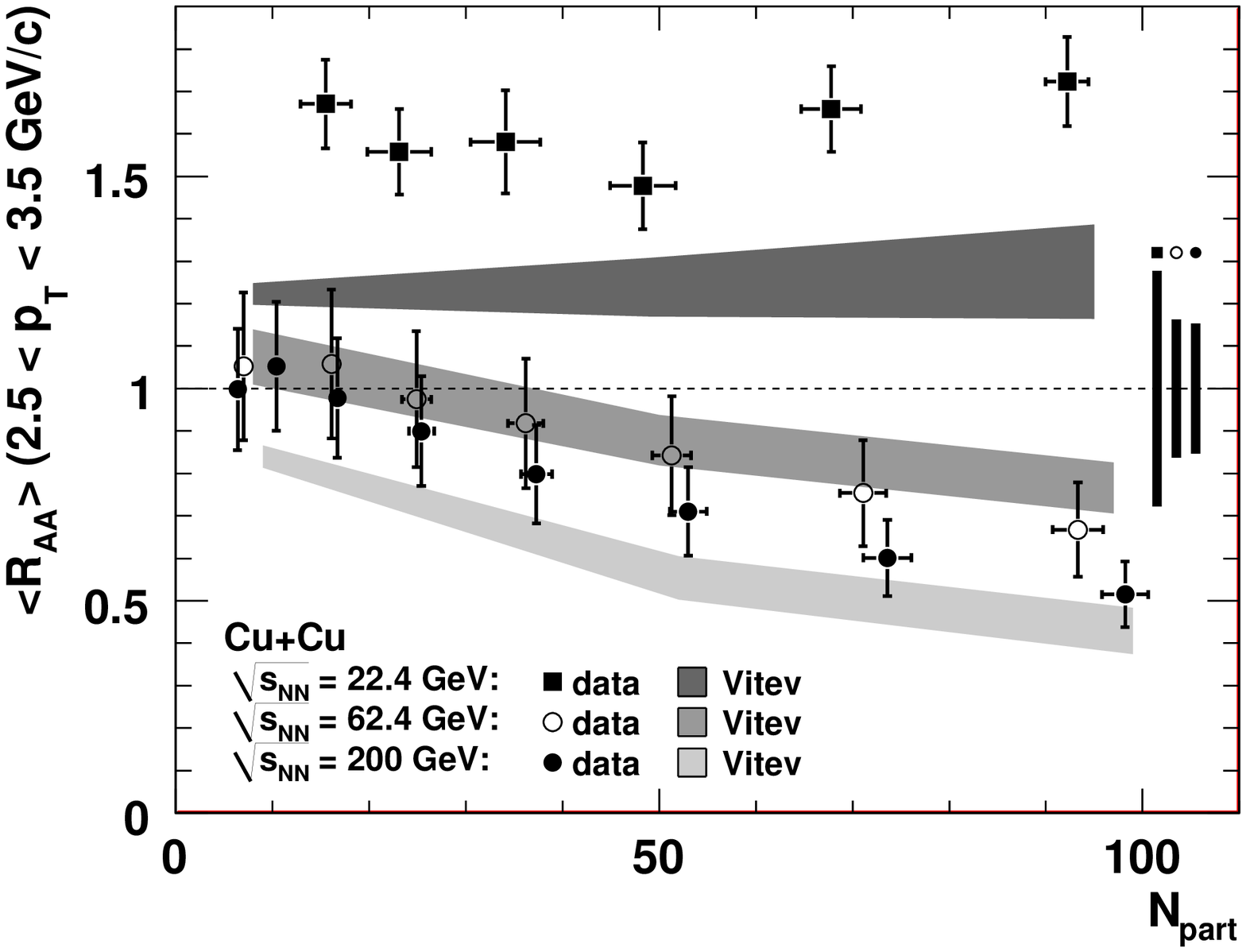}
\end{minipage}
\caption{Left: Measured $\pi^0$ $R_{AA}$ as a function of $p_\mathrm{T}$ for the
$0-10\%$ most central Cu+Cu collisions at $\sqrt{s_{NN}} =
22.4, 62.4, 200$~GeV in comparison to a jet quenching calculation.
The error bars in this figure
represent the quadratic sum of
the statistical uncertainties and the point-to-point uncorrelated and
correlated systematic uncertainties.  The boxes around unity indicate
uncertainties related to $\langle N_\mathrm{coll} \rangle$ and
absolute normalization.  The bands for the theory calculation
correspond to the assumed range of the initial gluon density
$\mathrm{d}N^\mathrm{g}/\mathrm{d}y$. The thin solid line is a
calculation without parton energy-loss for central Cu+Cu at
$\sqrt{s_{NN}} = 22.4$~GeV.
Right: The average $R_{AA}$ in the interval $2.5 < p_\mathrm{T} <
3.5$~GeV/$c$ as a function of centrality for Cu+Cu collisions at
$\sqrt{s_{NN}} = 22.4, 62.4$, and 200~GeV. The shaded bands
represent jet quenching calculations at three discrete centralities
($N_\mathrm{part} \sim 10, 50, 100$).
The boxes around unity represent the normalization
and $\langle N_\mathrm{coll}\rangle$ uncertainties for a typical
$N_\mathrm{coll}$ uncertainty of 12\,\%.
}
\label{fig:ppg084}
\end{figure}

	The RHIC-II project calls for a ten-fold increase in the luminosity
currently provided by RHIC, essential for full
exploration of rare probes and very high-$p_T$ signals that are
currently statistics-limited.  Examples include detailed heavy quarkonium
spectroscopy, more precise values and $p_T$-evolution of photon and hadron 
$R_{AA}$ at very high $p_T$, as well as photon
flow and HBT at high $p_T$.  Photon-tagged jets, particularly
if measured relative to the reaction plane, would give
further insight to energy loss mechanisms along with the complementary 
jet-dilepton correlations.  At the current luminosity,
it would take several years to collect
sufficient statistics for these measurements with just one species and
one energy setting: clearly inadequate for exploring some of the
key questions about the sQGP in depth and in a timely fashion.

There are, however, additional physics
opportunities.  The above measurements investigate the {\it properties} 
of the sQGP above $T_c$, certainly a necessary task. But, as emphasized 
in the Introduction, while the {\it existence} of a 
new state of matter has been reasonably well established at RHIC, 
key features of the phase {\it transition} itself, including
variations across the $T-\mu_B$ phase diagram, have not yet been mapped 
out.  The only way to do this is by a detailed energy/species 
scan, only possible over a reasonable time scale with RHIC-II luminosities.
The prospects are well illustrated by Fig.~\ref{fig:ppg084} which
shows $\pi^0$ $R_{AA}$ in Cu+Cu collisions for 22.4GeV (close to the
SPS range) to the highest RHIC energy, all measured by the same 
experiment (PHENIX).  The transition from $R_{AA}>1$ to strong
suppression occurs in this energy range.  Clearly this doesn't exclude
the possibility of {\it some} suppression already at SPS energies,
but the (Cronin) enhancement is still dominant - and essentially
constant - at all centralities, whereas at 62.4 GeV the suppression 
increases with centrality.

At current RHIC luminosities, even with the accelerator running for
$\sim 22-28$ weeks per year, the practical limit is at most two
species or three energies per run period.  The dominant part of
the run is spent on actual collisions and data 
taking rather than changing settings.  Since low $p_T$ observables
(bulk observables, thermal photons and dileptons, and hadron
ratios) explore the phase transition, long runs are not required to obtain
reasonable statistics for a given energy/species setting. Indeed, with 
the luminosity upgrade, sufficient data could be collected in
a few weeks at most settings, even after the luminosity loss with 
$\gamma^{-2}$ is accounted for.

The baryon chemical pontential,
$\mu_B$, can be varied by changing $\sqrt{s_{_{NN}}}$. RHIC
was designed to vary $\sqrt{s_{_{NN}}}$ and has successfully done so 
in several runs.  The
RHIC experiments have collected heavy-ion data at 
$\sqrt{s_{_{NN}}}=22$, 62, 130 and
$200$ GeV, covering several points in the $\sqrt{s_{_{NN}}}$ interval between
the SPS and the maximum RHIC energy.  Recent studies have shown that RHIC could
run collisions as low as $sqrt{s_{_{NN}}} = 5$ GeV --- the AGS range ---
to compare to previous AGS and SPS experiments, as well as fill in gaps in
earlier measurements. While this may at first appear to be a repetition 
of earlier work, we want to
re-emphasize the value of doing measurements over the entire energy range with 
{\it the very same apparatus}.  Since most systematic errors are the
same, the {\it evolution} of physics quantities characterizing the
transition can be traced to
higher precision and with more reliable systematics.

Lower energies, however, come at the price of an approximately quadratic 
luminosity decrease with $\sqrt{s_{_{NN}}}$.  Therefore,
as indicated at the end of Sec.~\ref{sec_star}, the RHIC-II luminosity
upgrade is critical for a meaningful
energy-scan program for low-mass dileptons.  For example, the RHIC-II 
luminosity at $\sqrt{s_{_{NN}}}=30$~GeV will be the same as the current
maximum-energy luminosity. Without upgrade, a factor of 40 or more luminosity
loss from $\sqrt{s_{_{NN}}} =200$~GeV
to $\sqrt{s_{_{NN}}} =30$~GeV would make an energy
scan program essentially useless for dileptons.  Based on
discussions in Secs.~\ref{sec_lmdilep}, \ref{sec_mix} and \ref{sec:na60},
a minimal $\sim 20$\% accuracy with a $\sim 2$\% mass resolution is
needed to develop sufficient discriminating power between even
rather distinct in-medium modification scenarios. Obviously,
future theoretical refinements will further raise
these requirements.
Direct photon flow measurements up to $p_T \sim 8-9$ GeV are needed to 
investigate contributions from parton
fragmentation, jet-thermal interactions and Bremsstrahlung off quarks.
Current luminosities limit the measurements to $\sim 5-6$ GeV,
see Sec.~\ref{sec:photon_elliptic}.  A tenfold luminosity increase
would extend the $p_T$-range by the necessary 2-3\,GeV.
Interferometry of pre-hadronic photons (emitted before the hadronic phase),
$p_T>2$ GeV~\cite{Bass04}, with the precision deduced in
Sec.~\ref{sec:gamma_hbt} requires about $7 \times 10^8$ events which would,
with current luminosities, allow measurements with only one system and/or
energy per year.  Also, many of the intermediate invariant mass and
transverse momentum, $M,p_T\sim 2$~GeV, signals, where sQGP radiation 
dominates (see Secs.~\ref{sec_kin} and \ref{sec_reso}),
can be triggered, making high luminosity extremely valuable
in this context as well.
Photon-based temperature measurements, described in Sec.~\ref{sec_denterria},
depend on detailed energy and species scans, including sufficient $pp$ 
reference data at the same energies as $AA$ collisions instead of relying on 
earlier results from other 
accelerators~\footnote{See Sec.~\ref{sec:photonraa} for an example of the
difference this might make.}.
Once again, the benefits of measuring
both spectra in the same detector are obvious: in many cases the systematic
errors decrease by $40-50$\%.

Colliding different ion species allows variation of the collision 
geometry at fixed participant number, $N_{\rm part}$.  
So far at RHIC several bulk observables
have been found to scale with $N_{\rm part}$
irrespective of collision geometry.  Recently it has been found
that even more subtle quantities such as the azimuthally-integrated nuclear
modification factor, $R_{AA}$, is similar in central Cu+Cu
collisions and mid-peripheral Au+Au collisions with the same $N_{\rm part}$.  
However, once azimuthal distributions are
studied, significant differences
emerge since the two collision geometries are not alike
(spherical Cu+Cu {\it vs} ellipsoidal Au+Au) and these different geometries 
are crucial for understanding processes such as in-medium energy loss.  Also, 
for sufficiently light ions, thermalization, apparent in the hydrodynamic 
behavior of the Au+Au and even Cu+Cu data, can no longer be achieved. 
Pinpointing the thermalization boundary could be crucial for determining the
behavior of probes near the phase boundary and the conditions needed for the
transition to occur.

We reiterate that higher luminosities can be used to
increase the statistics of a particular data set,
gaining access to rare probes, and also to increase the number
of ion species (symmetric or asymmetric collision systems) and/or the energy
settings by subdividing the RHIC runs since many
important signals do not require {\it very} high statistics. The
accelerator is the most flexible ever built and can run multiple system/energy
combinations per year if the physics justifies it.  In fact, RHIC
produced the very first collisions at $\sqrt{s_{_{NN}}}=9$GeV in a
12 hour test in June 2007.

\subsection{From Observables to Pertinent Theoretical Issues}

\begin{table}[!t]
\begin{centering}
{\tiny
  \begin{tabular}{|r|r|r|r|r|r|} \hline
Measurement & Signal    & Physics ques-  & Open issue & Open issue &
            Discussion \\
            & extracted & tion addressed & (exp)      & (theory) & 
	    or reference \\ \hline \hline
	    very low mass $e^+e^-$ & $dN/p_T$ & 
     direct/thermal $\gamma$ & normalization & sole source  & \\
                      &          & 
     (internal conv.)      & $pp$ reference   & of $e^+e^-$ ? & 
     Sec.~\ref{sec:internalconv} \\ \hline
     low mass     $l^+l^-$ & $dN/dM$ & many-body   & Dalitz  & 
     HTL {\it vs} lattice 
     &  \\ 
                      &            & effects  & rejection & 
     {\it vs} hadronic & Sec.~\ref{sec_lat} \\ \hline
     $\omega,\rho,\phi \rightarrow l^+l^-$ & mass
       & chiral  & Dalitz & mass shift?  
     & Secs.~\ref{sec_sps},~\ref{sec_frame},~\ref{sec_had} \\
     & width       & restoration  & rejection & broadening?  
     & \ref{sec_drop}, \ref{sec_lmdilep}, \ref{sec_csr},
       \ref{sec:na60} \\ \hline
       jet$-l^+l^-$  & yield {\it vs}  & medium      & rates     &
     energy loss  & Sec.~\ref{sec:scan} \\
       correlation   & reaction plane     & properties  & $pp$ ref. &
     mechanism   &  \\ \hline
     intermediate  & spectral & bound states / 
     & rates & degrees of freedom  & Sec.~\ref{sec_kin} \\ 
     mass dileptons & function & resonances above $T_c$  
     &       & temperature & \\ \hline
     $l^+l^-$ $q_t$ spectra,  & spectral & emission profile& rates &   
     thermalization        & Sec.~\ref{sec_mix} \\
     $v_2$                   & slopes & chiral mixing &       & 
     high-$q_t$ sources    & \\ \hline
     $\phi,\rho \rightarrow
     h^+h^-$ {\it vs} & BR {\it vs} system & all {\it vs} late     
     & low $e^+e^-$ & medium  & Sec.~\ref{sec:phi1020} \\
    $e^+e^-$ decays   &  centrality        & stages of coll.  & BR
     & modification   & \\ \hline
     open charm & yield & $E_{\rm loss}$ mechanism  & displaced & contamination
     & Sec.~\ref{sec:charm} \\
     & flow & thermal equilibration  & vertex & from $b$ decays & \\ \hline \hline
     low $p_T$ $\gamma$ & $T$ vs $d_{\rm of}$ & HG vs QGP & $\pi^0$ bkgd &
     flow/$T$  & Secs.~\ref{sec_denterria}, \ref{sec:internalconv}\\
                   &                &            & $\overline{n}$, $h^{\pm}$
     &     ambiguity & \\ \hline
     HBT $p_T^\gamma<1$ GeV/$c$ & size, lifetime &
     source radius $R(t)$ & resolution & other sources?  & 
     Sec.~\ref{sec:gamma_hbt} \\
      & low $p_T$ $\gamma$ yield &     & & & Ref.~\cite{wa98-lowpt} \\ \hline
     medium $p_T$ $\gamma$ & yield, $R_{AA}$ & jet-medium  & 
     $pp$ reference & jet conversion, & Sec.~\ref{sec:photonraa} \\
       &  & interactions   & & Bremsstrahlung & \\ \hline
     jet tagged $\gamma$ & assoc. yield & medium prop.     &
     rates,species & jet conversion, & Sec.~\ref{sec:scan} \\
                   & assoc. yield & $E$-scale, $k_T$  &
     acceptance    & Bremsstrahlung & \\ \hline
     high $p_T$ $\gamma$
     & spectrum & pQCD, scale & rates & fragmentation & Sec.~\ref{sec:scan} \\
     $\gamma$-jet & & jet energy
     scale & $AA$ {\it vs} $pp$ & & \\ 
     & & tomography, $k_T$ & & & \\ \hline
     HBT ($p_T^\gamma>2$ GeV) & size, lifetime & pre-equilibrium & rates, & hadronic
     & Sec.~\ref{sec:gamma_hbt}               \\
     &                & size         & low $q$ &
     dominates?  & \\ \hline
     all $p_T$ $\gamma$ flow & sign, strength &
     thermal/jet rad.  & rates & relative & Sec.~\ref{sec:photon_elliptic}\\
     (azimuthal asymm.) &             &
                  & isolation cut? & strength & \\ \hline
     $\gamma \pi$ corr. & $a_1$ mass & chiral restoration  & 
     small BR & final state &      Sec.~\ref{sec:a1} \\ 
      &  &   & 
     & absorption? &  \\ \hline
     all $\gamma$ at high  & $\gamma$-jet & CGC & forward  & 
     common $\gamma$, $\pi^0$ & Sec.~\ref{sec:ncc} \\
     rapidity             & $\chi_c$ &  & calorimetry & suppression? & \\ \hline \hline
\end{tabular}
}
\caption{ { \label{tb:signals}}   Bottom-up view of electromagnetic
probes in heavy-ion collisions: 
from observables to the
pertinent theoretical issues addrssing the underlying physics questions. }
\end{centering}

 \end{table}

In Table~\ref{tb:signals} we summarize how electromagnetic
probes can influence the theoretical understanding of relativistic heavy-ion
collisions.  In the first column we list specific measurements,
in the second the signal extracted either directly or with minimal model
assumptions, and the physics question addressed in the third column.  
The fourth column describes the most important
experimental issues and the fifth indicates open or ambiguous theory questions
that require significant further investigation.
Finally, the sixth column points either to a Section of this review or to a
reference where the issues are discussed in more detail.

While this table has a level of
subjectivity, it reflects our current thinking on the most
interesting issues related to EM probes at RHIC-II.  We do not list topics 
that are primarily or exclusively the domain of the LHC or the 
future Electron-Ion Collider (EIC).


\section{SUMMARY AND CONCLUDING REMARKS}
\label{sec_concl}
We now summarize our opinion of the most promising
developments in connection with EM probes at future RHIC experiments.

{\em Low-mass (axial-) vector-meson spectroscopy close to the chiral
transition}\\
Precision dilepton data will provide detailed information on medium
modifications of $\rho$, $\omega$ and $\phi$ mesons, and thus
illuminate the question of hadronic mass (de-)generation. ``Quality
control" of theoretical models via independent constraints from
symmetries, QCD sum rules and phenomenology is essential to limit the
scope of viable axial/vector ($A$/$V$) spectral functions.  An excitation
function measurement will discriminate temperature and net-baryon density 
effects to systematically map out in-medium effects
across a significant part of the QCD phase diagram. Information on
chiral symmetry should be inferred from a novel
measurement of the axial vector spectral function via $\pi^\pm \gamma$
invariant-mass spectra as well as through a well-defined combination
of effective chiral hadronic models, chiral sum rules and
finite-temperature QCD lattice computations of chiral order parameters such as
the pion decay constant and four-quark condensates.  

The required
theoretical tools are largely in place: chiral effective models for
realistic axial-/vector spectral functions at finite temperature can
be used to calculate the temperature dependence of order parameters
(moments of ``$V-A$" spectral functions).  Unquenched
lQCD evaluations of the latter should be pursued with high priority.
A convincing deconvolution of the vector correlator from the measured
spectra will require reliable space-time descriptions of
$AA$ collisions, expected to emerge from progress in
hydrodynamical simulations and complementary transport models. 

Footprints of chiral restoration are furthermore expected at dilepton
masses around the $a_1$ mass, $1-1.5$~GeV, due to ``chiral $V-A$
mixing".  To detect a continuum enhancement of less than a factor of
two requires accurate charm and background determinations providing
a signal with no more than 20\% total error, achievable with
the planned vertex-detector upgrades together with sufficient
statistics. In addition, dilepton elliptic flow and transverse momentum 
spectra will provide additional information on the space-time 
emission profile at different invariant masses.

{\em The highest temperatures of the matter formed at RHIC} \\
We have identified three promising regimes where the QGP is expected to 
be the dominant source of electromagnetic
radiation:
intermediate-mass dileptons ($1.5\leq M \leq 3$~GeV) and photons and
low-mass dileptons at intermediate transverse momentum, ($1 \leq q_t \leq
3$~GeV).
These measurements will be supplemented by $\gamma \gamma$ correlation
analyses which yield complementary information on the temperature and 
fireball size.

{\em QGP resonances} \\
The only direct way to experimentally determine the existence of hadronic bound
states/resonances in the sQGP is to search for a
resonant dilepton signal. Due to the nature of QGP emission, the
greatest sensitivity to pertinent vector states is in
the $M=2$~GeV mass region,  coinciding with current expectations
from lattice QCD and effective models. RHIC  provides optimal
conditions for this search since the initial temperatures are close to
the anticipated bound state dissolution temperatures.  

As is
the case for signatures of chiral mixing at lower masses,
an accurate determination of background and charm sources is
mandatory to restrict the total experimental error to below the $\sim 20$\%
level. If the resonance states exist at masses below $\sim 1.5$~GeV,
discriminating them from chiral-mixing effects will be more
involved, increasing the demand for accuracy and systematic
centrality and excitation-function studies.

{\em Detector and luminosity requirements} \\
To achieve the required background rejection and obtain high-precision charm
measurements, PHENIX needs the Hadron Blind Detector and vertex detector,
while STAR needs the Time-of-Flight detector, Heavy Flavor Tracker and Data
Acquisition System. Initial dilepton measurements can be made after these 
new detectors are in place. However, an
energy scan with statistics comparable to recent NA60 data
requires significant luminosity upgrades.
%

A summary is provided in Table~\ref{tb:signals}.

Based on the arguments given in this review, we believe that after the 
future RHIC detector and luminosity upgrades for heavy-ion collisions in the
$10 \leq \sqrt{s_{_{NN}}} \leq 200$~GeV regime,
electromagnetic probes will, in a combined experimental
and theoretical effort, result in decisive and unique new
perspectives of QCD matter at high (energy-) density.  In particular, new
insights into hadronic mass generation mechanisms related to chiral symmetry
restoration, thermal radiation at unprecedented
temperatures, microscopic properties of the sQGP and system size measurements
during the early matter evolution will be gained.

\section*{Acknowledgments}
We thank our colleagues J.~Casalderrey-Solana,
H.~van Hees, B.~M\"uller, P.~Petreczky, T.~Renk, J.~Ruppert, K.~Schweda,
R.~Seto, R.~Vogt, and C. Woody for valuable input and discussions.
We are particularly indebted to D.~d'Enterria for providing us
with Sec.~\ref{sec_denterria} on ``Direct Photons and Current RHIC Data'',
to J.~Jalilian-Marian for Sec.~\ref{sec_cgc} on ``Electromagnetic Signatures
of the Color Glass Condensate'', to J.~Sandweiss and his collaborators
for providing us with Sect.~\ref{sec:gamma_hbt} on
``Direct Photon Correlations" of this mansucript 
and to P.~Fachini for Sect.~\ref{sec:a1} (their write-ups have
been edited by the three authors to fit style and scope of the manuscript).
RR has been supported in part by a National Science Foundation CAREER
Award under grant PHY-0449489.  XZB has been supported in part by a
DOE Early Career Award and the Presidential Early Career Award for
Scientists and Engineers.


\end{document}